\documentclass[11pt,a4paper]{article}
\pdfoutput=1
\usepackage{jheppub}
\usepackage{enumitem}
\usepackage{amsmath}
\usepackage{amsfonts}
\usepackage{amssymb}
\usepackage{mathtools}
\usepackage{physics}
\usepackage{graphicx}
\usepackage{caption}
\usepackage{tensor}
\newcommand{\elec}{\mathcal{E}}
\newcommand{\p}{\begin{pmatrix}
1\\1
\end{pmatrix}}

\newcommand{\n}{\begin{pmatrix}
1\\-1
\end{pmatrix}}

\def\e{\mbox{{\rm e}}}
\pdfoutput=1

\title{Edge State Quantization: Vector Fields in Rindler}

\author[a]{Andreas Blommaert}
\author[a,b]{,Thomas G. Mertens}
\author[a]{, Henri Verschelde}
\author[c,d]{and Valentin I. Zakharov}
\affiliation[a]{Ghent University, Department of Physics and Astronomy\\
Krijgslaan, 281-S9, 9000 Gent, Belgium,}
\affiliation[b]{Joseph Henry Laboratories, Princeton University, Princeton, NJ 08544, USA} 
\affiliation[c]{ITEP, B. Cheremushkinskaya 25, Moscow, 117218 Russia,}
\affiliation[d]{School of Biomedicine, Far Eastern Federal University, Sukhanova str 8, 
Vladivostok 690950 Russia}

\emailAdd{andreas.blommaert@ugent.be}
\emailAdd{thomas.mertens@ugent.be}
\emailAdd{henri.verschelde@ugent.be}
\emailAdd{vzakharov@itep.ru}

\abstract{We present a detailed discussion of the entanglement structure of vector fields through canonical quantization. We quantize Maxwell theory in Rindler space in Lorenz gauge, discuss the Hilbert space structure and analyze the Unruh effect. As a warm-up, in $1+1$ dimensions, we compute the spectrum and prove that the theory is thermodynamically trivial. In $d+1$ dimensions, we identify the edge sector as eigenstates of horizon electric flux or equivalently as states representing large gauge transformations, localized on the horizon. The edge Hilbert space is generated by inserting a generic combination of Wilson line punctures in the edge vacuum, and the edge states are identified as Maxwell microstates of the black hole. This construction is repeated for Proca theory. Extensions to tensor field theories, and the link with Chern-Simons are discussed.}
\begin{document}
\maketitle
\section{Introduction}
\label{s intro}
One of the most important open problems in black hole physics is to provide for an understanding of black hole entropy in terms of microscopic degrees of freedom. Assuming that string theory is the fundamental theory of nature, the microscopic degrees of freedom responsible for black hole entropy must be of a stringy nature \cite{Susskind:1994sm}. Within the current understanding of entropy, it should be understood as arising from stringy entanglement across the horizon of the black hole.

An explicit calculation of the horizon entanglement entropy in string theory is complicated by two big obstacles.
\begin{itemize}
\item One would need to understand the string spectrum in a black hole background, or its near-horizon Rindler limit. So far the nonlinear nature of the string equations of motion in Rindler has stood in the way of an explicit quantization. 
\item Strings do not factorize across the horizon, making it highly nontrivial to define an entanglement entropy (see however work based on string field theory \cite{Balasubramanian:2018axm,Hata:1995di}). 
\end{itemize}
It seems at this point in time next to impossible to quantitatively address this nonfactorization problem directly in string theory (see however \cite{Donnelly:2016jet}). The same complication occurs in gauge theories and possibly also in generic QFTs. Understanding this feature in gauge theories has been studied intensively the last few years, mainly from a lattice perspective, see e.g. \cite{Buividovich:2008gq,Casini:2013rba,Casini:2014aia,Radicevic:2014kqa,Radicevic:2015sza,Radicevic:2016tlt,Donnelly:2014gva,Donnelly:2011hn,Donnelly:2012st,Huang:2014pfa,Ghosh:2015iwa,Hung:2015fla,Aoki:2015bsa,Soni:2015yga,VanAcoleyen:2015ccp,Casini:2015dsg,Soni:2016ogt,Michel:2016qge}. In this work we tackle this problem head-on and construct the Hilbert space of Maxwell and Proca theory in Rindler space. Several aspects of quantization of Maxwell theory have appeared in the literature before \cite{Crispino:2007eb,Higuchi:1992td,Crispino:2000jx,Moretti:1996zt,Lenz:2008vw}.
\\~\\
The lattice perspective on the issue gives rise to different possible definitions of entanglement. We adopt the most popular definition of entanglement entropy as replica trick entropy. It is our understanding that Euclidean replica trick calculations should be considered as guidelines to an unambiguous state counting interpretation of thermodynamics (i.e. a trace over a yet-to-be-determined Hilbert space). In recent years, many of these Euclidean calculations have been performed, both in QFT as in string theory \cite{Mertens:2013pza,Mertens:2014cia,Mertens:2014nca,Mertens:2014dia,Mertens:2015hia,Solodukhin:2011gn,Solodukhin:2015hma,He:2014gva,Mertens:2015adr,Mertens:2016tqv}.

For Maxwell theory, Kabat found a contribution to the partition function due to a coupling of non-scalar fields to the conical singularity introduced in the Euclidean calculation: the contact term. A negative sign in the regularized entropy led him to the statement that this term cannot come with a state counting interpretation \cite{Kabat:1995eq,Kabat:1995jq,Kabat:2012ns}. Donnelly and Wall have shown that there \textit{does} exist a statistical interpretation for the contact term as counting classical static backgrounds with nonzero horizon electric flux \cite{Donnelly:2014fua,Donnelly:2015hxa}. 

As part of this work, we lift this statistical interpretation to a genuine Hilbert space tracing computation. The classical backgrounds are in fact electric flux edge states created by large gauge transformations which are (isomorphic to) Wilson line punctures on the entangling surface. Wilson lines are the non-local gauge invariant observables that make up the physical algebra of Maxwell theory. Their non-locality is at the foundation of the non-factorization issues associated with the Maxwell Hilbert space. We find the correct algebra by investigating the effect on the canonical structure of inserting a dividing surface in flat space. At this point a direct link with 3d Chern-Simons theory can be made. \\
The resulting boundary canonical structure is the same as the asymptotic gauge symmetry - soft photon construction of Strominger et al. \cite{Hawking:2016msc,Hawking:2016sgy,Lust:2017gez}; the sole difference being a different choice of Cauchy surface (we will be working on surfaces of constant time i.e. spatial Cauchy surfaces), and is also identical to the boundary symplectic potential discussed by (among others) Donnelly and Freidel \cite{Donnelly:2016auv}. The edge states are identified as the Maxwell microstates of (large non-extremal) black holes. In explicitly linking these works together by an edge state quantization, we intend to fill a gap in the existing literature.

The generality of our procedure allows for multiple extensions. We discuss in full Proca theory and the state counting interpretation of its contact term. Though our procedure applies equally to Proca as it does to Maxwell, the interpretation of the edge sector is different: Proca has no gauge freedom and thus the concept of large gauge transformations creating the edge sector does not uphold here. Extensions to generic tensor fields are discussed without going into too much detail.
\\~\\
We will find that in quantizing Maxwell, a distinction must be made between $1+1$ dimensional Rindler and $d+1$ dimensional Rindler as the respective zero-mode sectors (which we will show to be the classical origin of the respective edge sectors) behave rather differently. In $1+1$, the zero-mode sector of the theory is all that remains in finite spatial volume, and we study it somewhat further in our framework. In particular we show that the edge states of the theory in $1+1$ dimensions are not normalizable, and as such cannot contribute to thermodynamic quantities. From the Wilson loop perspective this is readily appreciated: in $1+1$ dimensions, at fixed time, there are no non-trivial Wilson loops as they would have to retrace their steps in order to close. This is in unison with the well-known fact that $1+1$ dimensional CFTs have well-known formulae for the entanglement entropy, irrespective of the possible presence of a gauge symmetry. 

Before discussing these important aspects of Maxwell theory in Rindler, we first scrutinize an issue that arose in the literature for the $1+1$ case. Upon quantizing $1+1$ dimensional Maxwell theory in Rindler space in e.g. Weyl gauge or Coulomb gauge it is readily found that the theory is trivial due to a lack of degrees of freedom. Canonical quantization in Lorenz gauge however has proven to be somewhat more subtle. In \cite{Zhitnitsky:2010ji}, Zhitnitsky quantizes a theory similar to gauge-fixed $U(1)$ theory in Lorenz gauge, and finds that temporal and timelike photon polarizations, instead of canceling out, add up in the Unruh effect. This idea was later on picked up in \cite{Ohta:2010in,Urban:2009yg} in a cosmological context to model dark energy. This statement is in direct conflict with gauge invariance. We commence this work by presenting a canonical quantization of Maxwell theory in $1+1$ dimensional Rindler in Lorenz gauge and demonstrate that the Faddeev-Popov ghosts exactly cancel the two unphysical photon polarizations, as they should by construction. 
\\~\\
The concrete objectives of this work are subdivided into two main goals.
\begin{itemize}
\item Canonical quantization of Lorenz gauge Maxwell and Proca theory in $1+1$ dimensional Rindler space. For Maxwell, discuss the non-physical polarizations and ghosts and prove that they cancel out thermodynamically, except for the zero-mode sector.
\item Canonical quantization of Lorenz gauge Maxwell and Proca theory in $d+1$ dimensional Rindler space, construct the Hilbert space, including edge states and give physical meaning to this sector.
\end{itemize}

\begin{figure}[h]
\centering
 \includegraphics[width=0.35\textwidth]{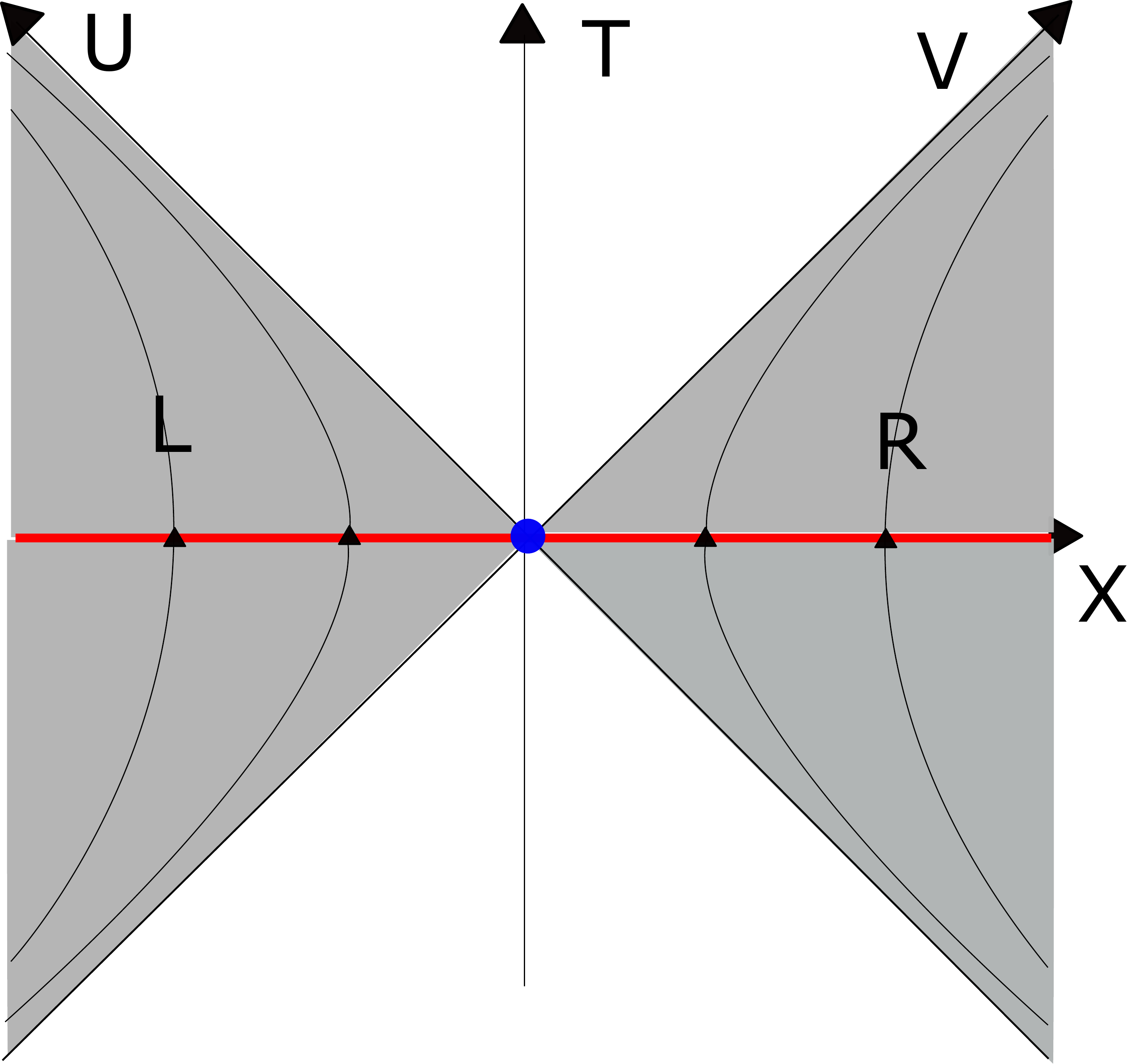}
\caption{Rindler wedges and coordinate systems. At $T=0$, the Minkowski Hilbert space splits into the left and right Hilbert space. The Rindler observer then evolves these states using $H_R$ along the hyperbolic trajectories. This is one way of making sure the right observer does not see anything originating from the left wedge. Likewise, the left Hilbert space evolves upwards with $H_L$.}
\label{rindlerwedges}
\end{figure}

In the sections concerning $1+1$ dimensional Rindler space, we mainly work in Rindler tortoise coordinates $(t,r)$. These are related to flat coordinates $(X,T)$ as 
\begin{equation}
ds^2= -dT^2 + dX^2 = -dUdV = e^{2r}\left(-dt^2+dr^2\right).
\end{equation}
where $U=T-X$ and $V=T+X$. The Rindler tortoise coordinates are conformally flat and thus especially convenient for studying wave equations in $1+1$ dimensions. In the R-wedge, they are related to the light cone coordinates $(U,V)$ as $U=-e^{r-t}$ and $V=e^{r+t}$, where $U<0$ and $V>0$.\footnote{Analogously, one uses $U=+e^{r-t}$ and $V=-e^{r+t}$ for the region $U>0$, $V<0$, which is the left Rindler wedge.} The Rindler horizon is located at $V=0$, or $r=-\infty$ in these coordinates (Figure \ref{rindlerwedges}).

In the sections concerning $d+1$ dimensional Rindler space, we mostly work in Rindler coordinates $(t,\rho,\mathbf{x})$. The radial coordinate $\rho$ is related to the tortoise coordinate $r$ as $\rho=e^r$. The $d-1$ flat directions are parameterized by $\mathbf{x}$. The metric is now $ds^2=-\rho^2dt^2+d\rho^2+d\mathbf{x}^2$ and the horizon is the hypersurface $\rho=0$.
\\~\\
This paper is organized in two main parts. The first part consists of sections \ref{s gauge} to \ref{s:edge11} and deals with Maxwell theory in $1+1$ dimensional Rindler space. In section \ref{s gauge} we discuss the canonical quantization, including longitudinal and temporal photon polarizations and calculate the Unruh effect. We pay special attention to the ghost sector and resolve an issue in work by Zhitnitsky. Section \ref{s:edge11} is a discussion on the edge sector that arises when gluing two Rindler wedges. We investigate the Minkowski ground state and conclude that edge states do not make a contribution to it, nor to thermodynamical quantities.

In the second part (consisting of sections \ref{s: action} and \ref{s:boundary and edge}) we present the detailed quantization of Maxwell theory in $d+1$ dimensional Rindler space. We discuss the canonical quantization of the theory in Lorenz gauge (section \ref{s: action}) including suitable boundary conditions. Section \ref{s:boundary and edge} contains the canonical quantization of the edge sector and the resulting thermodynamic quantities. In section \ref{s:extension} we present generalization to Proca theory and extensions to tensor field theories. Finally, in \ref{s:conclusion} we conclude on our findings and look ahead towards strings.
\\~\\
A complementary path integral perspective on the problem is given in \cite{Blommaert:2018oue}, where it is also generalized beyond the Abelian Rindler set-up.

\section{Edge States}
\label{s: hilbertspaces}
This section provides a summary of the main ideas followed in this work concerning the introduction and quantization of edge states, and the link with the contact term for bosonic QFTs.
\subsection{Origin of Edge States}
\label{ss:2.1}
Consider a generic classical field theory in the Hamiltonian formulation, characterized by a set of evolution equations (containing second-order time derivatives) and constraint equations (linking initial data on a Cauchy surface $\Sigma$). The latter decrease the number of independent canonical variables on $\Sigma$

Assume now there is a dividing surface separating $\Sigma$ into $\Sigma_1$ and $\Sigma_2$. If the constraints contain no spatial derivatives, i.e. they are ultralocal, then the Hilbert space factorizes across this dividing surface. Otherwise, this division introduces a matching constraint across the boundary, with associated edge degrees of freedom.
Moreover, from the point of view of an observer constrained to data from $\Sigma_1$ (i.e. observers restricted to $\Sigma_1$ resp. $\Sigma_2$ are causally disconnected), this boundary is an infinite redshift surface. Hence such an observer has no access to the boundary degrees of freedom. This in turn shows that the Hilbert space on $\Sigma$ does not factorize in the Hilbert spaces of observers restricted to $\Sigma_1$ resp. $\Sigma_2$. This ultimately leads to a contact term in the theory. Thus, \textit{a theory with generic, i.e. not ultralocal, constraint equations has edge states.} 

As an example, consider a scalar field with non-minimal coupling to the Ricci-scalar $\mathcal{L} = \frac{1}{2}\partial_\mu \phi \partial^\mu \phi + \xi R \phi^2$, which has a contact term $\sim \xi$. From the Lorentzian perspective, the scalar field is not a constrained Hamiltonian system. Therefore, the scalar has no edge states. The contact term for the non-minimally coupled scalar field is well known to be interpreted as a contribution to the generalized entropy, not to the entanglement entropy (which is manifestly independent of the choice of coupling). \\

From hereon we focus of Lorentzian flat space $\mathbb{R}^{1,d}$ with a planar dividing surface: Minkowski is separated into R-and L Rindler wedges. The advantage of investigating this setup is that it is easy to diagonalize the modular Hamiltonian of either region i.e. the Lorentz boost generator.

Constraint equations on the initial value formulation are translated into boundary contributions to $\delta S$, resulting in boundary conditions. Indeed, a general local Lagrangian $\mathcal{L}(A_\mu, \partial_\mu A_\nu)$ of a spin 1 field $A_\mu = A^L_\mu + A^R_\mu$ in Minkowski, splits into left (L) and right (R) parts only supported on their respective wedges, leads to the \emph{matching constraints} at the dividing surface $\partial \Sigma$:
\begin{equation}
\boxed{n_\mu \left.\frac{\partial \mathcal{L}}{\partial \partial_\mu A^R_\nu}\delta A^R_\nu\right|_{\partial \mathcal{M}}(\mathbf{x})=n_\mu \left.\frac{\partial \mathcal{L}}{\partial \partial_\mu A^L_\nu}\delta A^L_\nu\right|_{\partial \mathcal{M}}(\mathbf{x})\label{2.7}},
\end{equation}
with $\partial \mathcal{M}=\partial \Sigma \cross \text{time}$.

Similarly, from the point of view of an observer constrained to either wedge, variation of the action results in the boundary conditions:
\begin{equation}
n_\mu \left.\frac{\partial \mathcal{L}}{\partial \partial_\mu A^R_\nu}\delta A_\nu^R\right|_{\partial \Sigma}(\mathbf{x})=0.\label{2.5}
\end{equation}
A theory satisfies these boundary conditions if for instance it satisfies perfect electric conductor (PEC) or perfect magnetic conductor (PMC) boundary conditions:
\begin{alignat}{5}
\label{PEC}
&\text{PEC:}\,\, &&A_\nu^R\rvert_{\partial \mathcal{M}}(\mathbf{x})=0,\quad &&\nu=i,t, \quad\quad\quad &&n_\mu \left.\frac{\partial \mathcal{L}}{\partial \partial_\mu A^R_\rho}\right|_{\partial \mathcal{M}}(\mathbf{x})=0,\\
\label{PMC}
&\text{PMC:}\,\, &&n_\mu \left.\frac{\partial \mathcal{L}}{\partial \partial_\mu A^R_\nu}\right|_{\partial \mathcal{M}}(\mathbf{x})=0,\quad &&\nu=i,t, \quad\quad\quad &&A_\rho^R\rvert_{\partial \mathcal{M}}(\mathbf{x})=0.
\end{alignat}
The R-wedge bulk theory is defined to be a theory that solves the bulk EOM as well as obeys a consistent set of boundary conditions such as \eqref{PEC} or \eqref{PMC}. 
Afterwards one sums over boundary contributions, supplemented with the gluing condition \eqref{2.7}, to obtain the general solution of the problem in flat space.
It is interesting to compare this to the electromagnetic membrane paradigm, where boundary contributions are cancelled by horizon charges and currents \cite{Parikh:1997ma}. The logic is related, but not precisely the same. \\

An explicit example and the starting point of our discussion is the gauge-fixed version of free Maxwell theory in $d+1$ dimensions in Lorenz gauge:
\begin{equation}
S_M=\int_{\mathcal{M}} d^{d+1}x \sqrt{-g}\left(-\frac{1}{4}F^{\mu\nu}F_{\mu\nu}-\frac{1}{2}\nabla_\mu A^\mu \nabla_\nu A^\nu - \partial^\mu \bar{c} \partial_\mu c \right)\label{action 1},
\end{equation}
where $\mathcal{M}$ can either refer to the full Minkowksi space (with no relevant boundary for our purposes) or to either one of the Rindler wedges (with boundary the horizon $\partial \mathcal{M}$). This action is only properly defined on the subspace of field configurations satisfying Lorenz gauge 
\begin{equation}
\nabla^\mu A_\mu =0\label{2.2}.
\end{equation}
In a more general context this is referred to as the transversality constraint on $A$. This first class constraint equation will be imposed in the Gupta-Bleuler sense as a constraint on physical states in the theory. Imposing it at a classical level would be inconsistent with canonical quantization. Note that this transversality constraint appears as a second class constraint for the massive vector field (see section \ref{s:extension}), and is part of the Virasoro constraints in string theory at each level. For the moment, the Faddeev-Popov ghost action will be ignored, we come back to it in section \ref{s ghost}. \\

Variation of the action on $\mathcal{M}$ results in
\begin{equation}
\delta S_M=\int_{\mathcal{M}} d^{d+1}x\sqrt{-g}\,(\nabla_\mu \nabla^\mu A^\nu) \delta A_\nu + \int_{\partial \mathcal{M}} d^{d}x \sqrt{-h}\, n_\mu\left(F^{\mu\nu}+ g^{\mu\nu}\nabla^\sigma A_\sigma\right)\delta A_\nu,\label{action 3}
\end{equation}
with $n_\mu$ the inward pointing vector on $\partial \mathcal{M}$, and $h_{\mu\nu}$ the induced metric on $\partial \mathcal{M}$. Setting the variation of the action to zero, the first term in \eqref{action 3} results in the set of coupled equations:
\begin{equation}
\nabla^\mu\nabla_\mu A_\nu =0.\label{2.4}
\end{equation}
The boundary term is set to zero by choosing the boundary condition \eqref{2.5}. Within this gauge-fixed formulation, the Lorenz gauge \eqref{2.2} is a constraint on the initial data of the system (as in the beginning of this section). On solutions to the equations of motion, the Lorenz gauge constraint is equivalent to Gauss' law 
\begin{equation}
\nabla^\mu F_{\mu t}=0,\label{2.9}
\end{equation}
which is usually stated as the constraint on the Maxwell system. \\

On the horizon $\partial \mathcal{M}$, the constraints \eqref{2.2} and \eqref{2.9} become matching conditions from the Minkowski perspective:
\begin{equation}
n^\mu A_\mu^R\rvert_{\partial \mathcal{M}}(\mathbf{x})=n^\mu A_\mu^L\rvert_{\partial \mathcal{M}}(\mathbf{x}), \quad \text{and} \quad n^\mu F_{\mu t}^R\rvert_{\partial \mathcal{M}}(\mathbf{x})=n^\mu F_{\mu t}^L\rvert_{\partial \mathcal{M}}(\mathbf{x}).\label{2.8}
\end{equation}
From the point of view of an R-wedge observer however, the constraints \eqref{2.2} and \eqref{2.9} fix $n^\mu A_\mu^R\rvert_{\partial \mathcal{M}}(\mathbf{x})$ and $n^\mu F_{\mu t}^R\rvert_{\partial \mathcal{M}}(\mathbf{x})$ to a fixed value. The bulk R-wedge theory does not include the edge degrees of freedom \eqref{2.8}. These edge DOF are responsible for entanglement across the horizon that is not captured by the bulk R-wedge theory.

The natural way to implement these matching boundary conditions is to first set them to zero: $n^\mu A^R_\mu\rvert_{\partial \mathcal{M}}(\mathbf{x})=0$ and $n^\mu F^R_{\mu t}\rvert_{\partial \mathcal{M}}(\mathbf{x})=0$, and then sum over the different boundary contributions with the gluing condition \eqref{2.8}. This becomes especially manifest for the PMC-bulk theory: equation \eqref{PMC} contains the boundary conditions $n^\mu A_\mu^R\rvert_{\partial \mathcal{M}}(\mathbf{x})=0$ and $n^\mu F_{\mu t}^R\rvert_{\partial \mathcal{M}}(\mathbf{x})=0$. This shows PMC boundary conditions on the bulk R-wedge theory are the most natural to discuss entanglement.

The nature and implications of the boundary conditions $\nu=i$ in \eqref{PMC} are discussed in sections \ref{s: action} and \ref{s:boundary and edge}. For now it suffices to know that these do not generate additional entanglement.

There are two compelling reasons to look for edge modes in the static sector of the R-wedge theory:\footnote{For an example where the zero-mode sector for a scalar field becomes important, see \cite{Michel:2016fex}. }
\begin{itemize}
\item It is a general statement that static modes of a quadratic Lagrangian contribute to the Hamiltonian as boundary degrees of freedom. 
E.g. for a scalar field $\phi$ in flat space, specialized to the static solutions:
\begin{equation}
H = \frac{1}{2}\int_\Sigma d^{d}x\left((-\partial_t \phi)^2 + (\nabla \phi)^2+m^2\phi^2\right) = \frac{1}{2}\int_\Sigma d^d x\partial_i (\phi \partial^i \phi),
\end{equation}
which is a total derivative and hence reduces to a boundary term. This also holds for the improved Maxwell Hamiltonian in Rindler. For a boundary $\partial \Sigma: \rho=\epsilon$ in Rindler coordinates, we have:
\begin{align}
H = \int_\Sigma d^{d-1}\mathbf{x} d\rho \rho \left(\frac{1}{4}F^{\mu\nu}F_{\mu\nu} - F_{0\mu}F^{0\mu}\right) = -\frac{1}{2} \int_{\partial \Sigma} d^{d-1}\mathbf{x} \rho \left(F^{\rho i}A_i - F^{\rho 0} A_0\right).\label{2.12}
\end{align}
Furthermore, as discussed below, the static solutions do not satisfy the boundary conditions \eqref{2.5}, regardless of the regularization procedure, and hence these solutions do not appear in the respective bulk theories.
\item The horizon is an infinite redshift hypersurface. This implies that from the Rindler perspective there is no horizon dynamics and the horizon degrees of freedom are necessarily static. Mathematically, this follows from regularity of the field solution at Rindler radial coordinate $\rho=\epsilon \to 0^+$, just as happens with the origin in polar coordinates. A different but equally important consequence of the infinite redshift property of the horizon is that the fields associated with finite horizon charges (edge modes) have no radial extent away from the horizon. This also follows intuitively from the fact that there is no suitable scale in the theory with which they should decay away from the horizon. Note that these boundary-localized features of edge states are very explicit in Chern-Simons theory, as discussed below.
\end{itemize}

From the quantum point of view the need for edge states originates from an incompleteness of the R-and L-wedge bulk algebras in the Minkowski algebra. 

\begin{figure}[h]
\centering
 \includegraphics[width=0.35\textwidth]{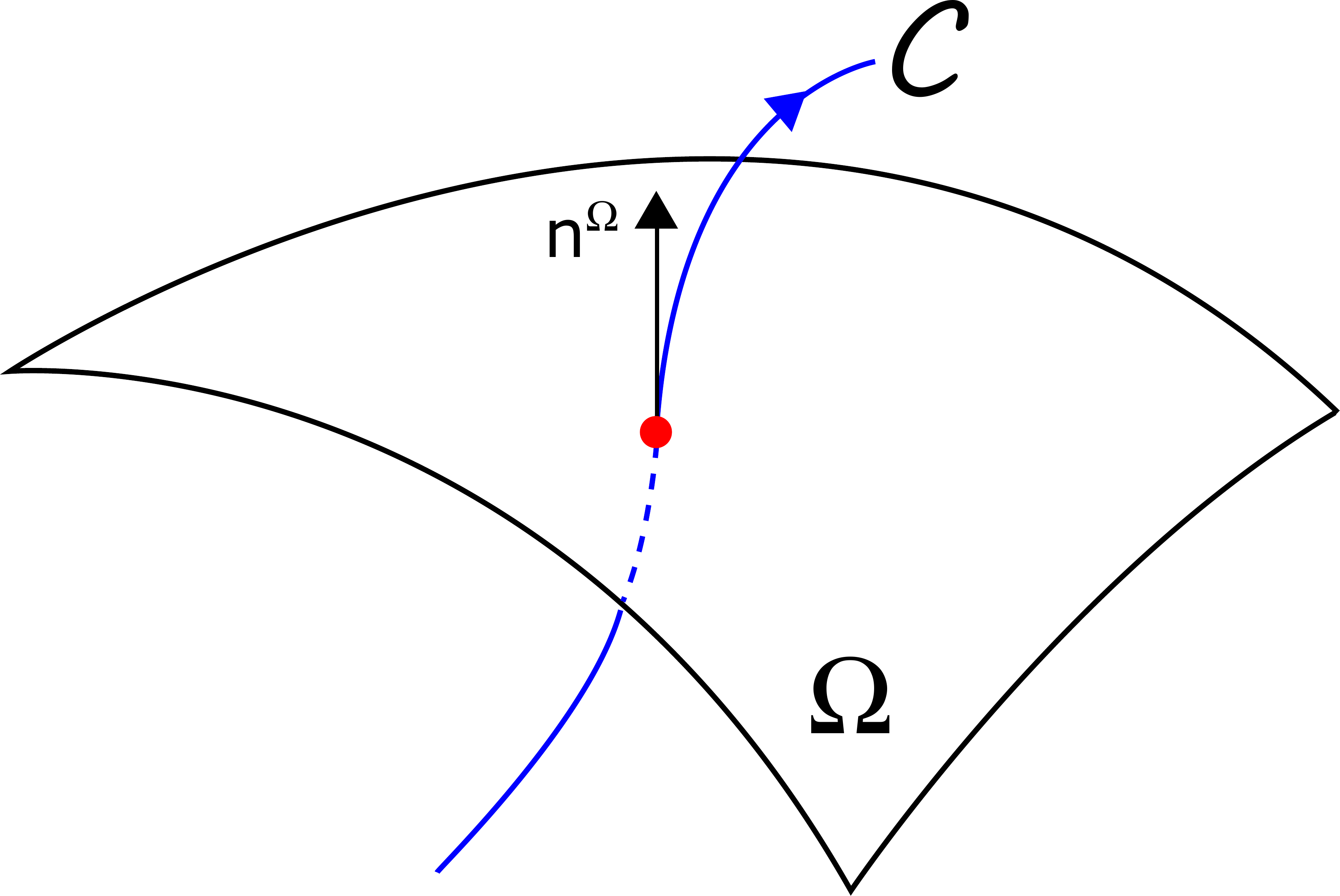}
\caption{Wilson line along $\mathcal{C}$ piercing a surface $\Omega$. The normal on the surface $n^\Omega$ and the orientation of the curve $\mathcal{C}$ determine whether the $\theta$-function evaluates to $+1$ or $-1$. }
\label{surfaceWilson}
\end{figure}

Let us elaborate. The Maxwell theory is quantized by imposing the canonical commutation relations (CCR) on the Maxwell  field $A$:
\begin{equation}
\comm{A_\mu (\rho,\mathbf{x})}{\Pi^\nu (\rho',\mathbf{y})}=i\delta^\nu_\mu\delta(\rho-\rho')\delta(\mathbf{x}-\mathbf{y}),\label{2.015}
\end{equation}
where 
\begin{equation}
\Pi^\nu=\frac{\partial \mathcal{L}}{\partial \partial_tA_\nu}
\end{equation}
is the conjugate field of $A$. Formula \eqref{2.015} includes quantization of the longitudinal and temporal polarization, whose DOF are projected out of the Hilbert space of the theory by the Gupta-Bleuler procedure. The physical Hilbert space of the theory, obtained by quotienting out the nullspace, is completely generated by the manifestly gauge-invariant algebra:
\begin{equation}
\boxed{\comm{\Phi_\Omega}{\mathcal{W}_\mathcal{C}}=\mathcal{W}_\mathcal{C}\theta(\Omega\cap \mathcal{C}), \quad \forall \Omega, \mathcal{C}}, \label{2.15}
\end{equation}
where $\Phi_\Omega$ represents the flux of $\Pi$ through a spatial co-dimension one surface $\Omega$
\begin{equation}
\Phi_\Omega=\int_\Omega n^\Omega_\nu \Pi ^\nu\label{2.22}
\end{equation} 
and $\mathcal{W}_\mathcal{C}$ is the Wilson line along the closed curve $\mathcal{C}$:
\begin{equation}
\mathcal{W}_\mathcal{C}=e^{i\int_\mathcal{C} A}.\label{5.9}
\end{equation}
The $\theta$-function introduced here evaluates to $\pm 1$ if the surface and line intersect, including a sign for their relative orientation: it is the intersection number. The angle of intersection is not important (figure \ref{surfaceWilson}).

The boundary conditions \eqref{2.7} destroy the possibility of flux through the horizon: $\Phi_{\Omega}=0$, $\Omega \in \partial \Sigma$. From the Rindler point of view this is not an issue: the horizon data is not accessible to the bulk Rindler observer and the entire bulk Hilbert space describing an arbitrary bulk EM-field configuration is generated by Wilson lines that reside entirely in the R-wedge. However, the Minkowski theory should cover the entire algebra \eqref{2.15} and in particular should thus contain configurations with nonzero horizon flux $\Phi_{\Omega}$, or equivalently the Minkowski Hilbert space contains states created by inserting Wilson lines through the horizon. This identifies edge states as configurations of horizon electric flux created by Wilson lines through the horizon. This is, of course, in complete agreement with the identification of constraint DOF on the boundary as the edge freedom in \eqref{2.8}.
\\~\\
As an alternative perspective, consider again a Wilson line along a curve $\mathcal{C}$. When restricting to closed curves, or curves that extend to infinity, this is invariant under small gauge transformations.\footnote{Indeed, performing the gauge transformation $A\to A+d\phi$, the Wilson line \eqref{5.9} transforms as
\begin{align}
\mathcal{W}_\mathcal{C} \, \to \, \mathcal{W}_\mathcal{C} e^{i\int_\mathcal{C} d\phi} \, = \, \mathcal{W}_\mathcal{C}.
\end{align}
}

Consider now a Wilson line piercing the horizon. When embedding Rindler space into Minkowski space, there is no reason to assume the gauge parameter $\phi(x)$ vanishes on the Rindler horizon, which means this operator is \textit{not} gauge-invariant from a Rindler perspective; it transforms under such a large gauge transformation as
\begin{equation}
\mathcal{W}_\mathcal{C} \, \to \, \mathcal{W}_\mathcal{C} e^{-i\phi(\mathcal{C}\cap \partial \Sigma)}.\label{9.7}
\end{equation}

The most general R-wedge theory thus naturally decouples into two separate theories. The first consists of the bulk R-wedge Hilbert space $\mathcal{H}_{\text{bulk},R}$ of states satisfying the boundary conditions \eqref{2.5}. The second consists of states associated with the horizon degrees of freedom, generalizing the constraint conditions to \eqref{2.8} and accounting for horizon piercing Wilson lines. Labeling the states associated with these edge degrees of freedom as $\ket{q}$ (they can be thought of as horizon charges, see section \ref{s:boundary and edge}), one can extend both left and right Hilbert spaces as:\footnote{Note that the actions of the R-wedge observer are limited to $\mathcal{H}_{\text{bulk},R}$. He will not be able to observe the presence of the fields associated with states in the Hilbert space $\mathcal{H}_{\text{edge},R}$ nor will he be able to change them or interact with them using his physical operators. This means $\ket{q}$ labels a superselection sector from the R-wedge perspective.}
\begin{equation}
\mathcal{H}_L=\bigoplus_{q\in\mathcal{H}_{\text{edge},L}}\ket{q}_L\otimes \mathcal{H}_{\text{bulk},L}, \quad\quad \mathcal{H}_R=\bigoplus_{q\in\mathcal{H}_{\text{edge},R}}\ket{q}_R\otimes \mathcal{H}_{\text{bulk},R},
\end{equation}
or
\begin{equation}
\mathcal{H}_L=\mathcal{H}_{\text{edge},L}\otimes \mathcal{H}_{\text{bulk},L}, \quad\quad \mathcal{H}_R=\mathcal{H}_{\text{edge},R}\otimes \mathcal{H}_{\text{bulk},R}.
\end{equation}
The most general Minkowski state then lives in the factorizable extended Hilbert space:
\begin{equation}
\mathcal{H}_{\text{ext}}=\mathcal{H}_L\otimes \mathcal{H}_R.
\end{equation}
The matching conditions \eqref{2.7} imply that the physical Minkowski Hilbert space consists of the diagonal subspace of this extended Hilbert space:
\begin{equation}
\mathcal{H}_{\text{bulk}}=\bigoplus_{q\in\mathcal{H}_{\text{edge}}}\left(\ket{q}_L\otimes \mathcal{H}_{\text{bulk},L}\right)\otimes \left(\ket{q}_R\otimes \mathcal{H}_{\text{bulk},R}\right).\label{2.11}
\end{equation}
The Minkowski vacuum state lives in this space as a state that is also diagonal in $\mathcal{H}_{\text{bulk},L}\otimes \mathcal{H}_{\text{bulk},R}$. As the latter decouple completely from the spaces $\mathcal{H}_{\text{edge},L/R}$, the total entanglement entropy in the Minkowski vacuum is simply the sum of the bulk entropy due to $\mathcal{H}_{\text{bulk},L/R}$ and the edge entropy due to $\mathcal{H}_{\text{edge},L/R}$:
\begin{equation}
S=S_{\text{edge}}+S_{\text{bulk}}.
\end{equation}
Both contributions are calculated in this paper from the canonical spectrum of the theory, for arbitrary $d$.

\subsection{Simplified Structure in Topological Theories}
Topological theories have no local bulk degrees of freedom. Introducing a boundary surface leads to an edge sector as the only DOF of the theory. In such a simplified situation, it is instructive to understand the intuitive picture of Wilson lines attached to the boundary surface in more detail. For Chern-Simons (CS) theories it can be made very explicit, see also \cite{Wong:2017pdm,Fliss:2017wop,Geiller:2017xad}. CS theory with a boundary reduces to chiral Wess-Zumino-Witten (WZW) theory on the boundary. For a cylindrical boundary, and focusing on the Abelian case, the $U(1)$ WZW Hilbert space consists of a primary of fixed charge $\left|q\right\rangle$ and all WZW descendants obtained by acting with $J_{-n}$, $n>0$:
\begin{equation}
\mathcal{H} = \left\{J_{-n}...J_{-m}\left|q\right\rangle, \,\, q\in \mathbb{R}\right\},
\end{equation}
and this for all values of the charge $q$, possibly discretized. This is the Fourier transform of states obtained by acting with local current operators, at a fixed timeslice:
\begin{equation}
\mathcal{H} = \left\{J(w_1)...J(w_m)\left|q\right\rangle, \,\, q\in \mathbb{R}\right\},
\end{equation}
for any choice of locations $w_i$ along the spatial circle.\footnote{One can extend this construction to the WZW theory on the infinite line (instead of a circle) to reach the standard situation of an extended entangling surface. This requires quantizing the CFT on the infinite line, which leads to continuous generalizations of the modes $J_\omega$, which satisfy a continuous version of the Kac-Moody algebra and Sugawara construction. Local operator insertions on the line can be Fourier transformed to the modes $J(x) = \int_{-\infty}^{+\infty}dx e^{i\omega x} J_\omega$.} Thus the edge sector Hilbert space is identified as the space of all punctures on the WZW plane. 
A more convenient basis is found by setting $J = \partial \phi$, and using the local vertex operators $e^{i q \phi}$:
\begin{equation}
\mathcal{H} = \left\{e^{iq_1 \phi(w_1)}...e^{iq_m \phi(w_m)}\left|q\right\rangle, \,\, q\in \mathbb{R}\right\},\label{boundaryspace}
\end{equation}
which creates a state that has fixed charge density at the locations $w_i$, as measured by $J_0 = \oint dz J(z)$. In $U(1)$ Chern-Simons, the equations of motion imply $A$ is flat: $A = \partial \phi$, which is identified with $J$ itself. A Wilson line that is anchored at the boundary at both endpoints, is then evaluated as
\begin{equation}
e^{i Q \int A} = e^{i Q\phi(\mathbf{x}_2) - i Q\phi(\mathbf{x}_1)},
\end{equation}
and is precisely found in \eqref{boundaryspace}, as a state with additional charge $+Q$ at $\mathbf{x}_2$ and $-Q$ at $\mathbf{x}_1$. Thus one can think of the Hilbert space as consisting of Wilson line endpoints.

\begin{figure}[h]
\centering
 \includegraphics[width=0.4\textwidth]{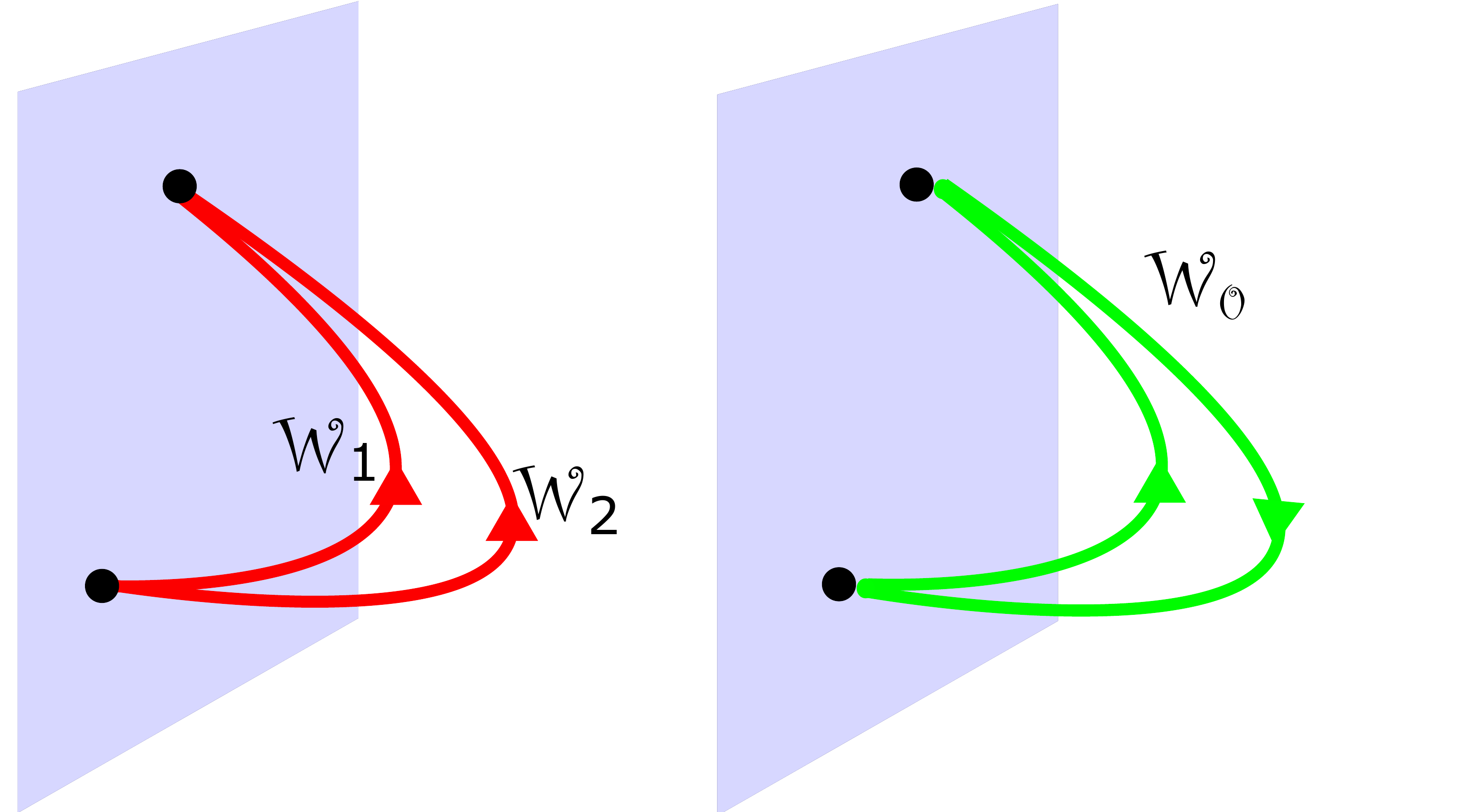}
\caption{Left: Two Wilson lines $\mathcal{W}_1$ and $\mathcal{W}_2$ at the same boundary point. Right: Closed Wilson loop $\mathcal{W}_{o} = \mathcal{W}_1\mathcal{W}_2^{-1}$.}
\label{Wilsonedge}
\end{figure}

The edge sector is insensitive to the specific shape of the Wilson line. In CS this is so due to the topological nature of the theory, but this is true more generally, and in particular also in Maxwell theory. The difference between two Wilson lines connected at the same boundary points is a closed Wilson loop in the bulk theory, and is part of the bulk sector (Figure \ref{Wilsonedge}). 
Explicitly, as the Hilbert space factorizes, and due to linearity, any Wilson loop operator acts only on the bulk part:
\begin{equation}
\mathcal{W}_{1}\left(\left|0\right\rangle \otimes \left|\text{bulk}\right\rangle\right) = \mathcal{W}_{o}\mathcal{W}_{2}\left(\left|0\right\rangle \otimes \left|\text{bulk}\right\rangle\right) = \mathcal{W}_{2}\left(\left|0\right\rangle \otimes \mathcal{W}_{o}\left|\text{bulk}\right\rangle\right)=
\mathcal{W}_{2}\left(\left|0\right\rangle \otimes \left|\text{bulk}\right\rangle\right),\label{2.27}
\end{equation}
where $\ket{\text{bulk}}$ is short for the entire bulk Hilbert space. We make this identification of the edge sector as the set of punctures of Wilson lines very explicit in section \ref{ss:6.2}. 

This picture is expected to generalize to other topological gauge theories, such as BF theories. As a first piece of evidence, dimensionally reducing 3d CS leads to 2d BF theory with a specific boundary term. This corresponds on the boundary surface to a dimensional reduction of 2d WZW to its zero-mode sector, reducing the Kac-Moody algebra to its zero-grade Lie algebra. The Hilbert space of 2d BF theory consists of the irreps of this underlying algebra.

As a remark, note that the standard holographic dictionary for 3d CS states:
\begin{equation}
\frac{\delta S_{\text{on-shell}}}{\delta A_\mu^{\text{bdy}}}=\left\langle J^\mu\right\rangle_{\text{bdy}}.\label{dic}
\end{equation}
Integrating by parts and keeping track of the boundary contributions, we find:
\begin{align}
\delta S=\int_{\mathcal{M}}d^{d+1}x\left(\frac{\partial \mathcal{L}}{\partial A_\nu}-\partial_\mu \frac{\partial \mathcal{L}}{\partial \partial_\mu A_\nu}\right)\delta A_\nu+\int_{\partial \mathcal{M}}d^d x\left(n_\nu \Pi^{\nu\mu}\right)\delta A_\mu,
\end{align}
where $n_\nu$ points outwards. Using the Euler-Lagrange equations and the dictionary \eqref{dic}, we find:
\begin{equation}
\frac{\delta S_{\text{on-shell}}}{\delta A_\mu^{\text{bdy}}}=n_\nu\Pi^{\nu\mu}\rvert_{\text{bdy}}=\left\langle J^\mu\right\rangle_{\text{bdy}}.
\end{equation}
This is the QFT analogue of $\frac{\partial S_{\text{on-shell}}}{\partial q_i}=p_i$, familiar from the Hamilton-Jacobi method. The timelike component becomes:
\begin{equation}
n_\nu\Pi^{\nu 0}\rvert_{\text{bdy}}=\left\langle J^0\right\rangle_{\text{bdy}}.\label{2.28}
\end{equation}
The radial canonical momentum of the $U(1)$ bulk theory evaluated at the boundary is a $U(1)$ charge in the holographic boundary theory. For CS, where the conjugate momentum is a component of the gauge field $A$ itself, this means that the bulk gauge field $A$ is sourced by $J(w)$ on the boundary \eqref{boundaryspace}. For Maxwell, we interpret \eqref{2.28} as the jump condition in Gauss' law, where $n_\nu\Pi^{\nu 0}$ becomes the radial electric flux \eqref{2.22} (see also section \ref{s:boundary and edge}).
 
\section{Bulk Maxwell Theory in $1+1$}
\label{s gauge}
We focus first on the special case $d=1$. 
\subsection{Canonical quantization}
\label{s: 3.1}
The first step in the canonical quantization of the theory is to find the complete set of bulk modes which span the solution space of the bulk EOM \eqref{2.4}. In tortoise coordinates and written out in components, the bulk EOM read:
\begin{align}
\label{3.1}
(\partial_r^2-\partial_t^2)A_r+2(\partial_t A_t - \partial_r A_r)&=0, \\
(\partial_r^2-\partial_t^2)A_t+2(\partial_t A_r - \partial_r A_t)&=0.\label{3.1b}
\end{align}
The field is decomposed in the basis of positive frequency Rindler modes:
\begin{equation}
\partial_tA_\mu = -i\omega A_\mu\label{3.2}.
\end{equation}

In order to determine an orthogonal set of solutions, an inner product must be defined on the solution space. We use the generalized KG inner product which in a general background reads 
\begin{equation}
(A,B)=i\int d\Sigma_\mu J^\mu.\label{3.3}
\end{equation}
$\Sigma$ is a spacelike hypersurface and $J$ is defined as
\begin{equation}
\label{inner product}
J^\mu = \frac{1}{\sqrt{-g}}\left(  {B_{\nu}} \Pi_{A^*}^{\mu\nu}-{A_{\nu }}^*{\Pi_B^{\mu\nu}}\right).
\end{equation}
We introduced the notation
\begin{equation}
\Pi^{\mu\nu}_A=\frac{\partial \mathcal{L} }{\partial \partial_\mu A_\nu}.\label{3.5}
\end{equation}
This represents a conserved current in the sense that
$\nabla_\mu J^\mu=0$,as can be explicitly checked using \eqref{3.5}, such that the inner product independent of the chosen hypersurface $\Sigma$. If one chooses a fixed time slice to evaluate this inner product, the conjugate momenta of the gauge field enters in \eqref{inner product}:
\begin{equation}
\Pi^{\mu}_A := \Pi^{t\mu}_A= \sqrt{-g}\left(F^{\mu t} -g^{t\mu} \nabla_\sigma A^\sigma\right).\label{3.6}
\end{equation}
Defining for notational convenience two scalar functions
\begin{align}
\label{3.7}
\phi^R_k = \frac{1}{\sqrt{16\pi k}}e^{-i k (t-r)}, \quad \phi^L_k = \frac{1}{\sqrt{16\pi k}}e^{-i k (t+r)},
\end{align}
for $\omega = k > 0$, which describe right- and left-moving massless scalar modes solving $\Box \phi = 0$, and a complex energy variable
\begin{equation}
\epsilon=\omega+i\label{epsilon},
\end{equation}
we find that the solution space of the sourceless Maxwell equations \eqref{3.1} for each value of $\omega =k > 0$ is spanned by the 4 sets of modes:\footnote{The equations decouple in tortoise light-cone components. This can be generalized to higher spin and to higher dimensions. Alternatively, the coupled differential equations can be reformed into a single fourth order ODE.}
\begin{equation}
\n \phi^R_k, \quad \p e^{2r} \phi^R_k, \quad \p \phi^L_k, \quad \n e^{2r} \phi^L_k,
\end{equation}
which can be reorganized into the modes
\begin{alignat}{3}
A^{(1)}_{\mu,k}&=\frac{\epsilon}{\abs{\epsilon}}\left(\p e^{2r}+\n\right)\phi^R_k, \quad\quad &&A^{(2)}_{\mu,k}=\frac{1}{\abs{\epsilon}}\left(\epsilon \p e^{2r}-\bar{\epsilon}\n\right)\phi^R_k, \nonumber \\
\label{3.10}
A^{(3)}_{\mu,k}&=-\frac{\bar{\epsilon}}{\abs{\epsilon}}\left(\p +\n e^{2r}\right)\phi^L_k, \quad\quad &&A^{(4)}_{\mu,k}=\frac{1}{\abs{\epsilon}}\left(\epsilon \p-\bar{\epsilon}\n e^{2r}\right)\phi^L_k.
\end{alignat}
These are mutually orthogonal w.r.t. the inner product \eqref{inner product}. The only non-zero inner products are
\begin{gather}
\label{3.11}
\begin{aligned}
\left(A^{(2)}_k, A^{(2)}_{k'}\right) &= \left(A^{(4)}_k, A^{(4)}_{k'}\right) = \delta(k-k'), \\
\left(A^{(1)}_k, A^{(1)}_{k'}\right) &= \left(A^{(3)}_k, A^{(3)}_{k'}\right) = -\delta(k-k').
\end{aligned}
\end{gather}
Notice that the modes $A^{(2)}$ and $A^{(4)}$ have positive $\delta$-norm and the modes $A^{(1)}$ and $A^{(3)}$ have negative norm. The quantum field is expanded in the complete set of modes \eqref{3.10} as
\begin{align}
A_\mu &=\int_0^{+\infty} dk \left( a^{(1)}_k A^{(1)}_{\mu,k} + a^{(2)}_k A^{(2)}_{\mu,k} + a^{(3)}_k A^{(3)}_{\mu,k} + a^{(4)}_k A^{(4)}_{\mu,k} + h.c. \right)\label{3.12}.
\end{align}
As usual, the coefficients $a^{(i)}_k$ of the individual modes become the quantum oscillators in the quantum theory such that we naturally impose the oscillator commutation relations\footnote{Note the minus signs which are appropriate for the oscillators of modes with negative norm.}
\begin{gather}
\label{3.13}
\begin{aligned}
\bigl[a^{(1)}_{k},{a^{(1)\dag}_{k'}}\bigr]&=-\delta(k-k'), \quad \bigl[a^{(2)}_{k},{a^{(2)\dag}_{k'}}\bigr]=\delta(k-k'),\\
\bigl[a^{(3)}_{k},{a^{(3)\dag}_{k'}}\bigr]&=-\delta(k-k'), \quad 
\bigl[a^{(4)}_k,{a^{(4)\dag}_{k'}}\bigr]=\delta(k-k').
\end{aligned}
\end{gather}
If the normalization of the modes is defined in a consistent manner using the Klein-Gordon inner product \eqref{inner product}, the oscillator commutation relations ought to imply the canonical equal time commutation relations and vice versa. Indeed, we find
\begin{align}
[A_\mu(t,r),\Pi^\nu(t,r')]&=\int_{0}^{+\infty} dk \left(A^{(2)}_{\mu,k} {\Pi_{k}^{(2) \nu}}^*-A^{(1)}_{\mu,k} {\Pi_{k}^{(1) \nu}}^* + ((1,2) \leftrightarrow (3,4)) - c.c.\right) \nonumber \\
&=i\delta^\mu_\nu \delta(r-r'),\label{3.14}
\end{align}
which proves the consistency of normalizing modes with the Klein-Gordon inner product \eqref{inner product}.
\subsubsection*{Physical Hilbert Space}
\label{s constraint}
Having established the nature of the expansion of the most general quantum field \eqref{3.12} satisfying the sourceless Maxwell equations \eqref{3.1} in the bulk of spacetime, we investigate the influence of the Lorenz gauge constraint \eqref{2.2} on the Hilbert space of the quantum theory. The states that survive this constraint make up the physical Hilbert space:
\begin{equation}
\nabla^\mu A_\mu^{(+)}\ket{\psi}=0\label{3.15}.
\end{equation}
Using the explicit mode expansion of the field \eqref{3.12}, this translates to
\begin{equation}
\int_{0}^{+\infty} dk \abs{\epsilon}\left[ \phi^R_k(\mathbf{x}) \left(a^{(1)}_k+a^{(2)}_k\right)\ket{\psi}-\phi^L_k(\mathbf{x}) \left(a^{(3)}_k+a^{(4)}_k\right) \right] \ket{\psi}=0.\label{3.16}
\end{equation}
Considering the fact that $\phi^R_k$ and $\phi^L_k$ represent mutually orthogonal sets, Lorenz gauge constrains the physical space to states $\ket{\psi}$ satisfying
\begin{gather}
\label{3.17}
\begin{aligned}
a^{(2)}_k\ket{\psi}=-a^{(1)}_k\ket{\psi}, \quad\quad a^{(4)}_k\ket{\psi}=-a^{(3)}_k\ket{\psi}.
\end{aligned}
\end{gather}
Introducing occupation numbers as $n^{(i)}_k={a^{(i)}_k}^\dag a^{(i)}_k$, the Lorenz gauge constraint imposes
\begin{equation}
\bra{\phi}n^{(2)}_k-n^{(1)}_k\ket{\psi}=0, \quad\quad \bra{\phi}n^{(4)}_k-n^{(3)}_k\ket{\psi}=0\label{3.18}.
\end{equation}
These constraints are solved by states of the form
\begin{equation}
\left|\psi\right\rangle = \prod_k\left({a^{(2)}_k}^\dag + {a^{(1)}_k}^{\dag}\right)^{n_k}\left({a^{(4)}_k}^\dag + {a^{(3)}_k}^{\dag}\right)^{m_k}\ket{0}.\label{3.19}
\end{equation}
All of these states are null and are hence pure gauge which can be checked explicitly. In terms of the field $A_\mu$, the physical field configurations associated with the null Hilbert space \eqref{3.19} are (residual) pure gauge:
\begin{equation}
A_\mu=\partial_\mu\phi\sim 0,\label{3.20}
\end{equation}
with $\Box \phi =0$. The result is that in $1+1$d Rindler Maxwell theory, the only state is the vacuum state. 

\subsubsection*{Hamiltonian}
\label{s hamiltonian}
The Hamiltonian of the theory is
\begin{align}
H =\int dr \mathcal{H} =\int dr \left(\partial_t A_\mu \frac{\delta \mathcal{L}}{\delta \partial_t A_\mu}-\mathcal{L}\right)\label{3.21}.
\end{align}
Inserting the Lagrangian \eqref{action 1}, this can be rewritten in the convenient form:
\begin{equation}
\label{hami}
H = \frac{1}{2}\int dr e^{-2r}\left[\left(\partial_tA_r - \partial_r A_t\right)\left(\partial_tA_r + \partial_r A_t\right) + \left(\partial_rA_r - \partial_t A_t\right)\left(\partial_tA_t + \partial_r A_r\right)\right].
\end{equation}
Plugging in the field expansion \eqref{3.12} and using the normalization of the scalar modes
\begin{equation}
\int dr \phi^R_k{\phi^R_{k'}}^*=\frac{1}{8k}\delta(k-k'), \quad \int dr \phi^R_k{\phi^L_{k'}}^*=0, \quad \int dr \phi^L_k{\phi^L_{k'}}^*=\frac{1}{8k}\delta(k-k'),
\end{equation}
we find 
\begin{align}
H &=\int_0^{+\infty} dk k \left(n^{(2)}_k-n^{(1)}_k+n^{(4)}_k-n^{(3)}_k\right),\label{3.24}
\end{align}
which is the expected result. The Lorenz gauge constraint \eqref{3.15} directly implies the vanishing of the Rindler Hamiltonian between physical states via \eqref{3.18}: $\bra{\psi}H\ket{\psi}=0$. This is obviously the only possible result for a theory which only consists of the vacuum. 

\subsection{Unruh Effect in Maxwell theory}
\label{s: 3.2}
In this subsection we derive the Unruh modes in $1+1$ dimensional Maxwell theory and calculate the Bogoliubov transformations linking the modes \eqref{3.10} to these Unruh modes.\footnote{In the left Rindler wedge, the energy of a mode is defined as the eigenvalue of $i\partial_\tau$, where $\tau$ is the proper time of the Rindler observer. The relation between proper time $\tau$ and coordinate time $t$ in the left wedge is $\tau=-t$. Consequently, the modes (as solutions of the equation of motion) are proportional to $e^{i \omega t}$. From the modes in the R-wedge \eqref{3.10}, one finds the L-wedge modes as:
\begin{equation}
A^{(i)L}_k = {A_k^{(i)R}}^*. \label{ansatz l}
\end{equation}
This results for example in: 
\begin{equation}
A^{(1)L}_{\mu,k}=\frac{\epsilon}{\abs{\epsilon}\sqrt{16 \pi \omega}}\left(\begin{pmatrix}
1\\1
\end{pmatrix}e^{2r}+\begin{pmatrix}
1\\-1
\end{pmatrix}\right)e^{i\omega(t-r)}.\label{p l u}
\end{equation}
The inner product used in the L-wedge is defined as minus the R-wedge inner product. Indeed, the minus sign is requires to account for the complex conjugation in \eqref{ansatz l} in order to obtain e.g.
\begin{equation}
\left(A^{(1)L}_{\mu,k},A^{(1)L}_{\mu,k'}\right)=-\delta(k-k'),\label{a.4}
\end{equation}
which implies the correct oscillator algebra:
\begin{equation}
\comm{a^{(1)L}_{k}}{{a^{(1)L}_{k'}}^\dag}=-\delta(k-k').
\end{equation}
In deriving this result we used the fact that in the left wedge the canonical conjugate momentum is defined as $\Pi^\mu=\frac{\delta \mathcal{L}}{\delta \partial_\tau A_\mu}$, where crucially $\tau$ is used instead of $t$.}

Introducing Minkowski light cone coordinates $U=T-X$ and $V=T+X$, and paying attention to the tensorial transformation of $A_\mu$, the modes \eqref{3.10} are rewritten in Minkowski coordinates as:
\begin{gather}
\label{3.26}
\begin{aligned}
A^{(1)R}_{\mu,k} &=\frac{\epsilon}{\abs{\epsilon}\sqrt{16 \pi \omega}}\left(-U\begin{pmatrix}
1\\1
\end{pmatrix}+\frac{1}{-U}\begin{pmatrix}
1\\-1
\end{pmatrix}\right)(-U)^{i\omega}\theta(-U),\\
A^{(2)R}_{\mu,k} &=\frac{1}{\abs{\epsilon}\sqrt{16 \pi \omega}}\left(-\epsilon U\begin{pmatrix}
1\\1
\end{pmatrix}-\frac{\bar{\epsilon}}{-U}\begin{pmatrix}
1\\-1
\end{pmatrix}\right)(-U)^{i\omega}\theta(-U),\\
A^{(3)R}_{\mu,k} &=-\frac{\bar{\epsilon}}{\abs{\epsilon}\sqrt{16\pi \omega}}\left(\frac{1}{V}\begin{pmatrix}
1\\1
\end{pmatrix}+V\begin{pmatrix}
1\\-1
\end{pmatrix}\right)V^{-i\omega}\theta (V),\\
A^{(4)R}_{\mu,k} &=\frac{1}{\abs{\epsilon}\sqrt{16\pi \omega}}\left(\frac{\epsilon}{V}\begin{pmatrix}
1\\1
\end{pmatrix}-\bar{\epsilon}V\begin{pmatrix}
1\\-1
\end{pmatrix}\right)V^{-i\omega}\theta (V),
\end{aligned}
\end{gather}
where we introduced appropriate Heaviside functions to reflect the limited support of the modes. The L-wedge modes are rewritten in a similar fashion. Following the standard argument from scalar QFT, one can combine L- and R-modes in such a way that they only contain positive frequency Minkowski modes:\footnote{We focus for simplicity on the (1)- and (2)-polarizations (the other polarizations are treated in exactly the same way).} 
\begin{gather}
\label{3.27}
\begin{alignat}{3}
U^{(1)R}_{\mu,k}&&=\frac{e^{\frac{\pi \omega}{2}}}{\sqrt{2\sinh \pi \omega}}\left(A^{(1)R}_{\mu,k}+e^{-\pi \omega}{A^{(1)L}_{\mu,k}}^*\right), \,\, U^{(1)L}_{\mu,k}&&=\frac{e^{\frac{\pi \omega}{2}}}{\sqrt{2\sinh \pi \omega}}\left(A^{(1)L}_{\mu,k}+e^{-\pi \omega}{A^{(1)R}_{\mu,k}}^*\right), \nonumber \\
U^{(2)R}_{\mu,k}&&=\frac{e^{\frac{\pi \omega}{2}}}{\sqrt{2\sinh \pi \omega}}\left(A^{(2)R}_{\mu,k}+e^{-\pi \omega}{A^{(2)L}_{\mu,k}}^*\right), \,\, U^{(2)L}_{\mu,k}&&=\frac{e^{\frac{\pi \omega}{2}}}{\sqrt{2\sinh \pi \omega}}\left(A^{(2)L}_{\mu,k}+e^{-\pi \omega}{A^{(2)R}_{\mu,k}}^*\right).
\end{alignat}
\end{gather}
We now associate oscillators $u^{(i)R}_k$ and $u^{(i)L}_k$ with these Unruh modes. The lowering operators will annihilate the Minkowski vacuum. The relation between the Rindler and the Unruh oscillators are the Bogoliubov transformations:
\begin{gather}
\label{3.28}
\begin{alignat}{3}
a^{(1) R}_{k}&&=\frac{e^{\frac{\pi \omega}{2}}}{\sqrt{2\sinh \pi \omega}}\left(u^{(1)R}_{k}+e^{-\pi \omega}{u^{(1)L}_k}^\dag\right),\,\, a^{(1) L}_{k}&&=\frac{e^{\frac{\pi \omega}{2}}}{\sqrt{2\sinh \pi \omega}}\left(u^{(1)L}_{k}+e^{-\pi \omega}{u^{(1)R}_k}^\dag\right), \nonumber \\
a^{(2) R}_{k}&&=\frac{e^{\frac{\pi \omega}{2}}}{\sqrt{2\sinh \pi \omega}}\left(u^{(2)R}_{k}+e^{-\pi \omega}{u^{(2)L}_k}^\dag\right),\,\,  a^{(2) L}_{k}&&=\frac{e^{\frac{\pi \omega}{2}}}{\sqrt{2\sinh \pi \omega}}\left(u^{(2)L}_{k}+e^{-\pi \omega}{u^{(2)R}_k}^\dag\right).
\end{alignat}
\end{gather}
The Minkowski vacuum state is defined as the state that is annihilated by all Minkowski annihilation operators. Since the Unruh modes are complete in the Minkowski Hilbert space and since they have a well-defined sign of Minkowski frequency, the Minkowski vacuum is equivalently defined as being annihilated by all Unruh annihilation operators:
\begin{equation}
u_k^{(i) R}\ket{M}=0, \quad u_k^{(i) L}\ket{M}=0,\quad i=1,2.\label{mink vac 1}
\end{equation}
The Unruh effect entails a thermal population of the Minkowski vacuum state, as perceived by an accelerating observer. One way to make this statement explicit is to calculate the Minkowski vacuum expectation value of the Rindler Hamiltonian. The R-wedge Hamiltonian \eqref{3.24} decomposes into a $U$-dependent contribution (polarizations 1,2) and a $V$-dependent contribution (polarizations 3,4). Using the oscillator commutation relations for the Unruh creators and the definition \eqref{mink vac 1} of the Minkowski vacuum we find:\footnote{$V$ is the entire spatial Minkowski volume.}
\begin{align}
\bra{M}n^{(2)R}_k - n^{(1)R}_k\ket{M}&=\frac{V}{2\pi}\frac{2}{e^{2\pi \omega}-1}.
\end{align}
Adding the (3) and (4) contributions, the definition of the Rindler Hamiltonian \eqref{3.24} implies
\begin{equation}
\bra{M}H\ket{M}=2\frac{V}{2\pi}\int_{-\infty}^{+\infty} dk  \frac{\omega}{e^{2\pi \omega}-1}\label{3.31},
\end{equation}
where $\omega=\abs{k}$. This is twice the expectation value of the scalar Hamiltonian in the Minkowski ground state discovered when deriving the scalar Unruh effect. The disturbing feature is that this arises from the contributions of the nonphysical longitudinal and timelike photon polarizations. In \cite{Zhitnitsky:2010ji} it was demonstrated that the energy \eqref{3.31} does \emph{not} represent a thermal particle density in Rindler (which is the interpretation of the scalar Unruh effect) but instead is an addition to the Casimir energy. As mentioned in the Introduction, no such contribution is found when working e.g. in Weyl gauge or Coulomb gauge. We conclude that the Casimir energy contribution \eqref{3.31} must be canceled by another effect. Fortunately, as we demonstrate in Section \ref{s ghost}, such a cancellation does happen on account of a correct treatment of the Faddeev-Popov ghost fields. This restores gauge invariance of the theory. \\

We argued in subsection \ref{s constraint} that the Lorentz gauge constraint \eqref{3.15} implies the vanishing of the expectation value of the Rindler Hamiltonian in physical states. The fact that formula \eqref{3.31} does not satisfy this constraint is not a contradiction, as this is only a constraint in the physical Rindler Hilbert space $\mathcal{H}_{\text{bulk},R}\subset \mathcal{H}_R$ constructed by the R-wedge observer, whereas the Minkowski vacuum in general contains states from the extended Hilbert space $\mathcal{H}_R$.\footnote{To illustrate this, we can write down an expression for the squeezed state (i.e. the Minkowski vacuum as an entangled state of L and R states): 
\begin{align}
\ket{M}&=\frac{1}{\sqrt{Z(\beta)}}\prod_\omega\sum_{m,l} e^{-\frac{\beta}{2}\omega(n^{(2)}-n^{(1)})}\frac{(-1)^m}{m!\,l!} \left({a^{(1)R}_k}^{\dag}\right)^{m}\left({a^{(2)R}_k}^{\dag}\right)^{l}\ket{R}\otimes \left({a^{(1)L}_k}^{\dag}\right)^{m}\left({a^{(2)L}_k}^{\dag}\right)^{l}\ket{L}.
\end{align}
Just as in scalar QFT, the Minkowski vacuum is a diagonally entangled state of L-and R-wedge states. The entangling states on their own are nonphysical in the sense of \eqref{3.15}.} In \cite{Zhitnitsky:2010ji}, it was likewise demonstrated that $\ket{M}$ is not annihilated by the R-wedge's BRST operator, which represents only half of the Minkowski BRST operator.
\subsubsection*{Ghost Sector}
\label{s ghost}
The ghost fields arising in the gauge fixed action are irrelevant for a description of any theory living only in a single Rindler wedge, as their presence was effectively taken into account by imposing the Gupta-Bleuler constraint \eqref{3.15}. What is less clear, is to what extent this is true once we go to Minkowski coordinates that cover both Rindler wedges. Recall that in the previous section we found a Casimir energy contribution due to nonphysical polarizations only. If we consider such effects, then there is a priori no reason for us to rule out similar effects for the ghost fields. And indeed, in Appendix \ref{app:ghost} we compute the Casimir contribution of the ghost sector in the Minkowski vacuum to be
\begin{equation}
\bra{M}H_{\text{gh}}\ket{M}=-2\frac{V}{2\pi}\int dk  \frac{\omega}{e^{2\pi \omega}-1}.\label{3.45}
\end{equation}
As anticipated, the ghost contribution looks thermal and exactly cancels the two bosonic contributions due to the bosonic nonphysical and pure gauge modes. So we finally end up with
\begin{align}
\bra{M}H\ket{M}=0.\label{3.46}
\end{align}
Due to a total lack of any degrees of freedom, it immediately follows that the bulk entanglement entropy across the horizon vanishes in the $1+1$ dimensional Rindler theory: $S_E=0$. 

\subsection{Boundary Conditions}
In section \ref{s: hilbertspaces} we discussed the importance of boundary conditions in Rindler space. Let us therefore briefly reflect on the effect of imposing boundary conditions \eqref{2.5} on both the gauge and ghost sector.
\subsubsection*{Maxwell Theory}
The PMC horizon boundary conditions \eqref{PMC} in $1+1$ dimensional Maxwell theory in tortoise coordinates reduce to:
\begin{gather}
\begin{aligned}
\tensor{F}{_r^t}\rvert_{\partial \mathcal{M}}=0\label{3.47}, \quad A_r\rvert_{\partial \mathcal{M}}=0.
\end{aligned}
\end{gather}
Using the modes \eqref{3.10} we find
\begin{gather}
\begin{aligned}
e^{-2r}F_{rt}^{(1)}=e^{-2r}F_{rt}^{(2)} &=2i\abs{\epsilon}\phi^R,\\ e^{-2r}F_{rt}^{(3)}=e^{-2r}F_{rt}^{(4)} &=2i\abs{\epsilon}\phi^L.
\end{aligned}
\end{gather}
Again using the modes \eqref{3.10}, and appreciating that the terms proportional to $e^{2r}$ drop out on $\partial \mathcal{M}:r=-\infty$, the second boundary condition in \eqref{3.47} is observed to be redundant to the first. This is by construction; an additional effective constraint on the modes at the horizon would destroy polarizations, and as such would not be acceptable boundary conditions: away from the horizon, the R-wedge theory is to be identical to the Minkowski theory i.e. insensitive to the boundary. This is an explicit check that in $1+1$ dimensions the PMC \eqref{PMC} represents a good set of boundary conditions.

The boundary condition \eqref{3.47} forces us to recombine the exponentials $\phi^{R/L}$ into sines and cosines. For this purpose modes on the same row can be recombined into the orthonormal sets
\begin{gather}
\label{3.51}
\begin{alignat}{3}
A_{\mu,k}^{(I)}&&=\frac{1}{\sqrt{2}}\left(A_{\mu,k}^{(1)} + A_{\mu,k}^{(3)}\right),\quad A_{\mu,k}^{(II)}&&=\frac{1}{\sqrt{2}}\left(A_{\mu,k}^{(2)} + A_{\mu,k}^{(4)}\right),\\
A_{\mu,k}^{(III)}&&=\frac{1}{\sqrt{2}}\left(A_{\mu,k}^{(1)} - A_{\mu,k}^{(3)}\right),\quad A_{\mu,k}^{(IV)}&&=\frac{1}{\sqrt{2}}\left(A_{\mu,k}^{(2)} - A_{\mu,k}^{(4)}\right).
\end{alignat}
\end{gather}
All results of sections \ref{s: 3.1} and \ref{s: 3.2} hold equally for these modes. 
Imposing \eqref{3.47} at $r=r_*$ discretizes the spectrum of the theory to
\begin{align}
\sigma_D &:=\qty\bigg{k| k =\frac{\pi(n+1/2)}{r_*}, n\in \mathbb{Z}},\label{3.52}
\end{align}
for the modes $A_{\mu,k}^{(II)}$ and $A_{\mu,k}^{(III)}$, and to
\begin{align}
\sigma_N &:=\qty\bigg{k| k =\frac{\pi n}{r_*}, n\in \mathbb{Z}},\label{3.53}
\end{align}
for the modes $A_{\mu,k}^{(I)}$ and $A_{\mu,k}^{(IV)}$. The quantum field subject to the boundary conditions \eqref{3.46} is thus expanded as
\begin{gather}
\begin{aligned}
A_\mu= &\sum_{k\in\sigma_D}\left(a^{(II)}_kA_{\mu,k}^{(II)}+a^{(III)}_kA_{\mu,k}^{(III)}+h.c\right) + \sum_{k\in\sigma_N}\left(a^{(I)}_kA_{\mu,k}^{(I)}+a^{(IV)}_kA_{\mu,k}^{(IV)}+h.c\right).\label{3.54}
\end{aligned}
\end{gather} 
Thus, the effect of imposing boundary conditions is a discretization of the spectrum, which turns the integrals in for example \eqref{3.31} into a sum. Eventually we will only be interested in taking the physical limit $r_*\to-\infty$ where the discrete spectrum becomes a continuum with a well-defined measure. This will be discussed in the more relevant case $d>1$ in section \ref{s: implementation bc}.
\subsubsection*{Ghost Theory}
Horizon boundary conditions in Rindler \eqref{2.5} are to be imposed on any theory, including the FP ghosts. The horizon boundary conditions on the FP ghost fields read:
\begin{gather}
\label{3.55}
\begin{aligned}
\partial_r c\rvert_{\partial \mathcal{M}}=0, \quad \partial_r \bar{c}\rvert_{\partial \mathcal{M}}=0.
\end{aligned}
\end{gather}
Comparing with the above discussion, the expansion of the ghost fields \eqref{3.35} in orthogonal modes shows that a similar recombination of the ghost modes into sines and cosines is required by the boundary conditions \eqref{3.55}. Consequently, exactly the same spectra \eqref{3.52} and \eqref{3.53} appear in the ghost field expansion.

This shows the integral over $k$ in \eqref{3.45} is discretized in exactly the same way as is the integral over $k$ in \eqref{3.31}, such that the vanishing of the total Rindler Hamiltonian in the Minkowski vacuum state \eqref{3.46} remains exact after including the effect of the horizon boundary conditions \eqref{2.5}.
\subsection{Summary}
We have analyzed Maxwell theory in $1+1$ dimensional Rindler spacetime in Lorenz gauge. After presenting the canonical structure of the theory, we discussed the Unruh effect. Upon scrutinizing the ghost sector, it was found that a perfect cancellation between FP ghosts and two unphysical polarizations of the Maxwell field takes place, even when the horizon boundary conditions in Rindler are taken into account.

\section{Edge Sector in $1+1$}
\label{s:edge11}
After analyzing the bulk sector in $1+1$ dimensions we turn to an analysis of the edge sector. As explained in section \ref{s: hilbertspaces}, edge states are associated with the horizon DOF \eqref{2.8}. In this section, we work in Rindler coordinates.
\\~\\
The R-wedge bulk theory which saturates the CCR \eqref{3.14} in the R-wedge, violates through the constraints \eqref{2.5} the horizon algebra \eqref{2.15}:
\begin{equation}
\comm{\int_\mathcal{C}A}{\Phi_{\partial\Sigma}}=i\theta (\mathcal{C}\cap \partial \Sigma)\label{4.1},
\end{equation}
or in local fields
\begin{equation}
\comm{A_\rho(\rho)}{\Pi^\rho\rvert_{\partial\Sigma}}=i\delta(\rho). \label{4.2}
\end{equation}
As discussed in section \ref{ss:2.1}, it is natural to expect that the edge mode sector is contained in the $\omega=0$ static sector of the theory. To confirm this we impose \eqref{4.1} on the zero-mode sector and investigate the resulting Hilbert space. 

\subsection{Canonical Quantization}
\label{s: 4.2}
The $\omega=0$ sector of the solution space of the EOM \eqref{2.4} and the transversality condition \eqref{2.2} is:
\begin{align}
A_t=c_t\rho^2+d_t, \quad A_\rho=\frac{d_\rho}{\rho}.\label{4.3}
\end{align}
The field strength is $F^{\rho t}=-\frac{2c_t}{\rho}$, hence:
\begin{align}
\Pi^\rho = -2c_t.\label{4.7}
\end{align}
The Minkowski electric field is denoted as $q$ and equals $q:= F_{TX} =-2c_t$. Thus the zero-modes sector contains constant electric field solutions, in line with \eqref{2.8}. 

For future convenience we redefine the expansion coefficient in \eqref{4.3}:
\begin{equation}
A_\rho = \frac{d_\rho}{\rho} = -a \partial_\rho \left(\frac{\ln \rho}{\ln \epsilon}\right).\label{4.9}
\end{equation}
Clearly, $a$ labels the pure gauge zero mode sector. In the limit $\epsilon\to 0^+$ the function between brackets behaves as
\begin{equation}
\lim_{\epsilon\to 0^+}\frac{\ln \rho}{\ln \epsilon}=
\begin{cases}
1,\quad &\rho=0,\\
0,\quad &\rho\neq 0.
\end{cases}\label{4.10}
\end{equation}
This shows that the $\epsilon\to 0^+$ limit of \eqref{4.9} is
\begin{equation}
A_\rho=a \delta(\rho) .\label{4.11}
\end{equation}
The quantity $a=\int d\rho A_\rho$ represents the expansion coefficient of a pure gauge mode which exponentiates to the radial Wilson line sourced, and localized, on the horizon. The zero mode sector is thus:
\begin{gather}
\begin{aligned}
\Phi=q, \quad \int_\mathcal{C}A=a.\label{4.8}
\end{aligned}
\end{gather}
Edge quantization is performed by imposing the horizon algebra \eqref{4.1} on \eqref{4.8}: 
\begin{equation}
\comm{a}{q}=i.\label{4.9}
\end{equation}
The Wilson line in $1+1$ dimensions is $\mathcal{W}_\mathcal{E}=e^{i \mathcal{E} a}$, this agrees with \eqref{2.15}:
\begin{equation}
\comm{\Phi}{\mathcal{W}_\mathcal{E}}=\mathcal{E}\mathcal{W}_\mathcal{E}.\label{4.15}
\end{equation}
A state with constant electric flux $\mathcal{E}$ is obtained by inserting a Wilson line $\mathcal{W}_\mathcal{E}$ in the vacuum
\begin{equation}
\ket{\mathcal{E}}=\mathcal{W}_\mathcal{E}\ket{0}.\label{4.16}
\end{equation}
Indeed, $q\ket{\mathcal{E}}=\mathcal{E}\ket{\mathcal{E}}$. These are the edge states of the $1+1$ dimensional theory. This is the first explicit example of the discussion on edge states in section \ref{s: hilbertspaces}. 
\\~\\
The boundary Hamiltonian is:\footnote{The canonical Noether Rindler Hamiltonian is $H=-V\frac{\mathcal{E}^2}{2}$, with the opposite sign. This wrong sign is cured by defining the stress tensor directly from the Maxwell action $S$ as
$$
T^{\mu\nu}\equiv -\frac{2}{\sqrt{-g}}\frac{\delta S}{\delta g_{\mu\nu}},
$$
The improved Hamiltonian receives an additional contribution $+V\mathcal{E}^2$ to result in \eqref{4.20}.}
\begin{gather}
\begin{aligned}
H=\int d\rho \sqrt{-g}\left(-\frac{1}{2}F_{\rho t}F^{\rho t}\right)
=\left(\int \rho d\rho\right) \frac{\mathcal{E}^2}{2}
=V\frac{\mathcal{E}^2}{2},\label{4.20}
\end{aligned}
\end{gather}
such that 
\begin{equation}
\hat{H}\ket{\mathcal{E}}=V\frac{\mathcal{E}^2}{2}\ket{\mathcal{E}}.\label{4.21}
\end{equation}
The constant nature of the edge state electric flux profile in $1+1$ dimensions implies a volume divergence in their energy. This shows that the edge state configurations do not contribute to thermodynamic quantities in Rindler, as their thermodynamic weight $e^{-\beta H} = e^{-\beta V\frac{\mathcal{E}^2}{2}}$ vanishes identically. Note that equivalently we can construct the eigenspace of the $a$ operator by working on the vacuum $\ket{a=0}$ as:\begin{equation}
\ket{\chi}:=e^{-i\chi q}\ket{a=0},\quad a\ket{\chi}=\chi\ket{\chi}. 
\end{equation}
The $q$ eigenspace is more natural in the present context since it diagonalizes the boundary Hamiltonian. Since $a$ and $q$ do not commute, summing over one of the sets of boundary DOF in \eqref{2.8} implies a summation over the other set. This is a manifestation of the on-shell equivalence of the transversality condition and Gauss' law.

\subsection{Gluing Rindler Wedges}
We discussed in section \ref{s: hilbertspaces} how both Rindler wedges are glued together in general. Here we apply this to our specific example. Transforming the R-wedge zero mode expansion \eqref{4.3} to Minkowski coordinates, one obtains
\begin{gather}
\begin{aligned}
A_T &= c_t X + \frac{1}{\rho^2}\left(d_t X - d_\rho T\right), \\
A_X &= - c_t T + \frac{1}{\rho^2}\left(d_\rho X - d_t T\right),\label{4.19}
\end{aligned}
\end{gather}
for $\left|X\right| > \left|T\right|$ and $X>0$. 

In the (a priori completely independent) L wedge, there is a similar expansion, with similar operators
\begin{gather}
\begin{aligned}
\Phi=\bar{q}, \quad \int_\mathcal{C}A=\bar{a}.\label{4.16}
\end{aligned}
\end{gather}
that can be analogously transformed to Minkowski coordinates resulting in identically the same expressions \eqref{4.19}, but this time valid in the region $\left|X\right| > \left|T\right|$ and $X<0$. 

The most general combination of L-wedge and R-wedge zero-modes does not result in a valid solution of the free Maxwell equations in Minkowski, as the electric field and the pure gauge mode is constrained by \eqref{2.8} to match on the horizon:
\begin{equation}
q=\bar{q},\quad a=\bar{a}.
\end{equation}
In other words, imposing Gauss' law on physical states constrains the expansion of a general physical Minkowski edge state to the form:
\begin{align}
\ket{\psi} = \sum_\elec f(\elec) \ket{\elec}_L \otimes \ket{\elec}_R=\sum_\chi g(\chi)\ket{\chi}_L\otimes \ket{\chi}_R,
\end{align}
such that $\mathcal{H}_{\text{bulk}} \subset \mathcal{H}_{\text{edge},L} \otimes \mathcal{H}_{\text{edge},R}$ is the diagonal combination, introducing maximal entanglement between L-and R-wedge edge states, conform \eqref{2.11}.

The elementary operators are extended trivially into the full Hilbert space, as e.g. $\bar{q}\to \bar{q} \otimes 1$, $q \to 1 \otimes q$. Electric fields are created by the continuous Wilson line operator
\begin{equation}
\mathcal{W}_\mathcal{E}:= \mathcal{W}_\mathcal{E} \otimes \mathcal{W}_\mathcal{E}
\end{equation}
and measured by either $q$ or $\bar{q}$. Due to the matching, the latter is a gauge invariant operator. Note that the boost operator $B=H_R-H_L$\footnote{$\tau=-t$ in the L-wedge.} commutes with the Wilson line operator on physical states
\begin{equation}
\left[B, \mathcal{W}_\elec\right] =  (q - \bar{q})V\elec\mathcal{W}_\elec = 0,
\end{equation}
such that adding electric flux retains the boost-invariance of the state, thus making these flux states admissible states for construction of the Minkowski vacuum.

\subsection{Minkowski Vacuum}
Next we explore the option of entanglement in the Minkowski vacuum due to the edge states with Hamiltonian \eqref{4.16}. 
\subsection*{Path Integrals}
We will use the Euclidean path integral formalism to define wavefunctionals and the relevant states in the theory. Let us first remind the reader how path integrals in this (free) theory are computed. We present the results using electromagnetic duality, which leads to an immediate final answer for path integrals with possible Wilson loop insertions.

For Abelian gauge theory, in $D$ dimensions, the path integral on a compact manifold (without insertions) can be dualized as follows
\begin{align}
Z &= \int \left[\mathcal{D}A\right]e^{-\frac{1}{2}\int dA \wedge *dA} \nonumber \\
&= \int \left[\mathcal{D}A\right] \left[\mathcal{D}F\right]\left[\mathcal{D}\lambda\right] e^{-\int \frac{1}{2} F \wedge *F + \lambda\wedge \left(F-dA\right)} \nonumber \\
&= \int \left[\mathcal{D}F\right]\left[\mathcal{D}\lambda\right] \delta(d\lambda)e^{-\int \frac{1}{2} F \wedge *F + \lambda \wedge F} \nonumber \\
&= \det(d)^{-1} \int \left[\mathcal{D}F\right]\left[\mathcal{D}f\right] e^{-\int \frac{1}{2} F \wedge *F + df \wedge F} \nonumber \\
&\sim \int \left[\mathcal{D}f\right] e^{-\int \frac{1}{2} df \wedge *df}.
\end{align}
In the first line, a Lagrange multiplier $\lambda$ is introduced. In the second line, $A$ is integrated out. In the third line, $\lambda = df$ is used, and in the final line $F$ is integrated out, and a field-independent determinant factor is dropped. If $A$ is a $p$- form, $F$ is a $p+1$-form, and $\lambda$ is a $D-p-1$-form. This is EM duality exchanging a $p+1$-form field strength for a $D-p-1$-form field strength. The situation we are interested in is a $1$-form in $2$ dimensions. For $D=2$ and $p=1$ the last line should be changed; $\lambda$ is now a scalar ($0$-form) that satisfies $\partial_\mu \lambda = 0$. On a compact manifold this is solved by $\lambda = \mathcal{E}$, a constant, the electric field. So
\begin{equation}
Z = \int d\mathcal{E} e^{-\frac{1}{2} \int \sqrt{g} \mathcal{E}^2} = \int d\mathcal{E} e^{- A\frac{\mathcal{E}^2}{2} }.\label{10.22}
\end{equation}
The area-preserving diffeomorphism invariance of 2d Yang-Mills (see e.g. \cite{Witten:1991we}\cite{Cordes:1994fc}) is very explicit here, and the path integral over $A$ has been traded into one over the electric field $\mathcal{E}$, which turns out to be an ordinary integral.\footnote{Note that this is the solution for any topology of the 2d manifold, we are not restricting to e.g. a disk or a sphere.}

Note immediately that, in terms of the spectrum discussed so far, the partition function reduces to an integral over \emph{only} the zero mode sector. 

\begin{figure}[h]
\centering
\includegraphics[width=0.5\textwidth]{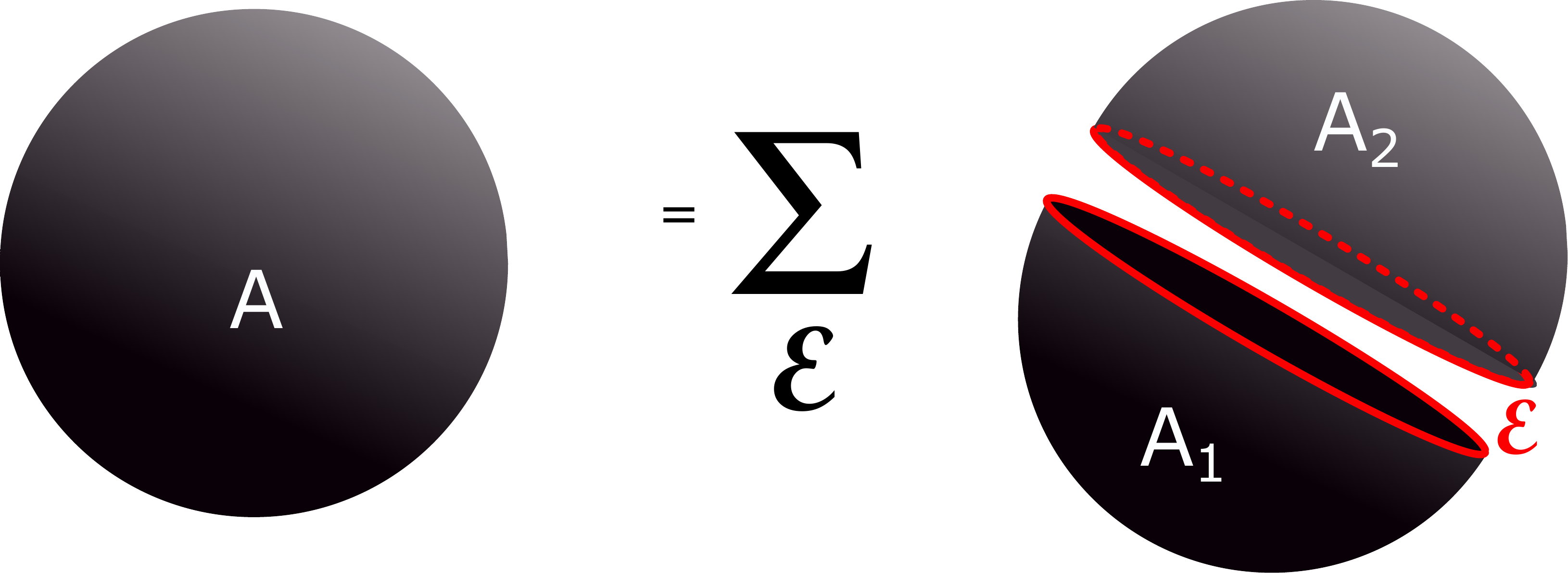}
\caption{Cutting open a path integral along a curve, requires summing over the intermediate value of $\mathcal{E}$. In formulas: $Z = \int d\mathcal{E} e^{-A_1\frac{\mathcal{E}^2}{2}}e^{-A_2\frac{\mathcal{E}^2}{2}}$. Fixing the value of $\mathcal{E}$ for a manifold with boundary gives $Z(\mathcal{E}) \sim  e^{-\frac{A_1}{2}\mathcal{E}^2}$.}
\label{cutting}
\end{figure}

When cutting open the path integral, and specifying a fixed value of $\mathcal{E}$ on the boundary curve (as will be done in the next subsection), one uses the same expression \eqref{10.22}, but without the $\mathcal{E}$-integral (Figure \ref{cutting}).
 On a non-compact manifold, there is no non-trivial solution, and $\mathcal{E}=0$. The path integral becomes trivial. In a theory with matter, this integral will turn into a sum due to the index theorem (cocycle condition on the wavefunction).

When including Wilson loop operators in the path integral, the modification arises when integrating out $A$. The scalar $\lambda$ now is determined by
\begin{equation}
\partial^\mu \lambda = \sum_i e\oint_{\mathcal{C}_i}dz^\mu \delta(z-x),
\end{equation}
which is solved by setting $\lambda= \mathcal{E} + e \sum_i \theta(\mathcal{C}_i)$, where the $\theta$ is a loose notation for being either outside or inside the loop. The electric field jumps when crossing the loop by an amount equal to $e$.
A simple example is a purely spatial Wilson line threading the entire spatial axis (as discussed in \eqref{10.18}). Insertion of such a Wilson line $\mathcal{W}_e$ in the path integral that generates a certain state, adds $e$ to the electric flux in said state. E.g. on the thermal state:
\begin{equation}
\mathcal{W}_e\left|\Psi\right\rangle = \mathcal{W}_e\frac{1}{Z}\sum_{\elec} e^{-A\frac{\elec^2}{2}}\left|\elec\right\rangle_L\otimes \left|\elec\right\rangle_R =\frac{1}{Z}\sum_{\elec} e^{-A\frac{\elec^2}{2}}\left|\elec+e\right\rangle_L\otimes \left|\elec+e\right\rangle_R.\label{10.18}
\end{equation}
\subsection*{Minkowski Wavefunctionals}
\begin{figure}[h]
\centering
\includegraphics[width=0.9\textwidth]{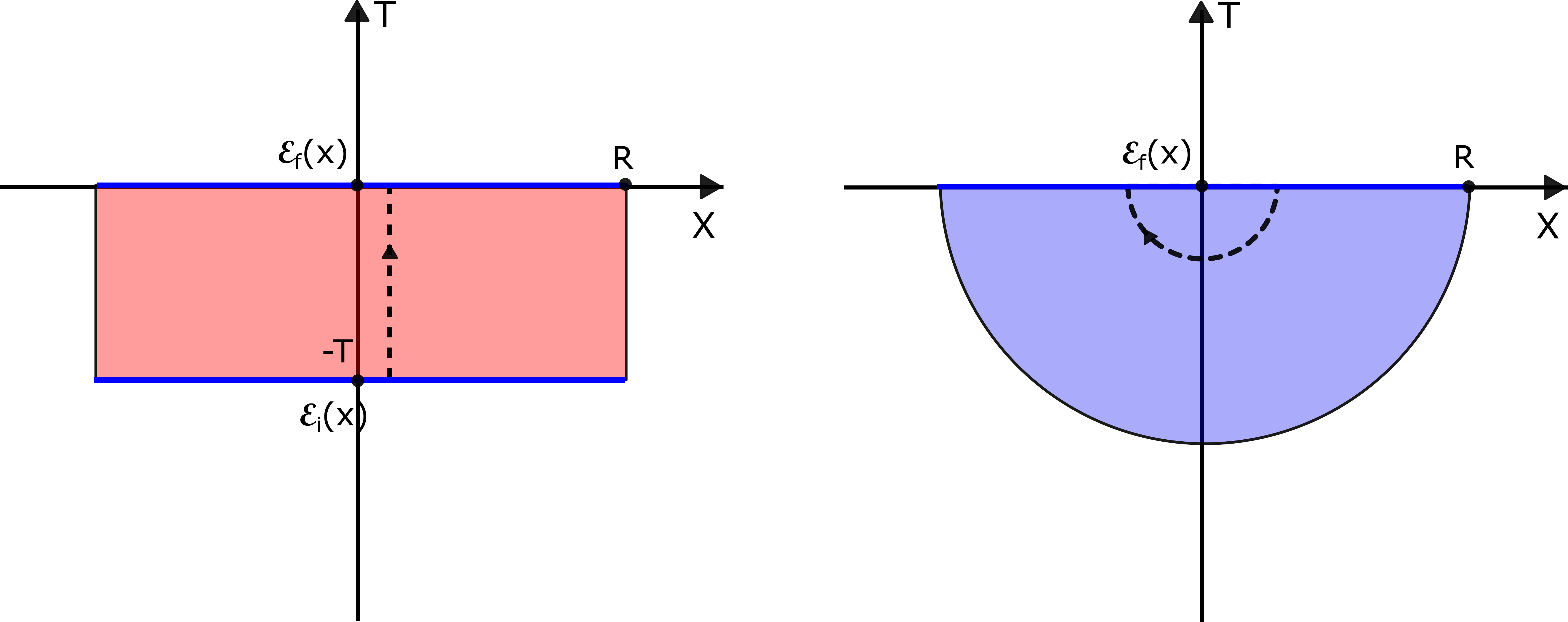}
\caption{Volume-regularized path integral. The blue semi-disk and the red rectangle have equal area, and hence the final state can be constructed using both regions.}
\label{eucl}
\end{figure}

Next, we construct the wavefunctionals of the relevant states. Consider the path integral on a strip of total Euclidean time $T$, evolving from an initial field $\elec_i(x)$ to the field $\elec_f(x)$ (see Fig. \ref{eucl} left).
This is written as
\begin{equation}
Z\left[\elec_i,\elec_f; T\right] = \left\langle \elec_f, 0\right|\left.\elec_i, -T\right\rangle = \sum_n \Psi_n\left[\elec_i(x)\right]^*\Psi_n\left[\elec_f(x)\right] e^{-E_n T}\label{path integral},
\end{equation}
where $\{\Psi_n\}$ forms a complete set of eigenfunctionals of the Hamiltonian. The wavefunctionals $\Psi$ are constrained by Gauss' law such that $\elec_i(x) = \elec_i$ and $\elec_f(x) = \elec_f$. The Hamiltonian for those field configurations is that of a free particle $H = \int dx \, \elec(x)^2/2$ such that the eigenbasis is simply the electric field basis $\Psi_n\left[\elec(x)\right] \sim \prod_x \delta(\elec(x)-\elec_n)$, with a proportionality factor independent of $n$. To prepare the wavefunctional for the field $\elec$ at $T=0$ one (path) integrates \eqref{path integral} over the initial field configurations $\elec_i$
\begin{equation}
\Psi_T\left[\elec\right] = \int [\mathcal{D} \elec_i] \, Z\left[\elec_i,\elec; T\right],
\end{equation}
such that
\begin{equation}
\Psi_T\left[\elec\right] = \sum_n \Psi_n \left[\elec\right] e^{-E_n T} \sim e^{-\frac{\elec^2}{2}LT}=e^{-A\frac{\mathcal{E}^2}{2}},\label{wavefunctional}
\end{equation}
where the normalization was left out. The result is a Gaussian distribution centered around $\elec=0$, or a thermal superposition of wavefunctionals that have a fixed electric field $\elec$. Taking $T$ to infinity, \eqref{wavefunctional} reduces to the ground state wavefunctional which has vanishing electric field. Indeed, $\Psi_\infty\qty[\mathcal{E}]=\delta(\mathcal{E})$.
\\~\\
It is possible to describe this same wavefunctional from the perspective of entanglement between L- and R-states. We thereto deform the integration region using the area-preserving diffeomorphism into a circular region,\footnote{In this process the total area over which is path integrated in preserved, such that $$A= LT=\frac{1}{2}\pi R^2.$$} which for convenience is chosen to be just as wide: $L=2R$ (see Fig. \ref{eucl} right). The wavefunctional \eqref{wavefunctional}, which is due to the area preserving diffeomorphism unaltered in this process, can now alternatively be written as
\begin{equation}
\Psi_T\left[\elec\right] = \left(\left\langle \elec\right|_L\otimes\left\langle \elec\right|_R\right) \left|\Psi \right\rangle \sim \left\langle \elec\right| e^{-\pi H}\left|\elec\right\rangle,
\end{equation}
where angular evolution is generated by the zero-mode Rindler Hamiltonian \eqref{4.20}. This Hamiltonian generates a thermal sum at inverse temperature $2\pi$, in the sense that the reduced density matrix of the state $\ket{\Psi}$ defined as $\rho_R = \text{Tr}_L \left|\Psi\right\rangle \left\langle \Psi\right|$ has the property
\begin{equation}
\left\langle \elec\right| \rho_R \left| \elec' \right\rangle = \left\langle \elec\right| e^{-2\pi H} \left| \elec' \right\rangle = \delta_{\elec\elec'}e^{-\frac{\pi R^2}{2}\frac{\mathcal{E}^2}{2}}.
\end{equation}
The state associated with the wavefunctional \eqref{wavefunctional} can thus be expanded as
\begin{equation}
\left|\Psi\right\rangle =\frac{1}{Z} \sum_{\elec}  e^{-\frac{\pi R^2}{2}\frac{\mathcal{E}^2}{2}} \left|\elec\right\rangle_L \otimes \left|\elec\right\rangle_R,
\end{equation}
and is a diagonally entangled state of L-and R-wedge zero mode states.

\begin{figure}[h]
\begin{minipage}{0.48\textwidth}
\centering
\includegraphics[width=0.8\textwidth]{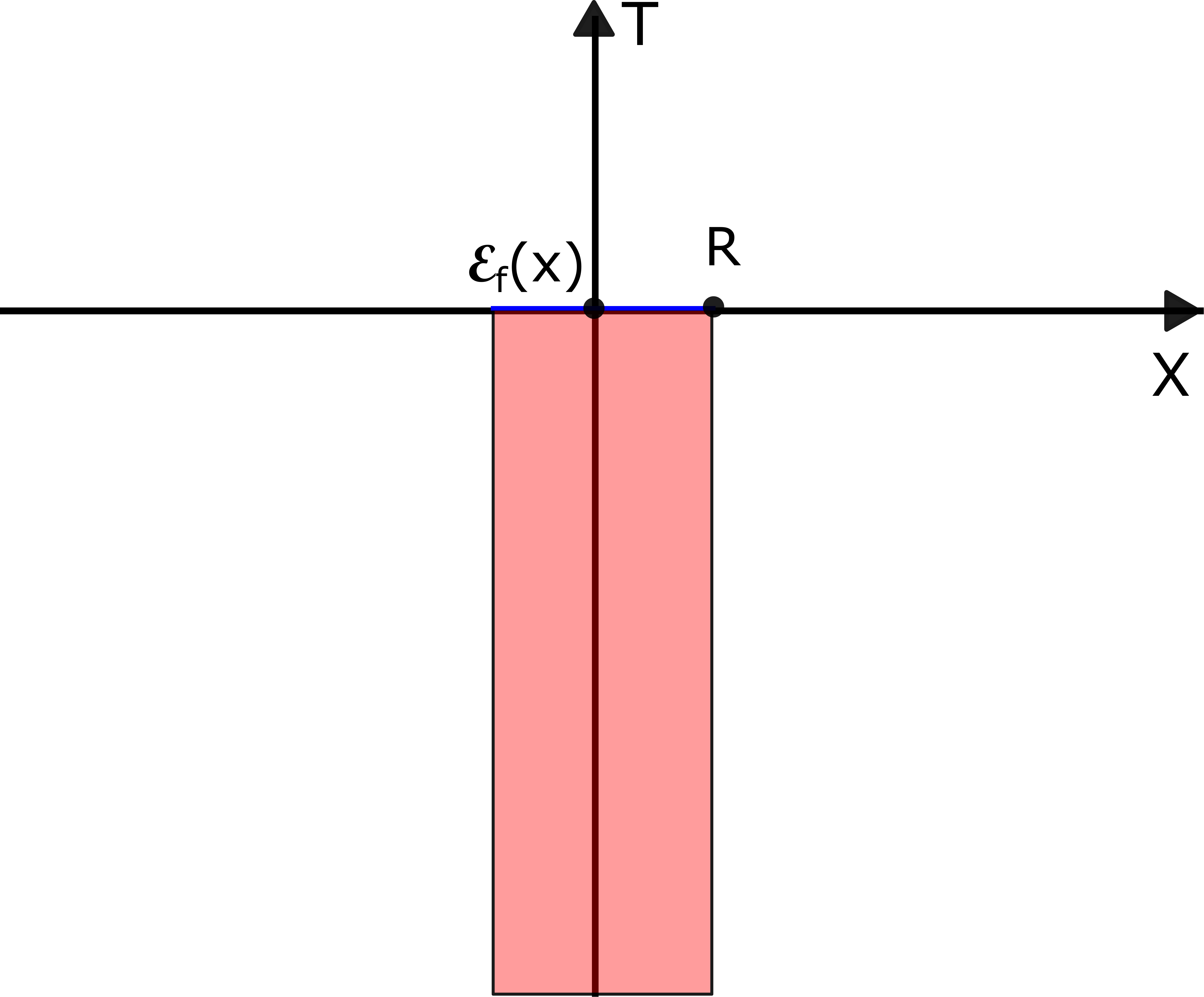}
\caption{Volume-regularized path integral. The vertical strip extends all the way to $-\infty$ making the area also infinitely large.}
\label{eucl2}
\end{minipage}
\hfill
\begin{minipage}{0.48\textwidth}
\centering
\includegraphics[width=0.8\textwidth]{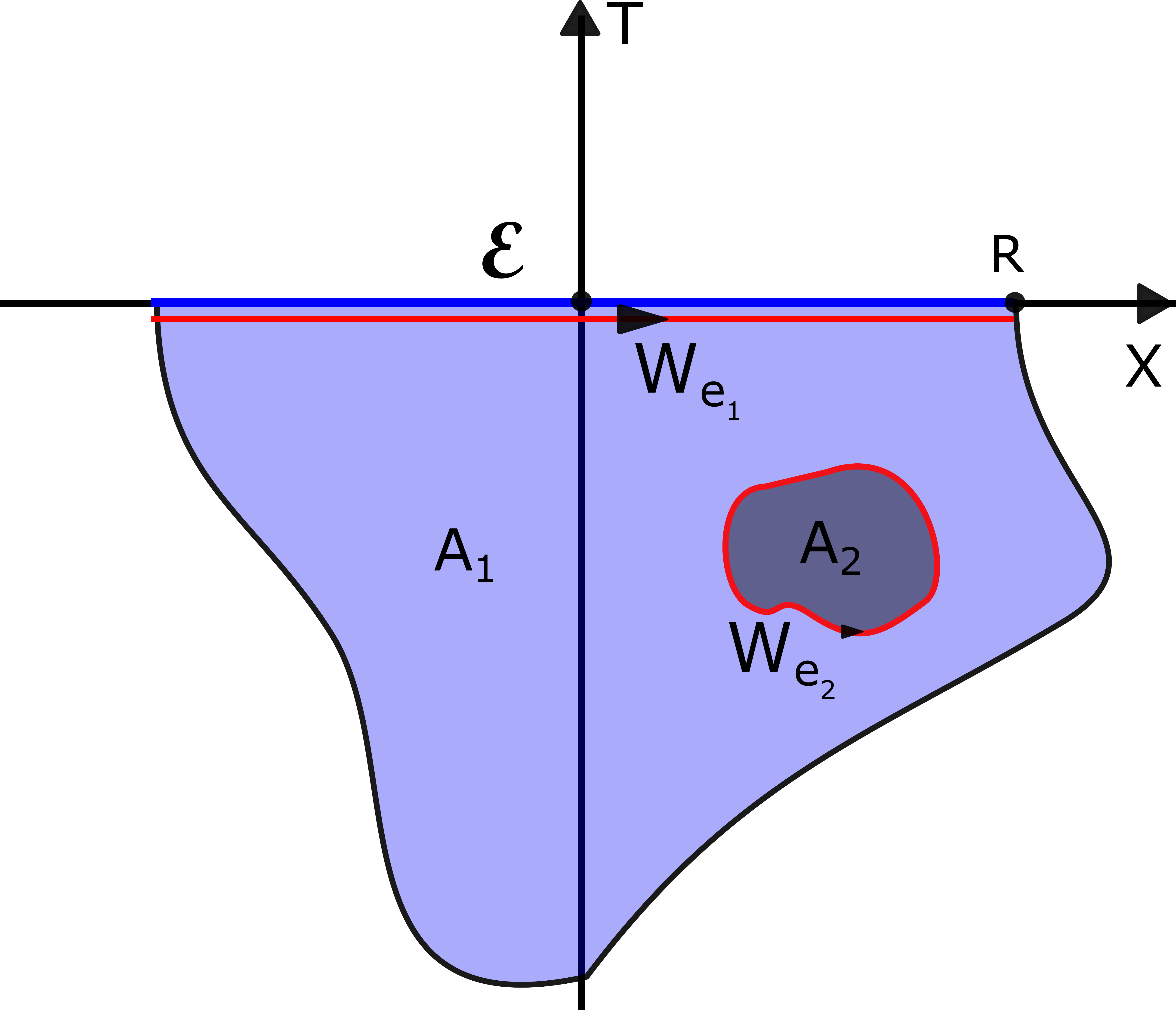}
\caption{Example of a general Euclidean path integral with Wilson loop and line insertions.}
\label{wilsonstate}
\end{minipage}
\end{figure}

One might naively argue that the procedure described above prepares the finite volume Minkowski vacuum. However, the Minkowski vacuum is only obtained in the limit $T\to\infty$ of \eqref{wavefunctional}. The area over which one needs to path integrate to obtain this vacuum wavefunctional is shown in figure \ref{eucl2}. The total area of this path integral diverges, even when preparing the vacuum of a volume-regularized theory.

Since the area-preserving diffeomorphism ensures the final thermal sum only depends on this area, the Minkowski state reduces to just the vacuum product state:
\begin{equation}
\boxed{
\left|M\right\rangle = \ket{0}_L \otimes \ket{0}_R},\label{mink vac zero}
\end{equation}
independent of any spatial volume regularization. This same argument generalizes immediately to $1+1$ dimensional non-Abelian Yang-Mills theory, as only the area-preserving diffeomorphism has been used.\footnote{For completeness, one can generalize the above construction to arbitrary Wilson loop and Wilson line insertions to produce other states. An example is given in Figure \ref{wilsonstate}, where the path integral prepares the wavefunctional:
\begin{equation}
\Psi\left[\mathcal{E}\right] \sim e^{-\frac{(\mathcal{E}-e_1)^2}{2}A_1}e^{-\frac{(\mathcal{E}-e_1-e_2)^2}{2}A_2}
\end{equation}
Upon taking $A_1\to+\infty$, this state becomes an electric field eigenstate:
\begin{equation}
\Psi\left[\mathcal{E}\right] \sim \delta(\mathcal{E} - e_1).
\end{equation}} 

The simple expression \eqref{mink vac zero} for the Minkowski vacuum state (only considering the zero modes) implies that the zero-mode contribution to the Rindler entropy vanishes. Combining this with our previous result from section \ref{s ghost}, we find that the total entanglement entropy of gauge fields in $1+1$ dimensional Rindler vanishes $S_E=0$, even when a volume regularization is applied to deal with the volume divergence in the zero mode energy \eqref{4.21}. 

Note that the zero mode sector discussed in this section is identical to the $k=0$ sector of the edge mode theory discussed in section \ref{s:boundary and edge}. As discussed in this section, these constant flux solutions are non-normalizable and therefore do not contribute to the thermodynamics. In fact, dropping the possibility of volume regularization, they do not even represent admissible solutions since they do not vanish at spatial infinity.
\subsection{Summary}
\label{s: conclusion part 1}
The static sector of $1+1$ dimensional Maxwell theory in Rindler spacetime leads to constant electric field solutions. As these carry infinite energy in infinite space, they are irrelevant for thermodynamic consideration. If one however regularizes space with periodic boundary conditions, this sector is present and represents the edge sector. The Minkowski vacuum has no such fields, and is completely trivial in terms of left and right Rindler wedges.

The main reason for this outcome is the constant electric flux profile associated with the edge states, pushing us to consider higher dimensions where zero modes with nontrivial spatial electric flux profiles exist.

\section{Bulk Maxwell Theory in $d+1$}
\label{s: action}

In this section we construct the R-wedge physical bulk Hilbert space in $d+1$ dimensions. Defining elementary scalar modes:\footnote{$\mathbf{x}$ denotes all coordinates parallel to the horizon. These modes are normalized as
\begin{equation}
\int_{0}^{+\infty} \frac{d\rho}{\rho}d\mathbf{x} \, \phi_{\omega, \mathbf{k}}(\rho,\mathbf{x}) \, \phi_{\omega', \mathbf{k}'}(\rho,\mathbf{x}) = \frac{1}{2\omega} \delta(\omega-\omega')\delta_{\mathbf{k},\mathbf{k}'}.
\end{equation}}
\begin{equation}
\label{elmode}
\phi_{\omega, \mathbf{k}}=\frac{\sqrt{\sinh(\pi\omega)}}{(2\pi)^{\frac{D-2}{2}}\pi}K_{i\omega}(k\rho)e^{i\mathbf{k}\cdot \mathbf{x}}e^{-i\omega t},
\end{equation}
solving $\Box \phi = -\frac{1}{\rho^2} \partial_t^2\phi + \frac{1}{\rho}\partial_\rho(\rho\partial_\rho\phi) - k^2\phi = 0$, and introducing unit vectors along the $d+1$ directions, as
\begin{alignat}{3}
\e^{(0)}_\mu = \left(\rho,0,\mathbf{0}\right), \quad\, \e^{(1)}_\mu = \left(0, 1 , \mathbf{0}\right), \quad\,
\e^{(k)}_\mu = \left(0,0,\frac{\mathbf{k}}{k}\right), \quad\, \e^{(a)}_\mu = \left(0,0,\mathbf{n}^a\right)
\end{alignat}
normalized as $e^{(\alpha)}\cdot e^{(\beta)}=\eta^{\alpha\beta} =\text{diag}(-,+^d)$, one can solve the bulk equations of motion $\nabla^\mu\nabla_\mu A_\nu=0$ by the orthogonal set of modes \cite{{Higuchi:1992td}}: 
\begin{gather}
\begin{aligned}
A_{\mu,\omega\mathbf{k}}^{(1)} &= \frac{1}{k}\left(\rho\partial_\rho, \frac{1}{\rho}\partial_t, \mathbf{0}\right)\phi_{\omega, \mathbf{k}} = \frac{1}{k}\left(\e^{(0)}_\mu \partial_\rho \phi_{\omega, \mathbf{k}} + \e^{(1)}_\mu\frac{1}{\rho} \partial_t \phi_{\omega, \mathbf{k}}\right), \\
A_{\mu,\omega\mathbf{k}}^{(0)} &= \frac{1}{k}\left(\partial_t, \partial_\rho,\mathbf{0}\right)\phi_{\omega, \mathbf{k}}=A_{\mu,\omega\mathbf{k}}^{(G)}-A_{\mu,\omega\mathbf{k}}^{(k)}, \\
A_{\mu,\omega\mathbf{k}}^{(k)} &= i\e^{(k)}_\mu \phi_{\omega, \mathbf{k}}, \\
A_{\mu,\omega\mathbf{k}}^{(a)} &= i\e^{(a)}_\mu \phi_{\omega, \mathbf{k}}.
\end{aligned}\label{5.5}
\end{gather}
These span the solution space of the equations of motion, normalizable as $\rho \to +\infty$, but with no boundary conditions imposed at $\rho=0$.\footnote{The structure of this solution space is made more manifest by working in tortoise light-cone components. We discuss this for tensor field theories in section \ref{s:hs}.} In a more naive treatment, one takes these modes as is to construct the bulk Rindler Hilbert space. The quantum field where horizon boundary conditions and constraint equations are neglected can thus be expanded as:
\begin{equation}
\label{QFgen}
A_\mu=\sum_{\omega,\mathbf{k}}\hat{\alpha}^{(1)}_{\omega\mathbf{k}}A^{(1)}_{\mu,\omega\mathbf{k}} + \hat{\alpha}^{(0)}_{\omega\mathbf{k}}A^{(0)}_{\mu,\omega\mathbf{k}} + \hat{\alpha}^{(k)}_{\omega\mathbf{k}}A^{(k)}_{\mu,\omega\mathbf{k}} + \hat{\alpha}^{(a)}_{\omega\mathbf{k}}A^{(a)}_{\mu,\omega\mathbf{k}} + (hc).
\end{equation}
The continuous modes are normalized in the Klein-Gordon norm, imposing oscillator commutation relations.\footnote{See also appendix \ref{a:bulk}.} For example:
\begin{equation}
\bigl[\hat{\alpha}^{(1)}_{\omega\mathbf{k}},\hat{\alpha}^{{(1)}\dagger}_{\omega'\mathbf{k}'}\bigr]=\delta (\omega-\omega') \delta (\mathbf{k}-\mathbf{k}').
\end{equation}
The bulk Rindler Hilbert space is obtained by imposing the boundary conditions \eqref{2.5} on this expansion, which constrains the range of $\omega$ in function of $\mathbf{k}$. The range of $\mathbf{k}$ depends on transverse boundary conditions, which will not be specified. 

The set of modes $A^{(G)}_{\mu,\omega\mathbf{k}} = \frac{1}{k}\nabla_\mu \phi_{\omega,\mathbf{k}}$ are pure gauge and orthogonal to the sets $A^{(1)}_{\mu,\omega\mathbf{k}},A^{(a)}_{\mu,\omega\mathbf{k}}$. The modes $A^{(k)}_{\mu,\omega\mathbf{k}}$ represent longitudinal modes that violate Lorenz gauge:
\begin{equation}
\nabla^\mu A^{(k)}_\mu = - k\phi \neq 0. 
\end{equation}
The other sets of modes do satisfy the transversality constraint. The physical R-wedge bulk Hilbert space $\{\ket{\psi}\}$ is then obtained by imposing $\hat{\alpha}^{(k)}_{\omega\mathbf{k}}\ket{\psi}=0$ and by quotienting out the null vectors.
\\~\\
The fact that $\omega$ and $\mathbf{k}$ are unrelated a priori is a consequence of a symmetry \cite{Lenz:2008vw}. The spatial scale transformation $\rho \to a\rho$, $\mathbf{x}_\perp \to a \mathbf{x}_\perp$ leads to a scale transformation in the full Rindler metric: $ds^2 \to a^2 ds^2$. It maps a hyperbolic trajectory into another one. This transformation is a symmetry of the Hamiltonian, provided one transforms the field itself as well using its standard scaling dimension.
An individual mode transforms as:
\begin{equation}
K_{i\omega}(k\rho) e^{i\mathbf{k}\cdot \mathbf{x}}e^{-i\omega t} \to K_{i\omega}(a k\rho) e^{i a\mathbf{k}\cdot \mathbf{x}} e^{-i\omega t},
\end{equation}
mapping $\psi_{\omega,\mathbf{k}} \to \psi_{\omega,a\mathbf{k}}$, with the same frequency. This, combined with rotational invariance, demonstrates that for any given $\omega$, any choice of $\mathbf{k}$ is a valid mode solution, leading to an infinite degeneracy of energy eigenstates, and the lack of a dispersion relation. Written in tortoise coordinates, this is merely translation invariance in $r$. 

This discussion applies equally to tensor field modes (to be discussed around equation \eqref{solhs}), mapping modes with $\mathbf{k}$ into modes with $a \mathbf{k}$. 
Inserting a brick wall (or any other boundary condition) at $\rho =\epsilon$ breaks this symmetry and leads to a dispersion relation $\omega(\mathbf{k})$.\footnote{String wave packets spread out in tortoise coordinates (as first argued by Susskind \cite{Susskind:1993aa}, see also \cite{Dodelson:2015toa,Dodelson:2017hyu}), demonstrating that their dynamics breaks this symmetry. For each higher spin field in the spectrum however, the symmetry is intact.}
\subsection{Implementation of Boundary Conditions}
\label{s: implementation bc}
The bulk R-wedge theory is defined to be subject to PMC boundary conditions:
\begin{align*}
n_\mu \left.\frac{\partial \mathcal{L}}{\partial \partial_\mu A^R_\nu}\right|_{\partial \mathcal{M}}(\mathbf{x})&=0,\quad \nu=t,i\\
A^\nu\rvert_{\partial \mathcal{M}}(\mathbf{x})&=0,\quad \nu=\rho.
\end{align*}
To make sense of these boundary conditions, we regularize the boundary surface from $\rho=0$ to $\rho=\epsilon$, and take the $\epsilon\to 0$ at the very end. In terms of the field strength the boundary conditions read:
\begin{align*}
\Pi^\rho\rvert_{\partial \mathcal{M}}(\mathbf{x})=0, \quad \rho F_{\rho i}\rvert_{\partial \mathcal{M}}(\mathbf{x})=0,\quad A^\rho\rvert_{\partial \mathcal{M}}(\mathbf{x})=0.
\end{align*}
In $3+1$ dimensions, these reduce to:
\begin{align}
\mathbf{E}_\bot\rvert_{\partial \mathcal{M}}(\mathbf{x}) =\mathbf{0}, \quad \rho\mathbf{B}_\parallel\rvert_{\partial \mathcal{M}}(\mathbf{x}) =\mathbf{0}, \quad A^\rho\rvert_{\partial \mathcal{M}}(\mathbf{x})=0\label{5.17}
\end{align}
$\mathbf{E}$ and $\mathbf{B}$ being the local field perceived by fiducial observers. Note the metric factor in the magnetic boundary condition. On the boundary $\partial \mathcal{M}:\rho=0$ this merely implies regularity of the transverse magnetic field. Since regularity is a sensible constraint, this presents with an intuitive explanation for why there are no edge modes associated with the $\nu=i$ boundary conditions.\footnote{More explicitly, the statement remains that edge modes are associated with the boundary DOF associated with constraint equations in the bulk; a claim that is confirmed by matching of the partition function with the replica trick partition function for various theories, see sections \ref{s:boundary and edge} and \ref{s:extension}. The absence of edge modes associated to tangential magnetic fields is explained more rigorously from a path integral perspective in \cite{Blommaert:2018oue}.} 

Appendix \ref{a:bulk} shows the mode decomposition of the field strength. It follows that the boundary conditions \eqref{5.17} that modes $A^{(1)}_{\mu,\omega\mathbf{k}}$ satisfy the boundary conditions if the associated Macdonald wavefunctions satisfy Dirichlet boundary conditions: 
\begin{equation}
\phi_{\omega,\mathbf{k}}\rvert_{\rho=\epsilon}=0.\label{5.34}
\end{equation}
Likewise modes $A^{(a)}_{\mu,\omega\mathbf{k}}$ satisfy the Rindler horizon boundary conditions if the associated Macdonald wavefunctions satisfy Neumann boundary conditions:
\begin{equation}
\rho\partial_\rho\phi_{\omega,\mathbf{k}}\rvert_{\rho=\epsilon}=0.\label{5.35 bis}
\end{equation}
The same is true for the modes $A^{(k)}_{\mu,\omega\mathbf{k}}$ and $A^{(G)}_{\mu,\omega\mathbf{k}}$. The relevant information about the physical R-wedge bulk Hilbert space is thus captured in the Dirichlet and Neumann spectra:
\begin{equation}
\sigma_D:=\{\omega|K_{i\omega}(k\epsilon)=0\},\quad \text{and} \quad \sigma_N:=\{\omega|\rho\partial_\rho K_{i\omega}(k\epsilon)=0\}\label{bc neumann}.
\end{equation}
The mode expansion of the physical part of the bulk quantum field $A_\mu$ (\ref{QFgen}) becomes:
\begin{equation}
A_\mu(t,\rho,\mathbf{x})=\sum_\mathbf{k}\left(\sum_{\omega\in\sigma_D}\hat{\alpha}^{(1)}_{\omega\mathbf{k}}A^{(1)}_{\mu,\omega\mathbf{k}}+\sum_{\omega\in\sigma_N}\sum_a\hat{\alpha}^{(a)}_{\omega\mathbf{k}}A^{(a)}_{\mu,\omega\mathbf{k}}\right) +(hc) \label{5.13}.
\end{equation}

Discretization of the spectrum as \eqref{bc neumann} does not affect the orthonormality and completeness of the solution space. In fact, orthonormality on the solution space of a set of differential equations requires an appropriate set of boundary conditions on its boundary. For Maxwell theory, this boils down to choosing for each polarization $(0),(1),(a),(k)$ separately either Neumann of Dirichlet boundary conditions. The specific choice \eqref{5.13} are the PMC boundary conditions, which we argued in section \ref{s: hilbertspaces} to be the natural ones. 
Orthonormality and completeness of the discretized sets \eqref{bc neumann} are readily proven from the continuous relations upon replacing:
\begin{equation}
\int d\omega \to \lim_{\epsilon\to 0}\sum_\omega \Delta^\epsilon, \quad \delta (\omega-\omega')\to\lim_{\epsilon\to 0}\frac{\delta_{\omega\omega'}}{\Delta^\epsilon}, \quad \phi_{k,\omega}\to \sqrt{\Delta^{\epsilon}}\phi_{k,\omega},
\end{equation}
where $\Delta^\epsilon$ is the regularized energy separation between two consecutive modes in the discretized spectrum. For example, the completeness relation is:
\begin{equation}
\lim_{\epsilon\to 0}\sum_\omega \Delta^\epsilon \frac{2\omega\sinh \pi \omega}{\pi^2}K_{i\omega}(k\rho)K_{i\omega}(k\rho')=\rho \delta (\rho-\rho')\label{4.38}.
\end{equation}
The modified Bessel function has the followings asymptotics:
\begin{equation}
K_{i\omega}(x) \approx \frac{1}{2}\Gamma(i\omega)e^{-i\omega \ln\frac{x}{2}} + cc, \quad x \ll 1, \quad\quad K_{i\omega}(x) \approx \frac{\sqrt{\pi}}{2}\frac{e^{-x}}{\sqrt{x}}, \quad x \gg 1\label{5.16}
\end{equation}
and exhibits uncontrollable oscillations near $\rho\approx 0$. For $k\epsilon\ll 1$ the spectra \eqref{bc neumann} become: 
\begin{equation}
\sigma_D:=\{\omega_n,|\omega_n=\frac{\pi n}{\ln \frac{2}{k\epsilon}}\},\quad \text{and} \quad \sigma_N:=\{\omega_n|\omega_n=\frac{\pi (n-1/2)}{\ln\frac{2}{k\epsilon}}\},\quad n\in \mathbb{N}_0\label{5.18}.
\end{equation}
\begin{figure}[h]
\begin{minipage}{0.48\textwidth}
\centering
\includegraphics[width=0.95\linewidth]{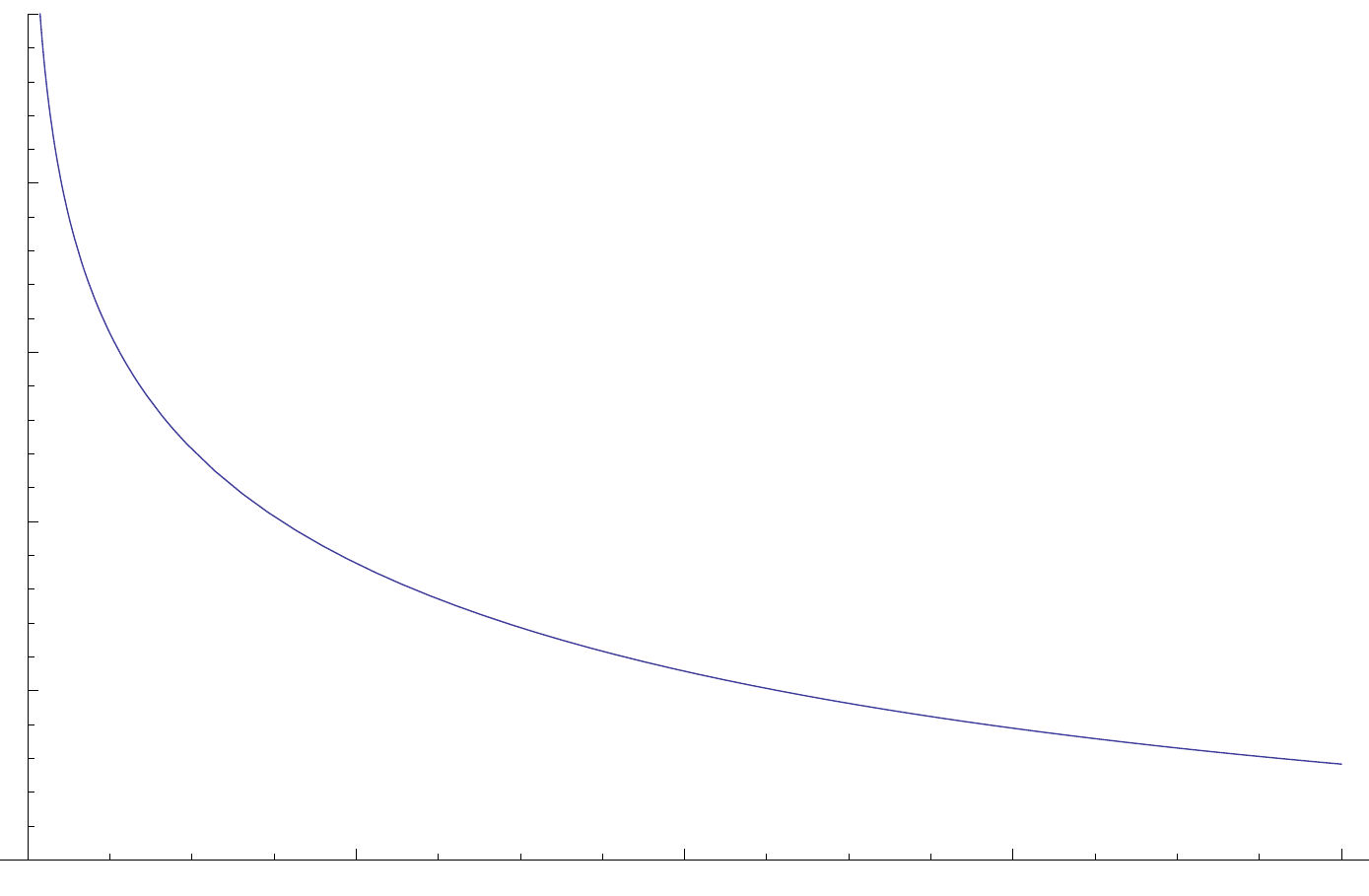}
\caption{$K_{0}(x)$ is a monotonically decreasing function.}
\label{pl:K0}
\end{minipage}
\hfill
\begin{minipage}{0.48\textwidth}
\centering
\includegraphics[width=0.95\linewidth]{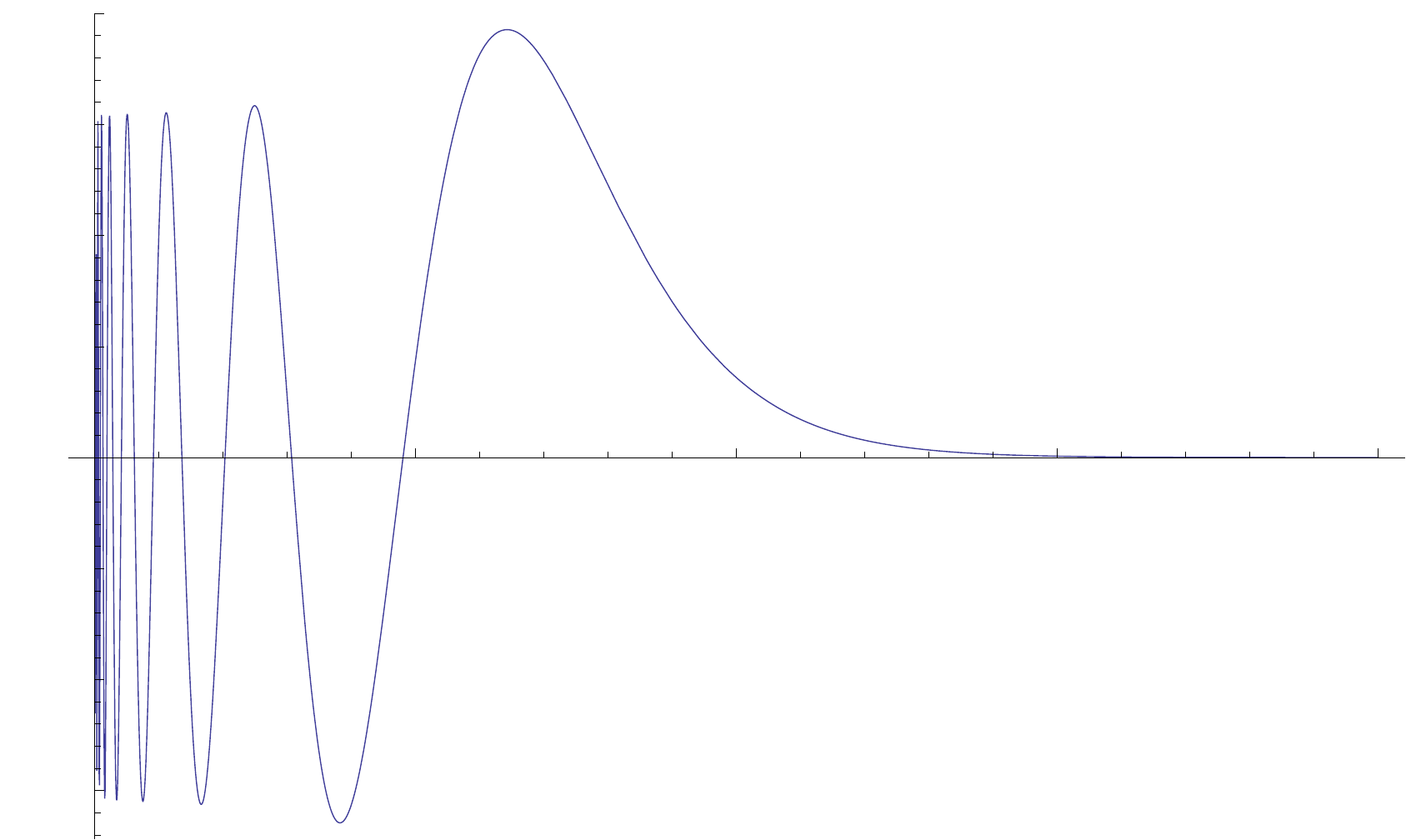}
\caption{$K_{i\omega}(x)$ oscillates wildly as $\sim e^{i\omega\ln(x)}$ for small $x$.}
\label{pl:Kim}
\end{minipage}
\end{figure}
These $k\epsilon\ll1$ spectra are equidistant, with energy difference: 
\begin{equation}
\Delta^\epsilon = \frac{\pi}{\ln\bigl( \frac{2}{k\epsilon}\bigr)} .
\end{equation}

\begin{figure}[h]
\centering
\includegraphics[width=0.4\textwidth]{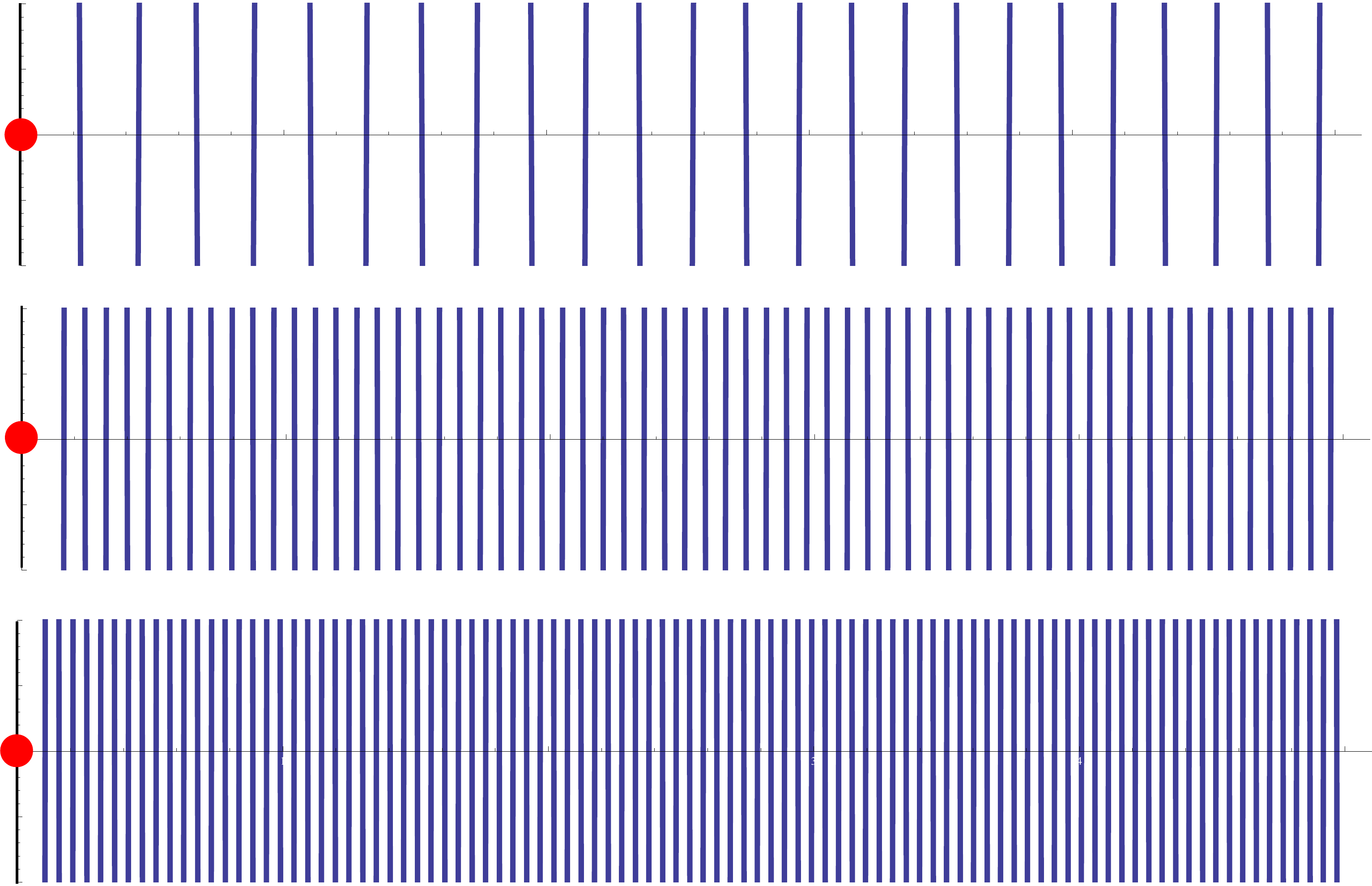}
\caption{Frequency spectrum $K_{i\omega}(\epsilon) =0$ as $\epsilon \to 0$. The spectrum becomes more and more dense, with the exception of $\omega=0$, which is never included in the physical spectrum with zero electric field on the horizon.}
\label{spectrumLimit}
\end{figure}

Obviously the physics of the system is only accurately described upon taking $\epsilon \to 0$. In this situation the bulk theory simplifies: the spacing between different solutions goes to zero $\lim_{\epsilon\to 0}\Delta^\epsilon=0$, such that the spectrum behaves effectively as a continuum (Figure \ref{spectrumLimit}).
All frequencies appear in the $\epsilon\to 0$ spectra \eqref{5.18}, with the exception of the zero mode $\omega=0$. For this zero-mode the small-$\epsilon$ approximation \eqref{5.16} is not valid, since $K_0(x)$ displays fundamentally different behavior near the origin compared to $K_{i\omega}(x)$. The latter oscillate out of control whereas the former blows up as $K_0(x)=\ln\frac{2}{kx}$. Such a wavefunction cannot satisfy the Dirichlet \eqref{5.34} nor the Neumann \eqref{5.35 bis} boundary condition. The associated quantum states are therefore not present in the bulk Hilbert space of the theory. 
\\~\\
The bulk theory \eqref{5.13} consists of one set of Dirichlet scalar field DOF and $d-2$ sets of Neumann scalar field DOF. In terms of partition functions, there is a subtle difference between a Neumann scalar DOF and a Dirichlet scalar DOF. In appendix \ref{a:d vs n} we prove that:
\begin{equation}
\frac{Z_D}{Z_N}=\prod_k\left(\frac{\beta}{\ln \frac{2}{k\epsilon}}\right)^{\frac{1}{2}}. \label{DvsN}
\end{equation}
The bulk entropy of Maxwell theory is \footnote{$S_{\text{scalar}}:=S_{N}$, see also the discussion at the end of section \ref{s:boundary and edge}.}
\begin{equation}
S_{\text{bulk}}=(d-1)S_{\text{scalar}}+(\Delta S)_{DN}.
\end{equation}
\subsection{Summary}
The theory splits in a bulk piece with vanishing electric flux on the horizon which is regularized to $\rho=\epsilon$, and an edge piece that accounts for the non-zero flux. The edge theory is accounted for in the zero-mode ($\omega=0$) sector. The bulk photon behaves for thermodynamical purposes as $d-1$ scalars, of which one has Dirichlet boundary conditions and the other $d-2$ have Neumann boundary conditions. 

\section{Edge Sector in $d+1$}
\label{s:boundary and edge}

\subsection{Canonical Quantization}
\label{ss: can quant}
Quantization of the edge sector is achieved in the same way as in $1+1$ dimensions. We introduce the zero mode DOF associated with pure gauge modes and radial electric flux and impose on the resulting expansion the commutator: 
\begin{equation}
\comm{\int_\mathcal{C}A}{\Phi_\Omega}=i\theta (\mathcal{C}\cap \Omega)\label{6.1},
\end{equation}
with $\Omega \subset \partial \Sigma$. This ensures the Wilson line algebra \eqref{2.15} is valid throughout $\mathbb{R}^{1,d}$.

\begin{figure}[h]
\centering
\includegraphics[width=0.5\textwidth]{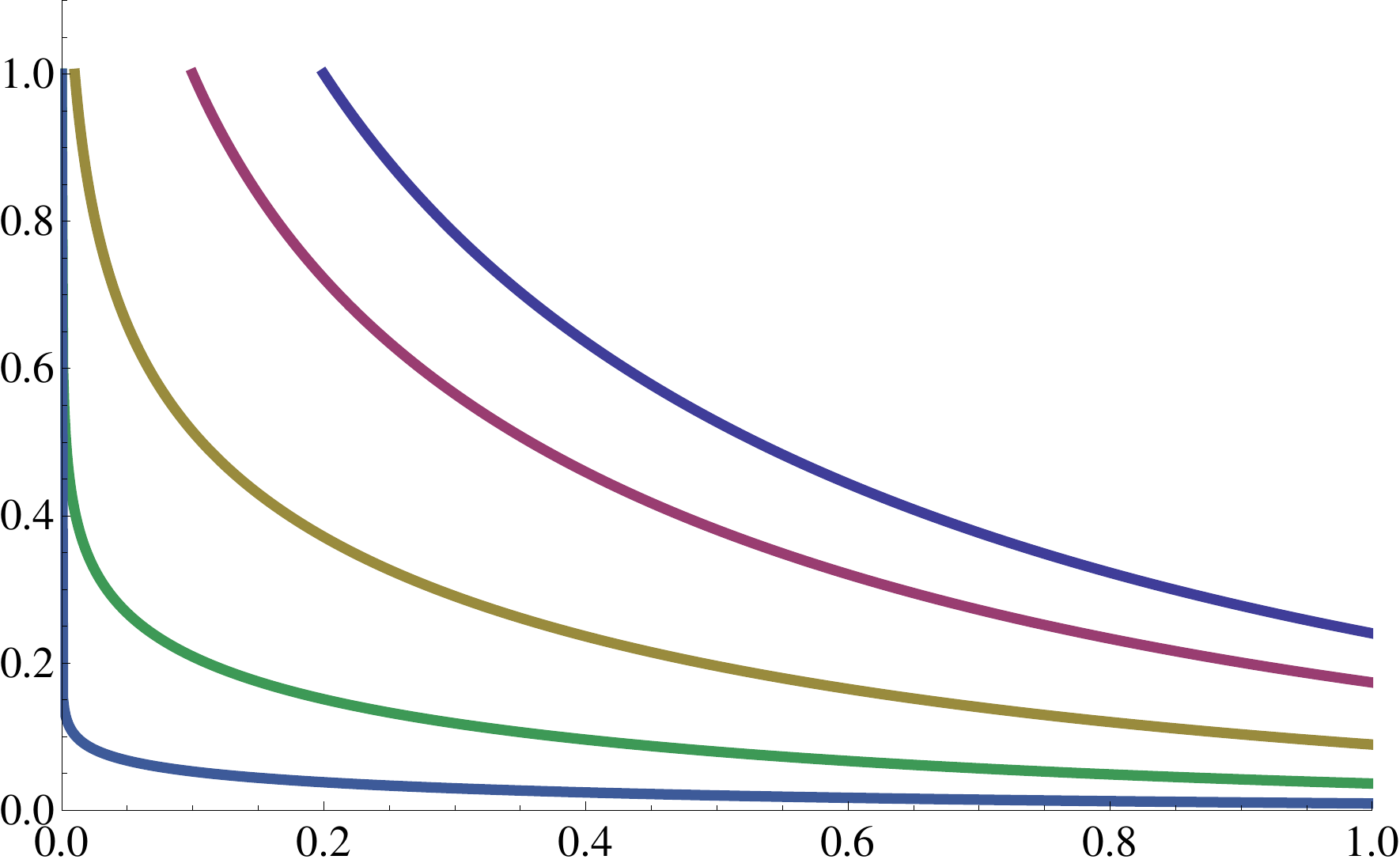}
\caption{Edge state mode function $\phi_{\mathbf{k}}(\rho)$ as $\epsilon \to 0^+$. $\epsilon$ decreases from top to bottom. The profile gets squeezed on the horizon as $\epsilon$ decreases.}
\label{edgeplot}
\end{figure}

The field expansion consisting of the pure gauge and radial flux $\omega=0$ modes is:
\begin{equation}
A=-\sum_\mathbf{k} \left(\frac{1}{k}q_\mathbf{k}A^{(1)}_\mathbf{k}+k a_\mathbf{k}A^{(G)}_\mathbf{k}\right),\label{6.3}
\end{equation}
where $A^{(1)}_\mathbf{k}$ and $A^{(G)}_\mathbf{k}$ are the $\omega=i\partial_t=0$ solutions \eqref{5.5}, with normalization of $\phi_\mathbf{k}$ defined below and the expansion coefficients defined in a convenient manner. The relevant components for the commutator \eqref{6.1} are:
\begin{equation}
A_\rho = -\sum_{\mathbf{k}} a_\mathbf{k}\partial_\rho \phi_{\mathbf{k}}, \quad \Pi^\rho = \sum_{\mathbf{k}}q_{\mathbf{k}}\phi_{\mathbf{k}}
\end{equation}
Hermiticity of these fields implies $q_{-\mathbf{k}} = q^{\dagger}_{\mathbf{k}}$ and likewise for $a_{-\mathbf{k}}$. The zero-mode scalar solution is:
\begin{equation}
\phi_{\mathbf{k}}(\rho,\mathbf{x}) = \frac{K_0(k \rho)}{K_0(k\epsilon)} e^{i \mathbf{k} \cdot \mathbf{x}},
\end{equation}
normalized such that $\left|\phi_{\mathbf{k}}(\epsilon)\right| = 1$. In the limit $\epsilon \to 0^+$, one finds
\begin{align}
\label{edgetheta}
\phi_{\mathbf{k}}(\rho,\mathbf{x}) \quad &\to \quad \left(1-\theta(\rho)\right) e^{i \mathbf{k} \cdot \mathbf{x}}, \\
\label{edgedelta}
\partial_\rho \phi_{\mathbf{k}}(\rho,\mathbf{x}) \quad &\to \quad -\delta(\rho)e^{i \mathbf{k} \cdot \mathbf{x}},
\end{align}
localizing this function on the horizon $\rho=\epsilon$ (Figure \ref{edgeplot}). The radial line integral of $A$ over a curve entering the R-wedge at position $\mathbf{x}$ is:
\begin{equation}
\mathcal{A}(\mathbf{x}):=\int_{\mathcal{C}}dx^\mu A_\mu = \int_{0}^{+\infty}d\rho A_\rho = \sum_{\mathbf{k}} e^{i \mathbf{k} \cdot \mathbf{x}} a_{\mathbf{k}},\label{6.8}
\end{equation}
and the radial flux becomes
\begin{equation}
\Pi^\rho(\rho,\mathbf{x}) = \sum_{\mathbf{k}} \phi_{\mathbf{k}}(\rho,\mathbf{x})q_{\mathbf{k}} \, \to \, \sum_{\mathbf{k}} \left(1-\theta(\rho)\right) e^{i \mathbf{k} \cdot \mathbf{x}} q_{\mathbf{k}}:=(1-\theta(\rho))\Phi(\mathbf{x}),\label{6.9}
\end{equation}
and is supported only on the horizon in the $\epsilon \to 0^+$ limit, as anticipated in section \ref{s: hilbertspaces}.\footnote{$\Phi(\mathbf{x})$ is the local flux density through the horizon and integrates over $\Sigma$ to $\Phi_\Sigma$.} This identifies the operators $a_{\mathbf{k}}$ and $q_{\mathbf{k}}$ as the Fourier expansion coefficients of respectively the residual gauge configuration on the horizon, and the electric flux $\Phi(\mathbf{x})$ through the horizon and thus as the edge DOF \eqref{2.8}.\footnote{This motivates the choice of normalization in \eqref{6.3}.}

From \eqref{6.1} we find
\begin{equation}
\comm{\mathcal{A}(\mathbf{x})}{\Phi(\mathbf{y})}=i\delta(\mathbf{x}-\mathbf{y})=\sum_{\mathbf{k}\mathbf{k}'}e^{i\mathbf{k}\cdot\mathbf{x}}e^{-i\mathbf{k}'\cdot\mathbf{y}}\comm{a_\mathbf{k}}{q_{-\mathbf{k}'}},
\end{equation}
such that:
\begin{equation}
\comm{a_{\mathbf{k}}}{q_{-\mathbf{k}'}}= i \delta_{\mathbf{k},\mathbf{k}'}.\label{6.10}
\end{equation}
A state with horizon electric flux $\Phi(\mathbf{x})=\sum_\mathbf{k} \epsilon_\mathbf{k} e^{i\mathbf{k} \cdot \mathbf{x}}$ is defined as
\begin{equation}
\ket{\Phi}=\prod_\mathbf{k} e^{i\epsilon_\mathbf{k}a_{-\mathbf{k}}}\ket{0}.\label{6.11}
\end{equation}
Indeed, 
\begin{align}
\hat{\Phi}(\mathbf{x})\ket{\Phi} =\sum_\mathbf{k} q_\mathbf{k} e^{i\mathbf{k} \cdot \mathbf{x}}\ket{\Phi} =\Phi(\mathbf{x})\ket{\Phi}.
\end{align}
Reality of the flux profile requires $\epsilon^{*}_\mathbf{k} = \epsilon_{-\mathbf{k}}$. This also sets the normalization $\left\langle \Phi\right|\left.\Phi\right\rangle =1$. Equivalently we can write $q_\mathbf{k}\ket{\epsilon_\mathbf{k}}=\epsilon_\mathbf{k}\ket{\varepsilon_\mathbf{k}}$.

Again, analogously, one could work with the conjugate operators, and write a state with holonomy profile: 
\begin{equation}
\ket{\chi}=\prod_\mathbf{k} e^{-i \chi_\mathbf{k}q_{-\mathbf{k}}}\ket{a_{\mathbf{k}} = 0}, \quad \chi(\mathbf{x}) = \sum_{\mathbf{k}}\chi_\mathbf{k}e^{i\mathbf{k} \cdot \mathbf{x}}.\label{5.28}
\end{equation}
This makes contact with the description of edge states as specified by Wilson line punctures on the entangling surface: one effectively integrates over all such profiles. This description is similar to the interpretation in Chern-Simons theory given in the Introduction.

It is instructive to compare our treatment of edge states with the general framework of Donnelly and Freidel \cite{Donnelly:2016auv}. There it was proposed to extend the canonical variables of Yang-Mills theory by the boundary gauge transformations and their conjugates: the normal electric field on the boundary, both of which are only supported on the boundary surface. This was deduced by demanding gauge invariance of the presymplectic potential. In our discussion, we imposed Lorenz gauge throughout the bulk. Time-independent residual gauge transformations
are of the form $A_\mu^{(G)} = \partial_\mu \chi$ with $\chi(\rho,\mathbf{x}) = \lim_{\epsilon\to 0}\sum_\mathbf{k} \chi_\mathbf{k} \frac{K_0(k\rho)}{K_0(k\epsilon)} e^{i\mathbf{k}\cdot \mathbf{x}}$ such that $\Box \chi = 0$. These gauge transformations are localized only on the horizon: $\chi(\rho,\mathbf{x})=(1-\theta(\rho))\chi(\mathbf{x})$. Thus we reach the same conclusion: we introduce pure gauge degrees of freedom on the boundary $a_\mathbf{k}$ as new canonical variables, conjugate to the horizon electric flux.

\subsection{Punctures and Electrostatics}
\label{ss:6.2}
\begin{figure}[h]
\begin{minipage}{0.48\textwidth}
\centering
 \includegraphics[width=0.95\textwidth]{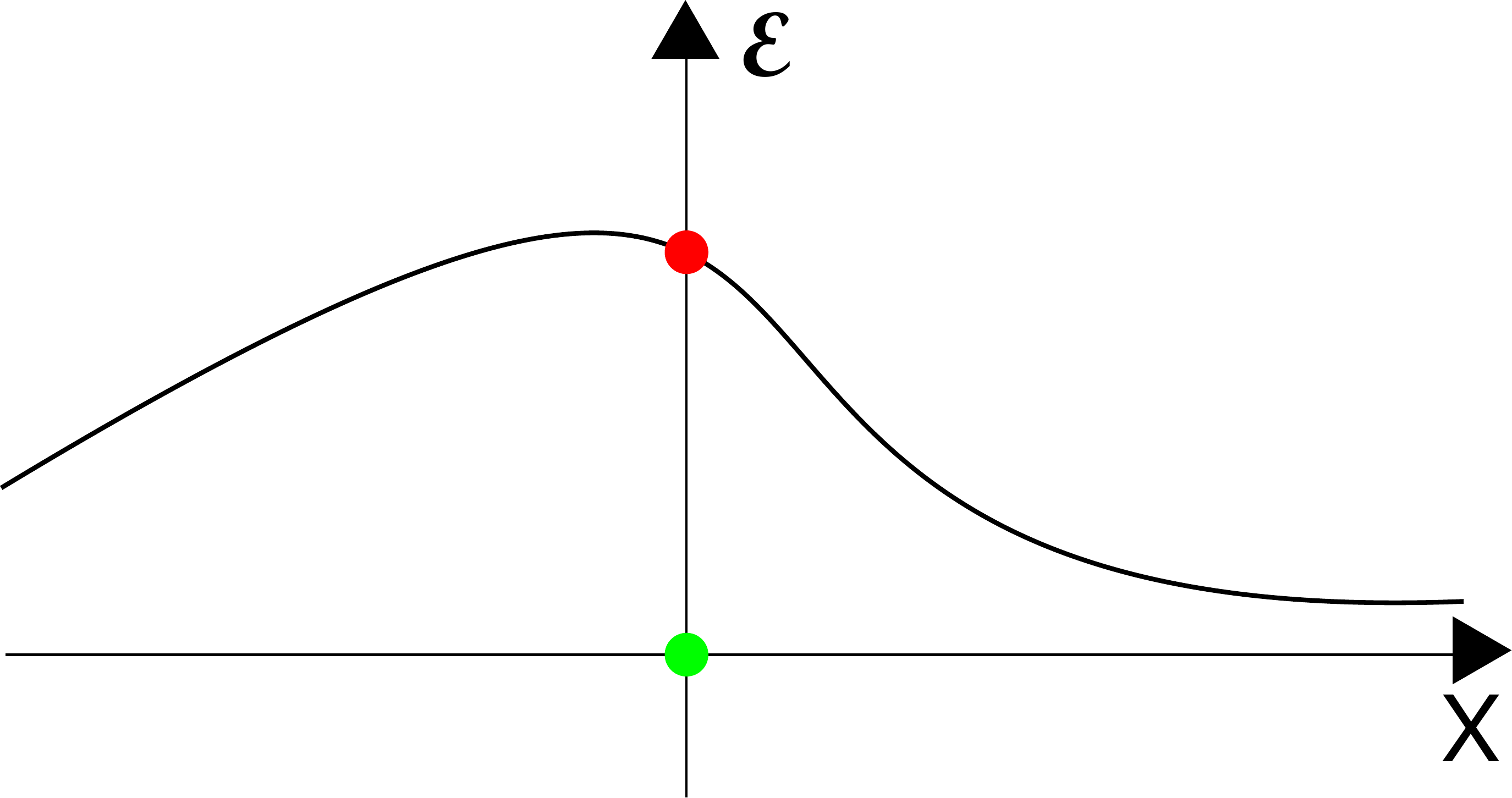}
\caption{Electric field profile in Minkowski space producible by Rindler bulk modes only. The Rindler bulk modes are only able to generate a profile with $\mathcal{E}(0)=0$.}
\label{MinkowskiE}
\end{minipage}
\hfill
\begin{minipage}{0.48\textwidth}
\centering
\includegraphics[width=0.24\textwidth]{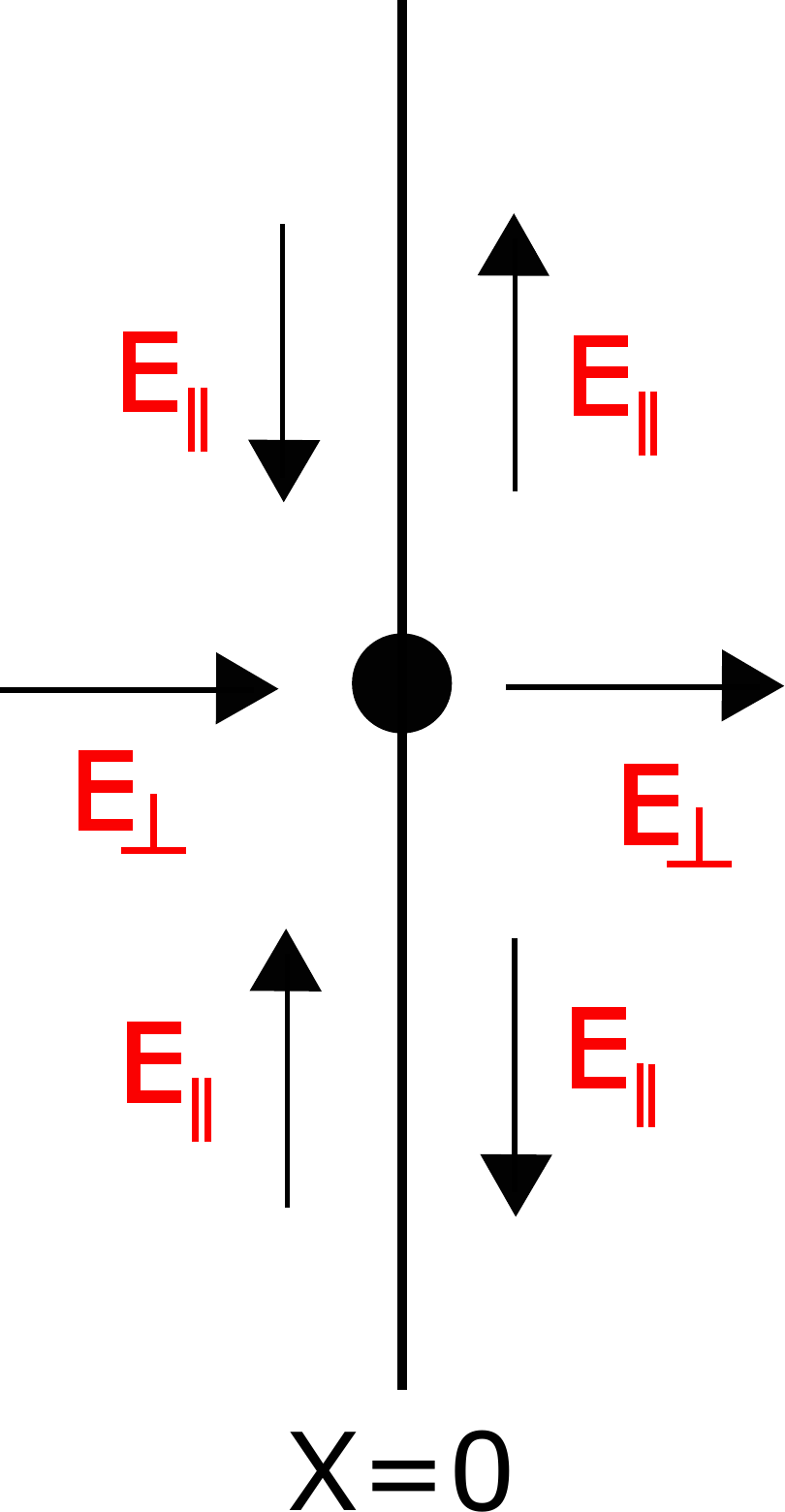}
\caption{Electric field on the horizon due to insertion of a singe Wilson line puncture from the L-wedge and R-wedge perspective. From Minkowski perspective, only the transverse flux exists.}
\label{minkflux}
\end{minipage}
\end{figure}

The total radial electric field in the R-wedge in Rindler coordinates, measured by both fiducial and inertial observers is:
\begin{align}
\mathcal{E}(t,\rho,\mathbf{x})&= (1-\theta(\rho-0^+))\Phi(\mathbf{x})+\mathcal{E}^{\text{bulk}}(t,\rho,\mathbf{x})\nonumber \\
&= \sum_{\mathbf{k}}q_{\mathbf{k}}\phi_{\mathbf{k}}(\rho,\mathbf{x})-\sum_{\mathbf{k}}k\alpha^{(1)}_{\omega,\mathbf{k}}\phi_{\omega,\mathbf{k}}\label{6.14} \\ 
&= \sum_\mathbf{k} q_\mathbf{k} \frac{K_0(k\rho)}{K_0(k\epsilon)}e^{i\mathbf{k}\cdot \mathbf{x}} -\sum_{\omega \in \sigma_D,\mathbf{k}} \alpha^I_{\omega,\mathbf{k}}k \frac{\sqrt{\sinh(\pi\omega)}}{(2\pi)^{\frac{D-2}{2}}\pi}K_{i\omega}(k\rho)e^{i\mathbf{k}\cdot \mathbf{x}}e^{-i\omega t}. \nonumber
\end{align}
The second term satisfies $\mathcal{E}^{\text{bulk}}\rvert_{\partial \mathcal{M}}(\mathbf{x})=0$, while the first term describes the (Rindler time-independent) electric flux through the horizon (Figure \ref{MinkowskiE}).
Upon including the zero-mode piece, an arbitrary Minkowski field with $\mathcal{E}(0) \neq 0$ can be produced by combining both Rindler wedges. This lack of continuity of $\mathcal{E}(0)$ when described using Rindler modes is explained by the pathological nature of $K_{i\omega}(k\rho)$ near $\rho \approx 0$.

Analogously, the radial line integral of $A_\mu$ on a curve $\mathcal{C}$ piercing the horizon only at $\mathbf{x}\in \partial \Sigma$ is:
\begin{equation}
\mathcal{A}_\mathcal{C}(\mathbf{x})=\mathcal{A}(\mathbf{x})+\mathcal{A}_\mathcal{C}^{\text{bulk}}(\mathbf{x}).\label{6.15}
\end{equation}
with $\mathcal{A}(\mathbf{x})$ the edge contribution defined in \eqref{6.8}. More generally, denote the set of point where a generic curve $\mathcal{C}$ goes out of the R-wedge as $\{\mathbf{x}_{out}\}$, and similarly define $\{\mathbf{x}_{in}\}$. Define the puncture operator \begin{equation}
V_q(\mathbf{x}):=e^{iq\mathcal{A}(\mathbf{x})}.
\end{equation}
From the R-wedge perspective this operator acts as the insertion of a finite $U(1)$ charge $q$ at position $\mathbf{x}$. This can be further illustrated by comparing \eqref{6.14} and \eqref{6.15} with the field of a point charge on the horizon as calculated in \cite{Alexander:1991qt}. We can now decompose the product of Wilson lines along $\mathcal{C}_i$, $\mathcal{W}_{\mathcal{C}_i}$, as
\begin{equation}
\mathcal{W}_{\mathcal{C}_i}=\left(\prod_{\mathbf{x}\in \{\mathbf{x}_{out}\}}\prod_{\mathbf{y}\in\{\mathbf{x}_{in}\}}V_{q}(\mathbf{x})V_{-q}(\mathbf{y})\right){\mathcal{W}_{\mathcal{C}_i}}^{\text{bulk}},\label{6.17}
\end{equation}
where ${\mathcal{W}_\mathcal{C}}^{\text{bulk}}$ is constructed using $\mathcal{A}_\mathcal{C}^{\text{bulk}}(\mathbf{x})$. 

The expression \eqref{6.17} allows us to prove \eqref{2.27} explicitly for Maxwell theory. Indeed, if $\mathcal{W}_{\mathcal{C}_i}$ and $\mathcal{W}_{\tilde{\mathcal{C}}_i}$ have the same set of oriented punctures, we get:
\begin{equation}
\mathcal{W}_{\mathcal{C}_i} (\ket{0}\otimes\ket{\text{bulk}})=\left(\prod_{\mathbf{x}\in \{\mathbf{x}_{out}\}}\prod_{\mathbf{y}\in\{\mathbf{x}_{in}\}}V_q(\mathbf{x})V_{-q}(\mathbf{y})\ket{0}\right)\otimes \ket{\text{bulk}}=\mathcal{W}_{\tilde{\mathcal{C}}_i} (\ket{0}\otimes\ket{\text{bulk}}).
\end{equation}
This expression proves that \textit{the edge sector - a generic configuration of $U(1)$ charges on the boundary - is generated by inserting all possible Wilson line punctures in the edge vacuum.} This is precisely the same feature as encountered in the CS edge sector discussed in section \ref{s: hilbertspaces}.

The horizon punctures also create electric flux tangential to the horizon from an R-wedge perspective. This can be concluded directly from the sourced Maxwell equations, or from the zero mode expansion $E^i=\sum_\mathbf{k} i \frac{k_i}{k^2}q_{\mathbf{k}}\partial_\rho\phi_{\omega,\mathbf{k}}$. From the Minkowski perspective, there is no such tangential electric flux, due to an equal but opposite L-wedge contribution. This can be appreciated from the definition $f(0):=\frac{f(0^+)+f(0^-)}{2}$ for the value of a function on a dividing surface. The situation is shown in figure \ref{minkflux}. The edge states only represent transverse horizon flux $\Phi(\mathbf{x})$, as they should: the vanishing of tangential horizon electric flux created by horizon punctures is required for consistency with \eqref{2.15}, which says essentially that a Wilson line creates an electric field line.

\subsection{Spectrum and Thermodynamics}
\label{ss: edge entropy}
In this final subsection concerning the edge sector, we calculate the energy spectrum and partition function. The computation we present is very similar to the one in \cite{Donnelly:2014fua}\cite{Donnelly:2015hxa}, but done here in an operator context.

The starting point is the edge mode Hamiltonian. The improved Hamiltonian is:
\begin{equation}
H_{\text{edge}}=\int d\mathbf{x} d\rho \rho \left(\frac{1}{4}F^{\mu\nu}F_{\mu\nu}-F^{t\mu}F_{t\mu}\right).
\end{equation}
Retaining only the relevant part, this simplifies to 
\begin{equation}
H_{\text{edge}}=\frac{1}{2}\left(\int d\mathbf{x} \int \frac{d\rho}{\rho}\left(F_{t\rho}\right)^2+\sum_i \int dx \int \frac{d\rho}{\rho}\left(F_{t i}\right)^2\right).\label{5.32}
\end{equation}
Plugging in the expansion \eqref{6.3}, one obtains:
\begin{equation}
H_{\text{edge}}=\frac{1}{2}\sum_\mathbf{k} q_\mathbf{k}q_{-\mathbf{k}}\frac{1}{k^2K_0(k\epsilon)^2}\left(k^2\int d\rho\rho K_0^2(k\rho)+\int d\rho \rho\left(\partial_\rho K_0(k\rho)\right)^2 \right)\label{5.35}.
\end{equation}
Using the equations of motion, this reduces to the $\rho=0$ boundary term \eqref{2.12}, and one finds the edge Hamiltonian
\begin{equation}
H_{\text{edge}}=\sum_\mathbf{k} q_\mathbf{k}q_{-\mathbf{k}}\frac{1}{2k^2K_0(k\epsilon)} = \sum_\mathbf{k} q_\mathbf{k}q_{-\mathbf{k}}\frac{1}{2k^2\ln \frac{2}{k\epsilon}}.\label{5.38}
\end{equation}
This is consistent: the Hamiltonian generates the correct classical EOM $\dot{a_{\mathbf{k}}}=0$ and $\dot{q_{\mathbf{k}}}=0$ for the static solutions in the physical situation where the cutoff is taken to zero:
\begin{equation}
\dot{q_{\mathbf{k}}}=-i\comm{H_{\text{edge}}}{q_\mathbf{k}}=0, \quad \dot{a_{\mathbf{k}}}=-i\comm{H_{\text{edge}}}{a_{\mathbf{k}}}=\lim_{\epsilon\to 0}\frac{1}{2k^2\ln \frac{2}{k\epsilon}}q_{\mathbf{k}}=0.
\end{equation}

Knowing both the edge Hilbert space \eqref{5.28} and the edge Hamiltonian \eqref{5.38}, it is possible to write down the edge spectrum. An electric flux eigenstate \eqref{6.11} is also an energy eigenstate:
\begin{equation}
H_{\text{edge}}\ket{\Phi}=\sum_{\mathbf{k}}E_\mathbf{k}\ket{\Phi}=E(\Phi)\ket{\Phi}\label{9.38}\\
\end{equation}
with
\begin{equation}
E_\mathbf{k}=\frac{\abs{\epsilon_\mathbf{k}}^2}{2k^2\ln \frac{2}{k\epsilon}},\label{6.23}
\end{equation}
which can be interpreted as the electrostatic energy of a charge distrubtion, including the redshift effect. Due to linearity of the Maxwell equations, this sector adds to the bulk sector to give the total Hamiltonian $H = H_{\text{edge}} + H_{\text{bulk}}$. This is the boost operator in Minkowski space. Upon using the standard Euclidean evolution argument, one finds the entanglement structure of the Minkowski vacuum state in terms of Rindler eigenstates as:\footnote{The flux path integral can be written in Fourier components as $\int \left[\mathcal{D}\Phi(\mathbf{x})\right] \sim \sqrt{A_{\perp}} \prod_{\mathbf{k}} \int d \epsilon_\mathbf{k}$. The precise prefactors in this transition are determined by Donnelly and Wall in \cite{Donnelly:2015hxa}.}
\begin{equation}
\label{626}
\boxed{
\ket{M} = \int \left[\mathcal{D}\Phi(\mathbf{x})\right]e^{- \pi E(\Phi)}\sum_n e^{- \pi \omega_n}\left|n,\Phi\right\rangle_L \otimes \left|n,\Phi\right\rangle_R }.
\end{equation}
The quantum number $n$ labels states in the bulk Hilbert spaces $\mathcal{H}_{\text{bulk},L/R}$. As usual, tracing over the left wedge yields a thermal partition function $Z =Z_{\text{bulk}}Z_{\text{edge}}$, whose edge part is
\begin{equation}
Z_{\text{edge}}=\mathrm{Tr}_{\mathcal{H}_{\text{edge}}}e^{-\beta H_{\text{edge}}} =\int\left[\mathcal{D}\Phi(\mathbf{x})\right]e^{-\beta E(\Phi)}.\label{6.25}
\end{equation}
With the probability distribution $p(\Phi) = e^{- 2\pi E(\Phi)}/Z_{\text{edge}}$, the edge entanglement entropy can be written as 
\begin{equation}
S_{\text{edge}} = - \int \left[\mathcal{D}\Phi(\mathbf{x})\right]p(\Phi) \ln p (\Phi).
\end{equation}
Evaluating the Gaussian path integral, the edge mode partition function \eqref{6.25} is:
\begin{equation}
Z_{\text{edge}}= A_\perp^{\frac{1}{2}}\left(\prod_\mathbf{k}\left(\frac{\beta}{2\pi \ln \frac{2}{k\epsilon}}\right)\frac{1}{k^2}\right)^{-\frac{1}{2}}, \label{6.28}
\end{equation}
The full partition function of Maxwell theory, including the edge sector is:
\begin{equation}
Z=Z_N^{d-1}\left(\frac{Z_D}{Z_N}Z_{\text{edge}}\right),
\end{equation}
using \eqref{DvsN} this becomes:
\begin{equation}
\boxed{Z=Z_N^{d-1}\left(A_\perp \prod_\mathbf{k} k^2\right)^{\frac{1}{2}}}.\label{6.29}
\end{equation}
The second factor is exactly the contact term found in the replica trick calculation.
\begin{itemize}
\item Note that this construction is rather sensitive to the choice of bulk boundary conditions. Had we chosen to adopt the PEC boundary conditions \eqref{PEC} for instance, the partition function would become instead $Z=Z_N^{d-1}\left(\left(\frac{Z_D}{Z_N}\right)^{d-2}Z_{\text{edge}}\right)$. This dependence on $(d-2)$ in the bulk partition function is unavoidably transfered to the total partition function, regardless of the prescription for $Z_{\text{edge}}$, such that the correct contact term \eqref{6.29} cannot be recovered. This is related to the duality anomaly investigated in \cite{Donnelly:2016mlc}. Thus, the PMC boundary conditions on the bulk theory are not only natural as pointed out in section \ref{s: hilbertspaces}, they are enforced by the equality of the entanglement entropy and the replica trick entropy. 
\item The contact piece can be regularized using the Schwinger proper time in the well-known heat kernel form:
\begin{equation}
\label{hkreg}
\ln Z_{\text{contact}} \sim \frac{1}{2} \ln \prod_\mathbf{k} k^2 = -\frac{1}{2} \int_{0}^{+\infty}\frac{ds}{s}\left(\text{Tr}\,e^{s\Delta_\perp}-e^{-s}\right) = -\frac{1}{2} \int_{\tilde{\epsilon}}^{+\infty}\frac{ds}{s}\text{Tr}\,e^{s\Delta_\perp} + \hdots,
\end{equation}
with the famous minus sign a heat kernel regularization artifact as explained in \cite{Donnelly:2014fua,Donnelly:2015hxa}. The factor $1/2$ represents the single degree of freedom contributing to the contact piece.
\item Although both the edge sector and the bulk sector contribute to the entanglement entropy, there is a sense in which both are different. The former can be shown not to be part of the distillable entanglement whereas the latter is. For more details on this lattice concept, we refer the reader to \cite{Soni:2015yga}\cite{VanAcoleyen:2015ccp}.
\end{itemize}

\subsection{Black Hole Microstates and Asymptotic Symmetries}
The $\epsilon$-regularization has allowed us to make sense of boundary conditions at the horizon, and to identify a proper state counting for entanglement purposes \eqref{6.29}. At this point we set $\epsilon=0$: \eqref{6.23} shows that $E(\Phi)=0$, in agreement with the presumed static nature of the solutions. The Rindler ground state becomes infinitely degenerate due to the edge sector. This degeneracy suggests a natural interpretation of \emph{edge states as black hole microstates}: the total energy in a generic state is not affected by the edge configuration, and for the Rindler observer the edge configurations are indistinguishable.

In quantum gravity, one expects that one cannot resolve the flux profile $\Phi(\mathbf{x})$ any smaller than the Planck scale $\ell_p$, reducing the infinite number of horizon degrees of freedom to only $\frac{A}{\ell_p^{d-1}}$, and directly leading to an entanglement entropy of the Bekenstein-Hawking form:
\begin{equation}
S \sim \frac{A}{G_N}.\label{6.39}
\end{equation}
Such an interpretation has an obvious species problem, as we \emph{chose} to study Maxwell fields in the vicinity of the black hole. One could add any number of theories to the action, each with their own edge sector  resulting in a different prefactor in \eqref{6.39}. The only rescue to this issue seems to be string theory, where the field content is universally fixed. In the above argumentation, one has to ad hoc reintroduce a UV cutoff parameter $\epsilon\sim l_p$. Such a procedure can only be made rigorous in a UV-complete theory such as string theory, where the range of integration in the would-be-divergent integral is naturally and unambiguously constrained by consistency requirements on the theory, i.e. modular invariance.
\\~\\
For a different interpretation of the edge sector, consider the basis \eqref{5.28}. Transferring to coordinate space and define $\hat{Q}(\tilde{\chi})=\int_{\partial \Sigma}d^{d-1}x \, \hat{Q}(\mathbf{x})\,\tilde{\chi}(\mathbf{x})$, where 
\begin{equation}
\hat{Q}(\mathbf{x})=\hat{\Phi}(\mathbf{x}).\label{6.33}
\end{equation}
We find:
\begin{equation}
e^{i\hat{Q}(\tilde{\chi})}\ket{\chi}=\ket{\chi+\tilde{\chi}}.
\end{equation}
For $\tilde{\chi}=\epsilon$ infinitesimal this becomes:
\begin{equation}
i\comm{\chi(\mathbf{x})}{\hat{Q}(\epsilon)}=\epsilon(\mathbf{x})=\delta \chi(\mathbf{x}).\label{AST}
\end{equation}
Since $\chi(x)$ labels the on-shell large gauge DOF of the theory, \eqref{AST} proves that $\hat{Q}$ generates a gauge transformation on the boundary, identifying it as a genuine $U(1)$ charge in the boundary theory. The identification \eqref{6.33} of the radial canonical momentum $\Phi(\mathbf{x})$ of Maxwell theory evaluated at the boundary as a $U(1)$ charge in the boundary theory is the equivalent of the holographic statement \eqref{2.28} in Chern-Simons theory.

Summarizing, large gauge transformations generate the edge sector which consists of static photons. With the sole difference being a choice of Cauchy surface (null instead of timelike), these are the asymptotic symmetries and soft photons argued by Hawking, Perry and Strominger \cite{Hawking:2016msc,Hawking:2016sgy} to provide electromagnetic hair to black holes, and open a path to black hole microstate counting and information restoration.\footnote{For a similar discussion on gravitational edge DOF see e.g. \cite{Hawking:2016sgy,Lust:2017gez}.} This further solidifies our identification of edge states as Maxwell black hole microstates.
\subsection{Summary}
The edge sector of Maxwell theory is found by quantizing (part of) the zero-mode sector of the theory, resulting in additional states living on the edges of the system. The partition function of these zero energy states accounts for the contact term. Depending on the choice of basis, these edge states represent $U(1)$ charge configurations, or large gauge transformations on the boundary (the horizon). We have derived the need for this term directly from the canonical algebra of Maxwell theory in Minkowski, as discussed in section \ref{s: hilbertspaces}. The interpretation of edge states as black hole microstates in QFT naturally emerges.

\section{Extension to Proca and Tensor Field Theories}
\label{s:extension}
The main message of this work is that edge states appear in all theories that have constraint equations. This implies a natural extension of the above discussion to Proca theory and to tensor field theories. Lorenz-gauge Maxwell theory is particularly suitable to generalize to the Proca field, or to tensor fields, since the Lorenz gauge constraint generalizes to the transversality condition in these theories. \textit{Even in the absence of gauge symmetry, a constraint on the initial value problem that includes spatial derivatives causes non-factorization across the entangling surface and a contact term.} This is manifestly present in all generalizations that we discuss here.

\subsection{Proca Theory}
\label{s:Proca}
The massive vector field is an important example to validate our conjecture that constraint equations are the telltale for contact terms in a theory. And indeed, Euclidean conical computations demonstrate that a contact term is present (see e.g. \cite{Solodukhin:2015hma}). The bulk sector of the Proca field has been studied before \cite{Castineiras:2011zz}. Explicitly, the Proca Lagrangian is:
\begin{equation}
\frac{1}{\sqrt{-g}}\mathcal{L} = -\frac{1}{4}F_{\mu\nu}F^{\mu\nu} - \frac{1}{2}m^2 A^\mu A_\mu,
\end{equation}
with equations of motion:
\begin{equation}
\label{iniform}
\nabla^\mu F_{\mu\nu} -m^2 A_\nu = \left(\nabla_\mu \nabla^\mu - m^2\right) A_\nu - \nabla^\mu \nabla_\nu A_\mu = 0.
\end{equation}
If $R_{\mu\nu\rho\sigma} = 0$, one obtains ($m\neq 0$):
\begin{equation}
\left(\nabla_\mu \nabla^\mu - m^2\right) A_\nu = 0, \quad \nabla^\mu A_\mu =0.
\end{equation}
When formulated in this way, the initial value problem has the transversality constraint $\nabla^\mu A_\mu =0$ which leads to a non-factorization of the Hilbert space when splitting the Cauchy surface in half. Note that in the original form \eqref{iniform}, the constraint is Gauss' law: $\mathbf{\nabla}\cdot \mathbf{E} = -m^2 A_0$. Both formulations are equivalent. We conclude that edge states will be present, in spite of the absence of gauge symmetry. This is confirmed from the Euclidean perspective: the Proca field has a contact term contribution to the entanglement entropy and hence must have an edge mode sector. 

The equation $(\nabla^\mu\nabla_\mu-m^2)A_\nu = 0$ is solved by the following set of generalizations of the Maxwell modes:
\begin{gather}
\begin{aligned}
\label{7.4}
A_{\mu,\omega\mathbf{k}}^{(1)} &= \left(1+\frac{m^2}{k^2}\right)^{-\frac{1}{2}}\frac{1}{k}\left(\rho\partial_\rho, \frac{1}{\rho}\partial_t, \mathbf{0}\right)\phi_{\omega, \mathbf{k}} \\
A_{\mu,\omega\mathbf{k}}^{(0)} &= \left(1+\frac{m^2}{k^2}\right)^{-\frac{1}{2}}\frac{1}{k}\left(\partial_t, \partial_\rho,\mathbf{0}\right)\phi_{\omega, \mathbf{k}}:=A_{\mu,\omega\mathbf{k}}^{(G)}-A_{\mu,\omega\mathbf{k}}^{(k)}, \\
A_{\mu,\omega\mathbf{k}}^{(k)} &= \left(1+\frac{m^2}{k^2}\right)^{-\frac{1}{2}}i\e^{(k)}_\mu \phi_{\omega, \mathbf{k}}, \\
A_{\mu,\omega\mathbf{k}}^{(a)} &= \left(1+\frac{m^2}{k^2}\right)^{-\frac{1}{2}}i\e^{(a)}_\mu \phi_{\omega, \mathbf{k}}.
\end{aligned}
\end{gather}
Define now the massive extension of the pure gauge mode as $A_{\omega\mathbf{k}}^{(\tilde{2})}=A_{\omega\mathbf{k}}^{(G)}+\frac{m^2}{k^2}A_{\omega\mathbf{k}}^{(k)}$. This is the combination that satisfies the transversality constraint $\nabla^\mu A_\mu=0$. These modes $A_{\omega\mathbf{k}}^{(\tilde{2})}$ are Klein-Gordon normalized to $\frac{m^2}{k^2}$, such that in the limit $\frac{m^2}{k^2} \to 0$ they reproduce the null pure gauge solutions of Maxwell theory. Defining $A_{\omega\mathbf{k}}^{(\tilde{2})}=\frac{m}{k}A_{\omega\mathbf{k}}^{(2)}$, we obtain the orthonormal solution space of bulk Proca theory as the set of modes $A_{\omega\mathbf{k}}^{(1)},A_{\omega\mathbf{k}}^{(2)},A_{\omega\mathbf{k}}^{(a)}$. Following the same reasoning as for Maxwell theory we find that the set $A_{\omega\mathbf{k}}^{(1)}$ obeys Dirichlet boundary conditions, and the other $d-1$ sets obey Neumann boundary conditions.
\\~\\
The edge algebra \eqref{6.1} is identical for Proca theory. The realization in terms of edge states, and the associated interpretation is different. Proca has no gauge symmetry and there are no null states, therefore we will not recover the interpretation that edge states represent large gauge transformations. The interpretation of edge states as boundary charges of (massive) electromagnetism does uphold, and we are again able to choose a basis of radial canonical momentum, as is obvious from \eqref{6.1}. These charges now do \emph{not} generate a boundary gauge transformation. In the following we explicitly construct the edge sector of Proca theory. 

The improved static boundary Hamiltonian for Proca theory is identical to the Maxwell boundary Hamiltonian \eqref{2.12}:
\begin{equation}
H_{\text{edge}}=\frac{1}{2}\int_{\partial \Sigma} d^{d-1}x\sqrt{-g}n_\mu\left(F^{\mu 0}A_0-F^{\mu i}A_i\right).\label{hbdyproca}
\end{equation}
In Maxwell theory the contribution to the boundary Hamiltonian from the pure gauge modes vanishes, resulting in $H_{\text{edge}}(q_\mathbf{k})$ in \eqref{5.38}: the electric flux modes diagonalize the energy. In Proca theory though, it is clear the modes $A^{(2)}_{\mathbf{k}}$ \emph{do} contribute to the boundary Hamiltonian \eqref{hbdyproca}. Along the lines of \eqref{6.10}, one could try to quantize the Proca edge sector by imposing a commutator between $q^{(1)}_\mathbf{k}$ and $q^{(2)}_\mathbf{k}$, but this turns out to be incorrect as it does not result in the correct counting of edge DOF. 

One way to quantize the edge sector that results in the correct final partition function \eqref{7.7} is as follows: we further extend the Hilbert space of the theory to allow states associated with a radial canonical flux that varies linear in time, which is still a solution of the classical scalar equations of motion of the field $\phi_{\omega,\mathbf{k}}$ composing the vector field \eqref{7.4}. In the quantum theory this results in the introduction of two sets of operators: $q_\mathbf{k}^0$ and $\dot{q}_\mathbf{k}^0$, normalized as in \eqref{6.3} but with $k\to \sqrt{k^2+m^2}$. Classically they represent the solution $q_\mathbf{k}\to q_\mathbf{k}(t)=q_\mathbf{k}^0+\dot{q}_\mathbf{k}^0t$. The canonical algebra \eqref{6.1} is now readily shown to determine the canonical conjugate to $q_\mathbf{k}$ as:
\begin{equation}
p_\mathbf{k}=\frac{\ln \frac{2}{\sqrt{k^2+m^2}\epsilon}}{2(k^2+m^2)}\dot{q}^0_{-\mathbf{k}},
\end{equation}
with
\begin{equation}
\comm{q_\mathbf{k}}{p_{\mathbf{k}'}}=\comm{q^0_\mathbf{k}}{p_{\mathbf{k}'}}=i\delta_{\mathbf{k}\mathbf{k}'}.
\end{equation}
This is consistent: the Hamiltonian generates the correct classical EOM $\frac{d}{dt}q_\mathbf{k}=\dot{q}^0_\mathbf{k}$, and (in the physical situation where the cutoff is taken to zero) $\frac{d}{dt}p_\mathbf{k}=0$. The Proca Hamiltonian in general depends on both sets of canonical variables $H_{\text{edge}}(q_\mathbf{k},p_\mathbf{k})=H_{\text{edge}}(q_\mathbf{k})+H_{\text{edge}}(p_\mathbf{k})$, and is thus non-diagonal in either the (generalized) coordinate or the momentum basis. 

An important point is to realize that states with $p_\mathbf{k}\neq 0$ are nonphysical: they represent configurations for which the radial canonical flux grows linear in time (this is in conflict with energy conservation). The physical edge sector of the theory is thus obtained by imposing 
\begin{equation}
p_\mathbf{k}\ket{\psi}=0\label{7.8}
\end{equation}
on states in the physical edge Hilbert space. This construction is very similar to the (bulk) Gupta-Bleuler construction employed in Maxwell theory. In the edge theory, the physical spectrum \eqref{7.8} consists of only one state: $\ket{p_\mathbf{k}=0}$. The edge partition function written as a trace over the edge Hilbert space is:
\begin{equation}
Z_{\text{edge}}=\bra{p_\mathbf{k}=0}e^{-\beta H_{\text{edge}}(q_\mathbf{k},p_\mathbf{k})}\ket{p_\mathbf{k}=0}=\bra{p_\mathbf{k}=0}e^{-\beta H_{\text{edge}}(q_\mathbf{k})}\ket{p_\mathbf{k}=0},
\end{equation}
where $H_{\text{edge}}(q_\mathbf{k})$ is identical to the Maxwell edge Hamiltonian \eqref{5.38} with $k\to \sqrt{k^2+m^2}$. Introducing twice the completeness relation in the coordinate basis, using $\bra{p_\mathbf{k}=0}\ket{q_\mathbf{k}} \sim 1$ and gathering the coordinates $q_\mathbf{k}$ in the field $\Phi(\mathbf{x})$ results in: 
\begin{equation}
Z_{\text{edge}}=\int\left[\mathcal{D}\Phi(\mathbf{x})\right]e^{-\beta E(\Phi)},
\end{equation}
with again $k\to \sqrt{k^2+m^2}$ in \eqref{6.23}. Evaluating this, one gets:
\begin{equation}
Z_{\text{edge}}=A_\perp^{\frac{1}{2}}\left(\prod_\mathbf{k}\left(\frac{\beta}{2\pi \ln \frac{2}{\sqrt{k^2+m^2}\epsilon}}\right)\frac{1}{k^2+m^2}\right)^{-\frac{1}{2}}.
\end{equation}
Retracing the steps of appendix \ref{a:d vs n}, one readily finds 
\begin{equation}
\frac{Z_D}{Z_N}=\prod_\mathbf{k}\left(\frac{\beta}{\ln \frac{2}{\sqrt{k^2+m^2}\epsilon}}\right)^{\frac{1}{2}}.
\end{equation}
The total Proca partition function in $d+1$ dimensional Rindler now becomes:
\begin{equation}
\boxed{Z=Z_N^d\left(A_\perp \prod_{\mathbf{k}}(k^2+m^2)\right)^{  \frac{1}{2}}},\label{7.7}
\end{equation}
from which we read off the correct contact term for Proca theory:
\begin{equation}
\ln Z_{\text{contact}} \sim \frac{1}{2} \ln \prod_\mathbf{k} (k^2 + m^2) = -\frac{1}{2} \int_{\tilde{\epsilon}}^{+\infty}\frac{ds}{s}\text{Tr}\,e^{-s(-\Delta_\perp + m^2)} + \hdots ,
\end{equation}
of an almost identical form as Maxwell, except for the additional factor $e^{-sm^2}$ in the heat kernel regularization \eqref{hkreg}. Note that taking the limit $m^2\to 0$ in Proca theory is smooth in the final formulas, except for the fact that one bulk polarization becomes pure gauge and is quotiented out of the bulk Hilbert space, replacing $d\to d-1$ in \eqref{7.7}. Again we emphasize that \emph{the Proca edge sector cannot be interpreted as large gauge transformations at the boundary}, simply because Proca has no null states.
\\~\\
As a nice extra, investigation of Proca theory explains the origin of the alleged contact term in $1+1$ dimensions for Maxwell theory. Indeed, looking back at section \ref{s:edge11}, it is seen that a non-zero mass will cause the electric field to decay away from the horizon, effectively replacing $\ln 1/\rho\to K_0(m\rho)$ and $\ln 1/\epsilon\to K_0(m\epsilon)$ everywhere in that section. So the Proca zero-mode in $1+1$ dimensions decays infinitely fast away from the horizon as in figure \ref{edgeplot}. This removes the volume divergence in the Hamiltonian \eqref{4.20} and instead results in the spectrum \eqref{7.7}, restricted to $\mathbf{k}=\mathbf{0}$. The Proca partition function in $1+1$ dimensions is:
\begin{equation}
Z=Z_N\left(m^2\right)^{\frac{1}{2}},
\end{equation}
so there is a contact term $\sim m$ reminiscent of the $d+1$ dimensional contact term. In Kabat's contact term calculation \cite{Kabat:1995eq}, a mass regulator is used; the above suggests that in doing so, one effectively changes the edge sector of the Maxwell field into that of the Proca field: it is most likely the Proca contact term that is recovered in \cite{Kabat:1995eq} and \textit{not} the one from Maxwell, which is trivial in $1+1$ dimensions as captured in \eqref{mink vac zero}. Alternatively, the Maxwell contact term in $1+1$ dimensional \emph{compact} space is non-zero and, using \eqref{4.20}, behaves as $\sim 1/\sqrt{V}$. The IR mass regulator effectively plays the role of a volume cutoff and reintroduces edge states in a theory that would ordinarily not have them.

We note that the logic of \cite{Donnelly:2016auv} to deduce edge states for theories with a gauge symmetry (demanding gauge invariance of the presymplectic potential) is not straightforwardly extended to this case.
\subsection{Tensor Fields}
\label{s:hs}
Generic tensor fields $\phi_{\mu\nu\rho...}$ are used to model general higher spin fields. A bosonic higher spin $s>1$ field is a symmetric tensor field that is divergenceless, traceless, and satisfies an equation of the form:
\begin{equation}
\left(\nabla^\mu \nabla_\mu - m^2\right) \phi_{(\nu\rho\sigma...)} = 0.\label{eomhs}
\end{equation}
It satisfies the constraint:
\begin{equation}
\nabla^\mu\phi_{\mu(\rho\sigma..)} =0,\label{7.9}
\end{equation}
and is traceless:
\begin{equation}
\tensor{\phi}{^\mu_{\mu(\sigma..)}}=0.
\end{equation}
A completely antisymmetric tensor field ($p$-form field) satisfies these same equations, now obviously with anti-symmetrization of the indexes.

In the special case $m^2=0$ the transversality constraint \eqref{7.9} is to be interpreted as a gauge choice that generalizes Lorenz gauge and is known as de Donder gauge. In string theory, this constraint arises as part of the Virasoro constraints at level $n=s$. The bosonic higher spin theory is thus described by a constrained Hamiltonian system with associated edge states and a contact term. This is confirmed by the replica trick calculations: massive $s>1$ bosonic field theories are known to have contact terms \cite{Solodukhin:2015hma}, demonstrating an edge sector must indeed exist in these theories. 

Surprisingly, the equation of motion \eqref{eomhs} is readily solved generically, so the Rindler modes for any tensor field can be written down immediately. Defining $x^\pm = t \pm r$, and diagonalizing the tensor components in $SO(2)$ spin of the $t-r$ plane as $s_\alpha=\pm 1$ for a $\pm$ tensor index $\alpha$ and zero otherwise, the equations decouple in the tortoise light-cone components:
\begin{equation}
\left(e^{-2r}\partial_r^2 -2 \left(\sum_\alpha \left|s_\alpha\right|\right)e^{-2r}\partial_r - e^{-2r}\partial_t^2 + 2 \left(\sum_\alpha s_\alpha \right)e^{-2r}\partial_t  + \partial_i^2 -m^2 \right)\phi_{\nu\rho\sigma...} = 0,
\end{equation}
with explicit mode solutions:\footnote{The labels $s_\alpha$ have to match with the spacetime $SO(2)$ spin indices corresponding to the indexes $\nu\rho\sigma...$.}
\begin{equation}
\label{solhs}
\phi^{(s_\alpha)}_{\nu\rho\sigma...,\omega\mathbf{k}} = \rho^{\sum_\alpha \left|s_\alpha\right|}K_{i\omega +\sum_\alpha s_\alpha}(\sqrt{m^2+k^2}\rho) e^{i\mathbf{k}\cdot \mathbf{x}}e^{-i\omega t}.
\end{equation}
Each such tensor component has an independent oscillator and the full field is just the most general linear combination of these:
\begin{equation}
\phi_{\nu\rho\sigma...} = \sum_{s_\alpha, \omega, \mathbf{k}} \hat{\alpha}^{(s_\alpha)}_{\omega\mathbf{k}}\phi^{(s_\alpha)}_{\nu\rho\sigma...,\omega\mathbf{k}} + (hc).
\end{equation}
As before, part of the zero-mode sector is expected to be associated with the edge sector. It would be interesting to expand the construction of this work to these theories and explicitly write down the edge sector and partition function. We leave this to future work.

\section{Conclusion}
\label{s:conclusion}
Throughout this paper we have highlighted different aspects of the thermodynamic properties of vector fields in Rindler space. In particular we have studied the structure of the Hilbert space contributing to the entanglement entropy across the black hole horizon. \\

In $1+1$ dimensional Lorenz gauge Maxwell theory, a puzzle appeared in the literature \cite{Zhitnitsky:2010ji} concerning an apparent addition (instead of cancellation) of unphysical polarizations when making the transition from Minkowski to Rindler coordinates, which is not present in Weyl gauge. Explicitly performing Lorenz gauge canonical quantization, we proved that the Faddeev-Popov ghost fields cancel exactly the thermodynamic contributions of the longitudinal and temporal photon polarizations. In doing so, the conflict with gauge invariance is resolved. \\
The contact term in $1+1d$ is subtle, and dependent on possible choice of regularization \cite{Donnelly:2012st}. Combining the results from previous sections, we can summarize the situation as follows.
\begin{itemize}
\item \emph{Maxwell in infinite volume.} \\
As we remarked in section \ref{s:edge11}: there is no vacuum entanglement in this case: $S_E = 0$, or
\begin{equation}  
\left|M\right\rangle = \left|0\right\rangle_L \otimes \left|0\right\rangle_R
\end{equation}
The thermal Rindler edge partition function $Z_{\text{th}} = \text{Tr}e^{-\beta H} = 1$ in this case, with $H$ given in \eqref{4.20}, because $H \sim V\to+\infty$, the spatial volume, and only the vacuum contributes. The thermal entropy vanishes as well $S_{\text{th}}=0$.
\item \emph{Maxwell in finite spatial volume regularization.} \\
Again $S_E = 0$, but now $Z_{\text{th}} = \sum_E e^{-\beta H } > 1$, which contains contributions from non-zero electric fields. There is accordingly a mismatch between the vacuum entanglement entropy $S_E$ and the thermal entropy $S_{\text{th}}$ in this case. The argument in section \ref{s:edge11} explains why this is so: one needs infinite spatial volume $V$ to be able to generate the Minkowski wavefunctional for these modes. The modes discussed in higher dimensions with $\mathbf{k} \neq \mathbf{0}$ or $m\neq0$ in other sections do not feel the extent of space all the way to the boundary and hence the standard Euclidean path integral rotation argument does apply for those modes.
\item \emph{Maxwell with a mass regulator.} \\
As studied in section \ref{s:Proca}, the contact contribution is in effect that of Proca theory, and $S_E = S_{\text{th}}$ as determined by the Proca edge partition function.
\end{itemize}

In comparing scalars, Maxwell, Proca and bosonic tensor field theories, we observed that any bosonic theory with non-ultralocal constraint equations has a contact term. This contact term has a statistical interpretation as counting edge states. These DOF account for the matching condition that accompanies the non-ultralocal constraint equations.

From the action principle, a theory on a space with a boundary comes with a set of boundary conditions. Correct implementation of these boundary conditions from the Rindler perspective requires the introduction of a regularization. Comparison with the replica trick partition function is the guideline throughout: it determines the appropriate set of boundary conditions on each independent polarization, as well as identifies the correct counting of edge DOF. In this work we have presented the canonical quantization of this edge sector from first principles: the canonical commutation relations \eqref{2.15} in $\mathbb{R}^{1,d}$.

For Maxwell in $d+1$ Rindler we find a bulk photon and an edge sector representing $U(1)$ charges on the boundary or alternatively large gauge transformations. This was inferred from the canonical algebra \eqref{2.15}. To reconstruct the Minkowski vacuum from the R-wedge perspective, the surface charge is not fixed and is in fact thermally populated with the Hawking temperature \eqref{626}. This is the edge sector equivalent of the bulk Unruh effect. \\
The edge states are the Maxwell microstates of the black hole or equivalently the soft hair of Hawking, Perry and Strominger labeling inequivalent Rindler vacua. The bulk photon is subject to PMC boundary conditions at the horizon. The latter being an infinite redshift surface, the bulk observer is unaware of these boundary conditions (and of the edge sector).
\\~\\
The introduction of edge states in Maxwell theory and the transition to an extended Hilbert space formulation solves the problem of factorizing a $U(1)$ Wilson line across an entangling surface. See also \cite{Harlow:2015lma,Lin:2017uzr}.

As discussed in the introduction, the nonfactorization problem is also present in string theory. By analogy, the stringy edge states are related to punctures of string endpoints on the horizon. A direct quantization of strings in Rindler would be invaluable if one wants to understand the nature of black hole entropy. In the context of this work, one could try to obtain information about string thermodynamics in Rindler space in a less direct manner by treating string theory as an infinite sum of higher spin theories and applying the reasoning of this work to all those higher spin theories, and then sum over the string spectrum. The Euclidean perspective on this idea was explored in \cite{Mertens:2015adr,He:2014gva}. To understand the limitations of such a procedure, a concrete avenue would be to further investigate strings on the $SL(2,\mathbb{R})/U(1)$ cigar background, which has Rindler space as a parametric limit (see e.g. \cite{Giveon:2013ica,Giveon:2014hfa,Mertens:2013zya,Mertens:2014saa,Mertens:2016tqv,Giveon:2015cma,Ben-Israel:2015mda,Giveon:2016dxe,Ben-Israel:2017zyi,Itzhaki:2018rld}).

When treating strings in curved space, we are assuming a classical geometry, which might not be a good approximation. It would therefore be interesting to investigate the nonfactorization problem and the notion of edge states holographically, in CFTs that are dual to string theory in a black hole background.
It is noteworthy though, that if one defines the entangling surface in string theory as a sharp surface (i.e. not smoothed out on the string scale $\ell_s^{-1}$), then the standard argument of defining accelerated point-like observers to cover the R-wedge, still applies here. And hence one needs to study quantum strings in a classical Rindler geometry.

\section*{Acknowledgements}
The authors thank Nele Callebaut, David Dudal and Gertian Roose for several valuable discussions. AB and TM gratefully acknowledge financial support from Research Foundation Flanders (FWO Vlaanderen). This work was partially supported by the European Commission through the grant QUTE. The work of VIZ was partially supported by RFBR grant 17-02-01108.

\appendix

\section{FP Ghost contribution to Unruh effect}
\label{app:ghost}
The Faddeev-Popov (FP) ghost Lagrangian is of the form 
\begin{equation}
\mathcal{L} = \sqrt{-g}\left(- \partial^\mu \bar{c} \partial_\mu c\right)\label{3.33}
\end{equation}
for two anticommuting scalar ghost fields $c$ and $\bar{c}$, where we take the convention of left-derivation for Grassmann variables. As the ghost fields are Grassmann fields, hermiticity of the Lagrangian requires $c$ to be Hermitian and $\bar{c}$ to be anti-Hermitian. Indeed
\begin{equation}
\left(\frac{\mathcal{L}}{\sqrt{-g}}\right)^\dag = - (\partial^\mu \bar{c} \partial_\mu c)^\dagger = -\partial_\mu c^\dagger \partial^\mu \bar{c}^\dagger = \partial^\mu c \partial_\mu \bar{c} = \frac{\mathcal{L}}{\sqrt{-g}}.
\end{equation}
The equation of motion for both fields is solved by 
\begin{gather}
\begin{aligned}
c(t,r) &= \int dk \left(c_k u_k + c^\dagger_k u^*_k\right), \\
\bar{c}(t,r) &= \int dk \left(\bar{c}_k u_k - \bar{c}^\dagger_k u^*_k\right),\label{3.35}
\end{aligned}
\end{gather}
with $u_k = \frac{1}{\sqrt{4\pi E}}e^{-i\omega t + i kr}$ and $\omega = \left|k\right|$. The conjugate momenta are $\pi_c = -\partial_t \bar{c}$ and $\pi_{\bar{c}} = \partial_t c$. The non-vanishing canonical anticommutation relations read
\begin{gather}
\begin{aligned}
\left\{c(t,r),\pi_c(t,r')\right\} &= - \int dk dk'\left(-\underbrace{\left\{c_k,\bar{c}^\dagger_{k'}\right\}}_{-\delta(k-k')}(i\omega) \frac{e^{ikr -ik'r'}}{4\pi \omega} + \underbrace{\left\{c^\dagger_k,\bar{c}_{k'}\right\}}_{-\delta(k-k')}(-i\omega) \frac{e^{-ikr +ik'r'}}{4\pi \omega}\right)\\& = -i \delta(r-r'),\\
\left\{\bar{c}(t,r),\pi_{\bar{c}}(t,r')\right\} &=  \int dk dk'\left(\underbrace{\left\{\bar{c}_k,c^\dagger_{k'}\right\}}_{-\delta(k-k')}(i\omega) \frac{e^{ikr -ik'r'}}{4\pi \omega} - \underbrace{\left\{\bar{c}^\dagger_k,c_{k'}\right\}}_{-\delta(k-k')}(-i\omega) \frac{e^{-ikr +ik'r'}}{4\pi \omega}\right)\\& = -i \delta(r-r').
\end{aligned}
\end{gather}
Note that due to the anti-hermiticity of $\bar{c}$, these commutation relations are consistent with taking the Hermitian conjugate.

This assignment of the sign of the anticommutation relations might seem non-standard, but it is in fact required for a consistent quantization of the ghosts \cite{Kugo:1979gm}. The sign is fixed by the choice of left-differentiation for products of Grassmann variables, together with the ordering in the Hamiltonian below. This occurs for any system of Grassmann variables, physical or not.\footnote{The textbook canonical quantization of the Dirac action uses right-differentiation convention, which therefore requires the opposite sign choice of the anti-commutation relation. Changing conventions, causes a sign flip in the conjugate momenta and hence in the canonical anticommutation relations, and requires a different ordering in the equation for $H$.

As an alternative argument, the evolution equation for $c$ is:
\begin{equation}
i \dot{c} = \left[c,H\right],
\end{equation}
while the standard anticommutation relations would imply
\begin{equation}
\left[c,H\right] = \left[c,-\pi_c \dot{c}\right] = -\left\{c,\pi_c\right\}\dot{c} = - i \dot{c},
\end{equation}
leading to the wrong-sign evolution equation.}

The canonical Hamiltonian density is defined conventionally as $\mathcal{H} = \dot{\phi}\pi-\mathcal{L}$ (the ordering of the first term is fixed by the convention of left-differentiation) and in terms of the ghost fields reads
\begin{equation}
\mathcal{H} = - \dot{c}\dot{\bar{c}} + \dot{\bar{c}}\dot{c} + \partial^\mu \bar{c} \partial_\mu c = \dot{\bar{c}}\dot{c} + \partial_r \bar{c} \partial_r c.
\end{equation}
Plugging in the mode expansions and integrating the Hamiltonian density over a spatial Cauchy surface, one finds
\begin{equation}
H = - \int dk \omega \left(\bar{c}^\dagger_k c_k + c^\dagger_k \bar{c}_k\right) - 2\int dk \frac{1}{2} \omega,\label{3.9}
\end{equation}
which is manifestly Hermitian. The vacuum energy (the second term in \eqref{3.9}) is negative and cancels two positive bosonic contributions. Indeed, the vacuum energy of both the longitudinal and the timelike polarization is positive and is thus canceled by the ghost contributions, leaving only the $D-2$ transverse polarizations to contribute to the vacuum energy. From here on we focus only on the first contribution in \eqref{3.9}.

The Hilbert space is constructed by applying the creation operators $c^\dagger$ and $\bar{c}^\dagger$ to the ghost vacuum $\ket{M}$. Due to the anticommutation relations, the resulting space is of indefinite metric signature, as implied by the spin-statistics theorem.\footnote{Note that the above discussion concerning the construction of the ghost Hilbert space is valid both in the Minkowski frame, and in the Rindler frame in tortoise coordinates as the $1+1$ dimensional scalar Klein-Gordon action is independent of the metric conformal factor.}

Following the flow of section \ref{s gauge}, we construct the Unruh modes and write down the Bogoliubov transformation. The only important difference with the scalar QFT treatment is a change in sign in the oscillator expansion of $\bar{c}$ due to its anti-hermiticity, which results in the Bogoliubov transformations of the R-wedge annihilators $c^R$ in terms of the Unruh operators $c^1$ and $c^2$:
\begin{gather}
\begin{aligned}
c^R_k &= \frac{1}{\sqrt{2\sinh(\pi \omega)}}\left(e^{-\pi \omega/2}c^{2\dagger}_{-k} + e^{\pi \omega/2}c^{1}_{k}\right),\\
\bar{c}^R_k &= \frac{1}{\sqrt{2\sinh(\pi \omega)}}\left(-e^{-\pi \omega/2}\bar{c}^{2\dagger}_{-k} + e^{\pi \omega/2}\bar{c}^{1}_{k}\right),
\end{aligned}
\end{gather}
with similar formulas for the L-wedge oscillators.\footnote{A consistency check for the minus sign is provides by the fact that the Bogoliubov transformation preserve the oscillator anticommutation relations. Consistency also requires $\left\{c^L, c^R\right\} = 0$.} 
The Minkowski vacuum is defined to be annihilated by all positive frequency Unruh annihilators $c$ and $\bar{c}$ leading to the set of conditions
\begin{alignat}{3}
c^R_k \ket{M} &= e^{-\pi \omega}c^{L\dagger}_k\ket{M}, \quad\quad \bar{c}^R_k \left|M\right\rangle &&= -e^{-\pi \omega}\bar{c}^{L\dagger}_k\left|M\right\rangle, \\
c^L_k\left|M\right\rangle & = e^{-\pi \omega}c^{R\dagger}_k \left|M\right\rangle, \quad\quad \bar{c}^L_k\left|M\right\rangle &&= -e^{-\pi \omega}\bar{c}^{R\dagger}_k\left|M\right\rangle,
\end{alignat}
which completely determine the Minkowski ground state. Indeed, after integrating we find the squeezed state expression (ghost part) for the Minkowski vacuum:
\begin{align}
\nonumber&\left|M\right\rangle =N\prod_{k}\text{exp}\left[e^{-\pi \omega}\left(c_k^{R\dagger}\bar{c}_{-k}^{L\dagger} - \bar{c}_{-k}^{R\dagger}c_{k}^{L\dagger}\right)\right]\ket{R} \otimes \ket{L} \\
&= N\left( \ket{R} \otimes \ket{L} + e^{-\pi \omega}\left(c_k^{R\dagger}\bar{c}_{-k}^{L\dagger} - \bar{c}_{-k}^{R\dagger}c_{k}^{L\dagger}\right)\ket{R} \otimes \ket{L} + e^{-2\pi \omega}\left(c^{R\dagger}_k\bar{c}^{R\dagger}\bar{c}^{L\dagger}c^{L\dagger}\right) \ket{R} \otimes \ket{L}\right),\label{ghost vacuum}
\end{align}
with the normalization $N = \frac{1}{1-e^{-2\pi \omega}}$.

Using the explicit form of the Bogoliubov transformations, the expectation value of the R-wedge ghost modes in the Minkowski ghost vacuum is found as
\begin{align}
\nonumber\left\langle M\right|\bar{c}_k^{R\dagger} c_k^{R} + c_k^{R\dagger} \bar{c}_k^{R}\left|M\right\rangle &= \frac{-1}{e^{2\pi \omega}-1}\left\langle M\right|\left\{\bar{c}_{-k}^{2},c_{-k}^{2\dagger}\right\} + \left\{c_{-k}^{2},\bar{c}_{-k}^{2\dagger}\right\}\left|M\right\rangle \\
&= \frac{V}{2\pi}\frac{2}{e^{2\pi \omega}-1},
\end{align}
such that the expectation value of the ghost Hamiltonian reads
\begin{equation}
\bra{M}H\ket{M}=-2\frac{V}{2\pi}\int dk  \frac{\omega}{e^{2\pi \omega}-1}.
\end{equation}

\section{Bulk Hilbert Space in $d+1$ Dimensions}
\label{a:bulk}
\label{sect:details}
With the scalar mode $\phi$ defined in \eqref{elmode}, a set of orthonormal modes (for $\omega\neq 0$) is found as:
\begin{align}
	A_\mu^{(0)} &= \frac{1}{k}(-i\omega \phi, \partial_\rho \phi, 0) \equiv A_\mu^{(G)} -A_\mu^{(k)},\\
	A_\mu^{(1)} &= \frac{1}{k}(\rho\partial_\rho \phi, -i\omega/\rho \phi, 0), \\
	A_\mu^{(k)} &= \frac{1}{k}(0,0, ik_i \phi), \\
	A_\mu^{(a)} &= (0,0, in^a_i \phi),
\end{align}
with $\nabla^\mu A^{(k)}_\mu = -k\phi = - \nabla^\mu A^{(0)}_\mu$, giving:
\begin{align}
	A_i &= \sum_{\omega,\mathbf{k}} i \frac{k_i}{k} \phi \alpha^{(k)}_{\omega\mathbf{k}} + in^a_i \phi \alpha^{(a)}_{\omega\mathbf{k}} + (hc),\\
	A_t &= \sum_{\omega,\mathbf{k}}- \frac{i\omega}{k}\phi \alpha^{(0)}_{\omega\mathbf{k}} + \frac{1}{k}\rho \partial_\rho\phi \alpha^{(1)}_{\omega\mathbf{k}}+ (hc),\\
	A_\rho &= \sum_{\omega,\mathbf{k}} \frac{1}{k} \partial_\rho \phi \alpha^{(0)}_{\omega\mathbf{k}} - \frac{i\omega}{k\rho}\phi \alpha^{(1)}_{\omega\mathbf{k}}+ (hc),
\end{align}
and
\begin{align}
	\Pi^i &= \sum_{\omega,\mathbf{k}}- \omega \frac{k_i}{k\rho} \phi \alpha^{(0)}_{\omega\mathbf{k}} - i \frac{k_i}{k} \partial_\rho \phi \alpha^{(1)}_{\omega\mathbf{k}} + \frac{k_i}{k \rho} \omega \phi \alpha^{(k)}_{\omega\mathbf{k}} + \frac{n^a_i}{\rho} \omega \phi \alpha^{(a)}_{\omega\mathbf{k}} + (hc),\\
	\Pi^t &= \sum_{\omega,\mathbf{k}}\frac{1}{\rho} k \phi (\alpha^{(0)}_{\omega\mathbf{k}} - \alpha^{(k)}_{\omega\mathbf{k}})+ (hc), \\
	\Pi^\rho &= \sum_{\omega,\mathbf{k}}-k \phi \alpha^{(1)}_{\omega\mathbf{k}}  + (hc).
\end{align}
One proves that these satisfy the correct canonical commutation relations if one imposes:
\begin{align}
	\bigl[\alpha^{(0)}_{\omega\mathbf{k}}, \alpha^{(0)\dagger}_{\omega'\mathbf{k}'} \bigr] &= -\delta_{\omega\omega'}\delta(\mathbf{k}-\mathbf{k}'), \\
	\bigl[\alpha^{(1)}_{\omega\mathbf{k}}, \alpha^{(1)\dagger}_{\omega'\mathbf{k}'} \bigr] &= \delta_{\omega\omega'}\delta(\mathbf{k}-\mathbf{k}'), \\
	\bigl[\alpha^{(k)}_{\omega\mathbf{k}}, \alpha^{(k)\dagger}_{\omega'\mathbf{k}'} \bigr] &= \delta_{\omega\omega'}\delta(\mathbf{k}-\mathbf{k}'), \\
	\bigl[\alpha^{(a)}_{\omega\mathbf{k}}, \alpha^{(b)\dagger}_{\omega'\mathbf{k}'} \bigr] &= \delta_{ab}\delta_{\omega\omega'}\delta(\mathbf{k}-\mathbf{k}'),
\end{align}
and all other combinations vanish. One also defines $\alpha^{(0)}_{\omega\mathbf{k}} = \alpha^{(G)}_{\omega\mathbf{k}} - \alpha^{(k)}_{\omega\mathbf{k}}$.

The magnetic field components are:
\begin{align}
	F_{\rho i} &= \sum_{\omega,\mathbf{k}} i\frac{k_i}{k}\partial_\rho \phi (\alpha^{(k)}_{\omega\mathbf{k}}-\alpha^{(0)}_{\omega\mathbf{k}}) + i n^a_i \partial_\rho \phi \alpha^{(a)}_{\omega\mathbf{k}} +\frac{\omega k_i}{k\rho} \phi \alpha^{(1)}_{\omega\mathbf{k}}, \\
	F_{ij} &= \sum_{\omega,\mathbf{k}} (n_i^a k_j-n_j^a k_i) \phi \alpha^{(a)}_{\omega\mathbf{k}}.
\end{align}
After applying the $n_\rho \frac{\delta \mathcal{L}}{\delta \partial_\rho A_\mu}\rvert_{\partial \mathcal{M}}=0$ boundary condition, the expansion of the quantum field in modes is
\begin{equation}
\small
	A_\mu(t,\rho,\mathbf{x})=\sum_\mathbf{k}\left(\sum_{\omega\in\sigma_D}\hat{\alpha}^{(1)}_{\omega\mathbf{k}}A^{(1)}_{\mu,\omega\mathbf{k}} + \sum_{\omega\in\sigma_N}\left(\sum_a\hat{\alpha}^{(a)}_{\omega\mathbf{k}}A^{(a)}_{\mu,\omega\mathbf{k}} + \sum_L\hat{\alpha}^{(k)}_{\omega\mathbf{k}}A^{(k)}_{\mu,\omega\mathbf{k}} + \sum_0\hat{\alpha}^{(0)}_{\omega\mathbf{k}}A^{(0)}_{\mu,\omega\mathbf{k}}\right)\right) + (hc) \label{QF}.
\end{equation}
The boundary conditions impose Dirichlet boundary conditions $\phi\rvert_{\partial \mathcal{M}}=0$ on the modes $A^I_{\mu,\omega\mathbf{k}}$ and Neumann boundary conditions $\partial_\rho\phi\rvert_{\partial \mathcal{M}}=0$ on all other sets of modes. 

\section{Dirichlet vs Neumann Scalar Partition function}
\label{a:d vs n}
We discuss the ratio of the partition functions of a Dirichlet and a Neumann scalar field in Rindler.
\subsection*{Direct Calculation of Functional Determinants}
To compute the partition function of a scalar field, we transform the scalar Lagrangian on the thermal manifold to tortoise coordinates. One needs the functional determinant associated to the eigenvalue problem
\begin{equation}
\left(-\partial_\tau^2-\partial_r^2+(k^2+m^2)e^{2r}\right)\psi(t,r) = \lambda \psi(t,r). \label{eig}
\end{equation}
We set $m=0$ here, but its dependence is readily restored as usual. The following are functional determinants of operators defined on the interval $r\in \left[\ln\epsilon,\ln R\right]$, with either Neumann (N) or Dirichlet (D) boundary conditions on the end points. The notation $\det (O) ^{AB}$ refers to the functional determinant of the operator $O$ with $A$ boundary conditions at $r=\ln\epsilon$ and $B$ boundary conditions at $r=\ln R$.
\\~\\
First setting $\partial_t=0$, the ratio of determinants with different boundary conditions at the horizon $r=\ln \epsilon$ is given by:
\begin{equation}
\frac{\det(-\partial_r^2+k^2e^{2r})^{ND}}{\det(-\partial_r^2+k^2e^{2r})^{DD}} = \frac{\psi_{(1)}(\ln R)}{\psi_{(2)}(\ln R)}, \label{d5}
\end{equation}
which is evaluated as the ratio of two solutions, $\psi_{(1)}$ and $\psi_{(2)}$, to the initial value problem $(-\partial_r^2+k^2e^{2r})\psi_{(i)}(r) = 0$, defined by
\begin{align}
\psi_{(1)}(\ln \epsilon) &= 1, \quad \psi'_{(1)}(\ln \epsilon) = 0, \nonumber \\
\psi_{(2)}(\ln \epsilon) &= 0, \quad \psi'_{(2)}(\ln \epsilon) = 1.
\end{align}
This can be interpreted as a generalization of the Gelfand-Yaglom theorem (see e.g. \cite{Dunne:2007rt}) to determinants of the same operator with different boundary conditions.\footnote{This formula can be proven by writing 
\begin{equation}
\frac{\det(-\partial_x^2+V(x))^{ND}}{\det(-\partial_x^2+V(x))^{DD}} = \frac{\det(-\partial_x^2+V(x))^{ND}}{\det(-\partial_x^2)^{ND}}\frac{\det(-\partial_x^2)^{DD}}{\det(-\partial_x^2+V(x))^{DD}}\frac{\det(-\partial_x^2)^{ND}}{\det(-\partial_x^2)^{DD}},
\end{equation}
where the first two factors are directly evaluated using the standard Gelfand-Yaglom theorem. The last factor is evaluated explictly as
\begin{equation}
\frac{\det(-\partial_x^2)^{ND}}{\det(-\partial_x^2)^{DD}} = \frac{\pi }{2\ln R/\epsilon} \prod_{n\geq 1} \left(1-\frac{1}{(2n)^2}\right)= \frac{1}{\ln R/\epsilon} = \frac{\psi^{V=0}_{(1)}(\ln R)}{\psi^{V=0}_{(2)}(\ln R)},\label{d1}
\end{equation}
and hence is the same as a naive application of GY again.} 
Explicitly:
\begin{align}
\psi_{(1)}(\ln R) &= -k\epsilon \left(K'_0(k\epsilon)I_0(kR)-I'_0(k\epsilon)K_0(kR)\right), \\
\psi_{(2)}(\ln R) &= K_0(k\epsilon)I_0(kR)-I_0(k\epsilon)K_0(kR).
\end{align}
In the limit $k\epsilon\to 0$ with $R$ fixed, \eqref{d5} thus becomes
\begin{equation}
\frac{\det(-\partial_r^2+k^2e^{2r})^{ND}}{\det(-\partial_r^2+k^2e^{2r})^{DD}}=\ln 2/k\epsilon.
\end{equation}
Generalizing to include discrete momentum $n\in \mathbb{Z}$ along the thermal circle, we find analogously
\begin{equation}
\frac{\det(-\partial_r^2+k^2e^{2r}+\left(\frac{2\pi n}{\beta}\right)^2)^{ND}}{\det(-\partial_r^2+k^2e^{2r}+\left(\frac{2\pi n}{\beta}\right)^2)^{DD}}=\frac{\delta_{n,0}}{\ln 2/k\epsilon}+\frac{2\pi |n|}{\beta},
\end{equation}
or using zeta-regularization for the infinite product $\prod_{n\in\mathbb{Z}_0}\frac{2\pi |n|}{\beta}=\beta$:
\begin{equation}
\prod_{n\in\mathbb{Z}}\frac{\det(-\partial_r^2+k^2e^{2r}+\left(\frac{2\pi n}{\beta}\right)^2)^{ND}}{\det(-\partial_r^2+k^2e^{2r}+\left(\frac{2\pi n}{\beta}\right)^2)^{DD}}=\frac{\beta}{\ln 2/k\epsilon}.
\end{equation}
This results in
\begin{equation}
\frac{Z_D}{Z_N}=\prod_k\left(\frac{\beta}{\ln 2/k\epsilon}\right)^{\frac{1}{2}},
\end{equation}
which is the result mentioned in \eqref{DvsN}.\footnote{Note that in fact this is only the exact result for $k\epsilon\ll 1$, whereas the partition function in general gets contributions from arbitrarily high $k$. The same is true for the edge partition function \eqref{6.28}. The cancellation of the explicit $\epsilon$ dependence in the total partition function \eqref{6.29} however, holds for arbitrary $k$.}

\subsection*{Radial Path Integration}

An alternative perspective on this difference was given by Donnelly and Wall \cite{Donnelly:2015hxa}. The thermal partition function is a trace on a cylinder in the tortoise near-horizon region. Generally, we can describe the thermal trace via channel duality as a matrix element between boundary states. We work here in the very near-horizon region, where we can also neglect the exponential potential in \eqref{eig}. This identifies the (new) IR cut-off $R^* \sim \frac{1}{k}$. For the zero-mode contribution ($\partial_t =0$), we then find, setting $\eta = \ln R^*/\epsilon$:
\begin{equation}
Z = \text{Tr} e^{-\beta \eta T_{00}} = \left\langle \psi_f \right| e^{-\eta \beta T_{rr}} \left|\psi_i \right\rangle,
\end{equation}
between suitable boundary states. The zero-mode sector is found by considering $\partial_r^2 \phi = 0$, solved by $\phi = p r + q$, for conjugate operators $p$ and $q$. One propagates the state in between in the $r$-direction using $T_{rr} = p^2/2$. We leave $\psi_f$ arbitrary and consider the ratio between Dirichlet and Neumann boundary states:
\begin{equation}
\label{toco}
Z_{D/N} = \frac{\left\langle \psi_f \right| e^{-\eta \beta T_{rr}} \left|\psi_D \right\rangle}{\left\langle \psi_f \right| e^{-\eta \beta T_{rr}} \left|\psi_N \right\rangle}.
\end{equation}
Using that $\psi_D(p) = \sqrt{\frac{1}{2\pi}}$ and $\psi_N(p) = \delta(p)$, one finds
\begin{equation}
e^{-\eta \beta p^2} \psi_D(p) = \frac{1}{\sqrt{\eta \beta}} \delta(p), \quad e^{-\eta \beta p^2} \psi_N(p) = \delta(p),
\end{equation}
and finally:
\begin{equation}
\frac{Z_D}{Z_N} \to \frac{1}{\sqrt{\eta \beta}}=\left(\frac{\beta}{\ln R^*/\epsilon}\right)^{-\frac{1}{2}},
\end{equation}
which agrees with the previous computation, provided we replace $R^* \sim 1/k$.
Alternatively, to compute \eqref{toco}, one can perform a Gaussian saddle point integration as:
\begin{align}
\text{Dirichlet:} &\quad \int_{-\infty}^{+\infty} dp \psi_f(p)^* e^{-\frac{\beta \eta}{2} p^2} \frac{1}{\sqrt{2\pi}} \approx \psi_f^*(0) \sqrt{\frac{1}{\beta \eta}}, \\
\text{Neumann:} &\quad \int_{-\infty}^{+\infty} dp \psi_f(p)^* e^{-\frac{\beta \eta}{2} p^2} \delta(p) \approx \psi_f^*(0).
\end{align}


\end{document}